\documentclass[12pt,preprint]{elsarticle}
\usepackage{hyperref,lineno,natbib,amsmath,amssymb,graphicx,multirow,algorithmic,color,microtype,fullpage,overpic}
\usepackage[skip=-2pt]{caption}
\usepackage{subcaption}
\usepackage{xcolor}

\usepackage{textcomp}

\journal{Journal of Computational Physics}
\definecolor{dblue}{rgb}{0.0,0.0,0.5}
\definecolor{dmag}{rgb}{0.831,0.165,1.0}
\definecolor{dred}{rgb}{1.0,0,0}
\definecolor{jade}{rgb}{0.1333,0.5647,0.4784}
\definecolor{lblue}{rgb}{0,0.6745,1.0}
\definecolor{pmag}{rgb}{0.580,0.129,0.572}
\definecolor{pgry}{rgb}{0.572,0.572,0.572}

\usepackage{hyperref}

\definecolor{webgreen}{rgb}{0,.35,0}
\definecolor{webbrown}{rgb}{.6,0,0}
\definecolor{RoyalBlue}{rgb}{0,0,0.9}
\definecolor{purp}{rgb}{0.4,0.2,0.8}
\definecolor{mywhite}{rgb}{1.0,1.0,1.0}

\hypersetup{
   colorlinks=true, linktocpage=true, pdfstartpage=3, pdfstartview=FitV,
   breaklinks=true, pdfpagemode=UseNone, pageanchor=true, pdfpagemode=UseOutlines,
   plainpages=false, bookmarksnumbered, bookmarksopen=true, bookmarksopenlevel=1,
   hypertexnames=true, pdfhighlight=/O,
   urlcolor=webbrown, linkcolor=RoyalBlue, citecolor=webgreen,
   pdfauthor={Nicholas M. Boffi and Chris H. Rycroft},
   pdfsubject={Parallel three-dimensional simulations of quasi-static elastoplastic solids},
   pdfkeywords={fluid mechanics, projection method, plasticity, elastoplasticity},
   pdfcreator={pdfLaTeX},
   pdfproducer={LaTeX with hyperref}
}

\newcommand{\bigO}{\mathcal{O}}

\newcommand{\half}{\frac{1}{2}}

\newcommand{\p}{\partial}
\newcommand{\Tr}{\text{Tr}}
\renewcommand{\vec}[1]{\mathbf{#1}}
\newcommand{\ten}[1]{\mathbf{#1}}
\newcommand{\tC}{\ten{C}}

\newcommand{\vx}{\vec{x}}
\newcommand{\vv}{\vec{v}}
\newcommand{\bPhi}{\boldsymbol{\Phi}}
\newcommand{\prx}[1]{\frac{\p #1}{\p x}}

\newcommand{\prt}[1]{\frac{\p #1}{\p t}}
\newcommand{\drt}[1]{\frac{d #1}{d t}}
\newcommand{\dt}{\Delta t}
\newcommand{\dx}{\Delta x}

\newcommand{\bx}{\mathbf{x}}
\newcommand{\bX}{\mathbf{X}}

\newcommand{\tD}{\ten{D}}

\newcommand{\Del}{\tD^\text{el}}
\newcommand{\Dpl}{\tD^\text{pl}}

\newcommand{\bsig}{\boldsymbol\sigma}

\newcommand{\sC}{\mathcal{C}}
\newcommand{\sR}{\mathcal{R}}

\newcommand{\bL}{\mathbf{L}}
\newcommand{\Trans}{\mathsf{T}}

\DeclareMathOperator{\tr}{tr}

\begin{document}
\begin{frontmatter}
  \title{Parallel three-dimensional simulations of quasi-static elastoplastic solids}
\author[SEAS]{Nicholas M. Boffi}
\ead{boffi@g.harvard.edu}
\author[SEAS,LBL]{Chris H. Rycroft}
\ead{chr@seas.harvard.edu}

\address[SEAS]{John A. Paulson School of Engineering and Applied Sciences, Harvard University, Cambridge, MA 02139}
\address[LBL]{Computational Research Division, Lawrence Berkeley Laboratory,
Berkeley, CA 94720}

\begin{abstract}
    Hypo-elastoplasticity is a flexible framework for modeling the mechanics of many hard materials under small elastic deformation and large plastic deformation. Under typical loading rates, most laboratory tests of these materials happen in the quasi-static limit, but there are few existing numerical methods tailor-made for this physical regime. In this work, we extend to three dimensions a recent projection method for simulating quasi-static hypo-elastoplastic materials. The method is based on a mathematical correspondence to the incompressible Navier–Stokes equations, where the projection method of Chorin (1968) is an established numerical technique. We develop and utilize a three-dimensional parallel geometric multigrid solver employed to solve a linear system for the quasi-static projection. Our method is tested through simulation of three-dimensional shear band nucleation and growth, a precursor to failure in many materials. As an example system, we employ a physical model of a bulk metallic glass based on the shear transformation zone theory, but the method can be applied to any elastoplasticity model. We consider several examples of three-dimensional shear banding, and examine shear band formation in physically realistic materials with heterogeneous initial conditions under both simple shear deformation and boundary conditions inspired by friction welding.
\end{abstract}
\end{frontmatter}

\section{Introduction}
\label{sec:intro}
\newlength{\subpanelwid}
\setlength{\subpanelwid}{0.46\textwidth}
Modeling the failure of materials is a fundamental problem in modern engineering and science. At small loads, the strain in a material is typically smooth, but as the loading is increased, the strain may become highly localized, which ultimately leads to failure~\cite{jirasek02a}. This localization of strain may be driven by geometrical effects such as necking~\cite{hutchinson74,ghosh77,tvergaard93,guo07}, or by instabilities in the material response, such as due to plastic yielding. In the latter case, positive feedback causes regions that have already plastically deformed to yield further, often resulting in a localized plane of strain called a shear band~\cite{rudnicki75,hutchinson81,harren88,conner04,bei06}. Shear bands have been studied analytically~\cite{asaro77,steif82}, but it remains challenging to model their formation in inhomogeneous, three-dimensional material samples, and numerical tools must be employed. Numerically modeling materials can be performed using the framework of elastoplasticity, and herein we develop a three-dimensional extension of a two-dimensional algorithm~\cite{rycroft15} suitable for simulating deformation in hard elastoplastic materials. As one application, we demonstrate the ability of the method to capture fine-scale features of shear banding in the deformation of amorphous materials.

For stress levels below the material yield stress, an elastoplastic material deforms purely elastically and returns to its undeformed state upon removal of the load. When the yield stress is reached, the material begins to deform plastically, leading to permanent, irreversible deformation that persists beyond load removal~\cite{gurtin10}. Elastoplasticity is complex, and it admits a number of mathematical descriptions~\cite{xiao06,rycroft15}, each of which amounts to a specification of the interaction of the elastic and plastic components of deformation at a microscopic level~\cite{prager60,reina14}. In this paper, we utilize the hypo-elastoplasticity model~\cite{truesdell55}, in which the Eulerian rate of deformation tensor is decomposed additively into elastic and plastic parts, $\tD = \Del + \Dpl$~\cite{hill58}. Hypo-elastoplasticity has some drawbacks: for example, it is generally only used for linear elastic deformation, because it is difficult to represent nonlinear elasticity through $\Del$ alone~\cite{dienes79, nagtegaal81}. The formalism is also based on the material velocity rather than displacement field, and hence can lead to accumulation of numerical integration errors~\cite{hughes80, eterovic90}. Despite these flaws, hypo-elastoplasticity is well-suited to problems with small elastic deformation and large plastic deformation, which makes it particularly appropriate for studying failure and shear banding in hard materials.

On the other hand, the hypo-elastoplastic formulation has several numerical advantages. Since it is based on the Eulerian rate of deformation tensor, it is well-suited for a fixed-grid framework. Fixed grids have simpler topologies than their Lagrangian counterparts, and are easier to program and parallelize. This is particularly important in three dimensions, where the computational expense mandates parallelization, and where the implementation difficulty increases relative to lower dimensional simulations. Fixed-grid methods are also the methods of choice for fluid simulation~\cite{bell89,almgren96,sussman99}, and they allow a wider range of numerical linear algebra techniques to be used, such as the geometric multigrid method~\cite{Briggs2000,demmel}.

The additive decomposition of $\tD$, coupled with the linear-elastic constitutive relation and a continuum formulation of Newton's second law, leads to a closed hyperbolic system of partial differential equations (PDEs) for the material velocity, stress, and variables intrinsic to the plasticity model. To properly resolve elastic waves, the timestep $\Delta t$ for an explicit numerical scheme is restricted by the well-known Courant--Friedrichs--Lewy (CFL) condition~\cite{courant67}. The CFL condition states that $\Delta t$ must obey a constraint for numerical stability: \smash{$\Delta t \leq \frac{h}{c_e}$}, where $c_e$ is a typical elastic wave speed in the medium and $h$ is the grid spacing.

In metals and other materials of interest, elastic waves can travel at kilometers per second, while in many practical scenarios the timescale of loading is much longer than the elastic wave travel time~\cite{bing-2005}. The CFL condition thus poses a prohibitive limit on the timestep for probing realistic timescales and system sizes, and the development of alternative simulation approaches which avoid resolving elastic waves is necessary. By looking at the limit of long times and small velocities, one can show that Newton's second law can be replaced by a constraint that the stresses must remain in quasi-static equilibrium. However, these equations are no longer a hyperbolic system, since the ability to time-integrate the velocity field is lost.

Recently, Rycroft \textit{et al.}~\cite{rycroft15} demonstrated a mathematical analogy between quasi-static elastoplasticity and the incompressible Navier--Stokes equations for fluid flow. The incompressible Navier--Stokes equations combine an equation for the velocity -- dependent on the fluid pressure -- with a requirement that the velocity must be divergence free. Much like in the case of quasi-static elastoplasticity, the divergence-free requirement on the velocity field is obtained as a limit of a PDE for the pressure. A well-known algorithm in this setting is the projection method of Chorin~\cite{chorin67, chorin68}. In Chorin's method, an intermediate step is taken for the velocity field, but this intermediate velocity does not obey the incompressibility constraint. An elliptic problem is solved that simultaneously enforces the incompressibility constraint and enables computation of the pressure. Rycroft \textit{et al.}~translated Chorin's projection method to quasi-static elastoplasticity by exploiting the mathematical analogy between the two sets of limiting equations. The new method was quantitatively tested on some simple two-dimensional systems using a multi-threaded single-processor implementation. It was shown to be practical and efficient, enabling the use of timesteps that are orders of magnitude larger than those required by the CFL condition.

Analogies between elastoplastic deformation and incompressible flow have been noted previously, and have been used to employ Vanka-type smoothers for solid mechanics problems~\cite{vanka, vanka2}. However, there are several important differences between this prior work and our own: we employ a formalism based on the material velocity field, while previous approaches solve for the displacement field. Similarly, to our knowledge, our work is the first projection-type method for elastoplasticity based on Chorin's method.

In a similar vein, our method has strong similarity to numerical algorithms for low-Mach flows. Such methods avoid the resolution of acoustic waves yet retain compressible effects in the fluid, and much like incompressible fluid dynamics, result in an elliptic equation for the pressure. They have been used to study Type Ia supernovae~\cite{almgren_super} and nuclear combustion~\cite{bell_nuclear}, and have their roots in an earlier approach due to Majda and Sethian~\cite{majda_seth}. Analytical justification for the limiting procedures taken has been provided by Klainerman and Majda~\cite{klein_maj1, klein_maj2}. It is possible that similar approaches could be developed for alternative limits of the equations of hypo-elastoplasticity.

In this paper, we extend the quasi-static projection method to three dimensions. Due to the greatly increased problem size, we develop a distributed parallel implementation, which creates algorithmic challenges. In particular, the projection method requires that we solve a coupled elliptic equation for the components of velocity. We develop a custom parallel geometric multigrid code to solve the resulting linear system.
As a physical testbed for our methodology, we employ an athermal formulation of the shear transformation zone (STZ) theory of Falk, Langer, Bouchbinder and coworkers as a plasticity model for a bulk metallic glass~\cite{falk98,bouchbinder07,langer08,bouchbinder09}. Metallic glasses naturally lend themselves to study through the hypo-elastoplasticity framework, as their elastic deformation is generally small and well-described by a linear theory, yet they can exhibit significant plastic deformation~\cite{bmg-struc}. Their elastic moduli are typically on the order of 10--100~GPa, and hence experimental loading conditions often place samples in the quasi-static regime~\cite{huf-def}. They also present interesting and poorly understood fundamental physics~\cite{manning07, manning09, sun-2013, antonaglia-2014}, which can be difficult to probe in dimensions greater than one without the numerical methods presented here.

Simulation studies of bulk metallic glasses have been essential to our understanding of their properties. The development of the original STZ theory was guided by observations of molecular dynamics simulations~\cite{falk_langer_rev}. Simulations of the necking instability in a bar under uniaxial tension using the STZ plasticity model highlight the interplay between elastic and plastic deformation~\cite{eastgate03, rycroft12, moriel18}. The physical mechanisms of fracture in BMGs were explored using the two-dimensional projection method~\cite{rycroft12b}, subsequently allowing the fracture toughness of BMGs to be predicted across a wide range of experimental conditions~\cite{vasoya16}. Later experimental measurements due to Ketkaew \textit{et al.}\ demonstrated that these simulation-based predictions were quantitatively correct~\cite{schroers2018}. Indeed, testable predictions for complex amorphous systems such as BMGs are rare, and the development of efficient numerical methods such as the ones presented here provide a way to generate them and to guide future experimental inquiry.

The STZ theory is a useful test case for our method both physically and numerically, but the numerical methodology is general and can be used for many plasticity models within the hypo-elastoplasticity framework. These could include free-volume based models of BMGs~\cite{argon79}, plasticity models based on the random first-order transition theory of the glass transition~\cite{Wisitsorasak1287}, hypo-elastic materials~\cite{hypo-elas1, hypo-elas2, hypo-elas3}, geophysical models~\cite{hajarolasvadi17,ma18}, and rate-independent plasticity models~\cite{rate-ind1, rate-ind2, rate-ind3, rate-ind4}. We also emphasize that Chorin's projection method represents a first step towards more sophisticated projection-based algorithms such as gauge methods~\cite{saye_dg1, saye_dg2, saye_gauge} and pressure-Poisson methods~\cite{brown01, min06}, and that we have laid the groundwork here to generalize these algorithms to the case of hypo-elastoplasticity.

Under loading, BMGs exhibit shear bands~\cite{zhang-2006,greer-2013}, which rapidly lead to material failure~\cite{bing-2005, schuh-2007,maass15} and are one of the primary limitations in employing BMGs in applications~\cite{maas-2014}. Analytical work probing shear bands in amorphous materials is difficult, particularly in two or three dimensions, which highlights a need for computational investigations. The development of our method enables the study of shear-banding in three dimensions at large scale and high resolution without excessive computational expense. Our simulations demonstrate that this scale and resolution is indeed necessary, and expose interesting fine-scale and uniquely three-dimensional features of shear banding. Our methodology opens the door to future studies probing the shape, structure, and topology of shear bands, as well as the mechanism and statistical properties of their formation.

The structure of this paper is as follows. In Section~\ref{sec:proj}, we describe the relation between Chorin's projection method and the projection algorithm for hypo-elastoplasticity employed here. In Section~\ref{sec:numerics}, we describe a finite-difference implementation of our projection method, and describe a forward-Euler based explicit method for solving the hypo-elastoplastic equations in the non-quasi-static limit. We also discuss our parallel multigrid implementation used for the stress projection. In Section~\ref{sec:examples}, we demonstrate convergence between the explicit and projection methods in a regime in which the two are expected to produce similar results, and study several interesting examples of shear banding dynamics in a metallic glass.

\section{Projection methods for fluid dynamics and hypo-elastoplasticity}
\label{sec:proj}
\subsection{Hypo-elastoplasticity}
We consider an elastoplastic material with Cauchy stress tensor $\bsig(\vx, t)$ and velocity field $\vv(\vx, t)$ at a position $\bx$ and time $t$. The total rate of deformation tensor $\tD$ is defined as the symmetric part of the velocity gradient, \smash{$\tD = \frac{1}{2}(\bL + \bL^\Trans)$} with $\bL = \nabla \vv$. For any field $f(\vx, t)$, we define the advective time derivative by \smash{$\drt{f} = \prt{f} + \left(\vv\cdot\nabla\right){f}$}. The fundamental assumption of hypo-elastoplasticity is that the rate of deformation tensor can be additively decomposed into a sum of elastic and plastic parts, $\tD = \Del + \Dpl$.

For stiff elastoplastic materials with small elastic deformation, the linear elastic constitutive law provides an accurate description,
\begin{equation}
    \frac{\mathcal{D} \bsig}{\mathcal{D} t} = \tC : \Del = \tC : \left(\tD - \Dpl\right).
    \label{eqn:lnr_elas}
\end{equation}
In Eq.~\ref{eqn:lnr_elas}, $\tC$ is the fourth-rank stiffness tensor, taken to be homogeneous and isotropic. With Lam\'e's first parameter $\lambda$ and shear modulus $\mu$, the components of $\tC$ are given by $C_{ijkl} = \lambda\delta_{ij}\delta_{kl} + \mu\left(\delta_{ik}\delta_{jl} + \delta_{il}\delta_{jk}\right)$~\cite{lubliner08}. The time derivative $\frac{\mathcal{D}\bsig}{\mathcal{D} t} = \drt{\bsig} - \bL \bsig - \bsig \bL^T + \Tr(\bL)\bsig$ denotes the Truesdell objective stress rate.

From Newton's second law, the material velocity obeys the equation
\begin{equation}
    \rho \drt{\vv} = \nabla\cdot\bsig
    \label{eqn:nwtn_law}
\end{equation}
where $\rho$ denotes the material density. Equations~\ref{eqn:lnr_elas}~\&~\ref{eqn:nwtn_law} form a hyperbolic system of equations for the stress and velocity fields, which can be solved explicitly using standard finite-difference simulation methods. This hyperbolic system will resolve elastic waves, and so the timestep $\Delta t$ and grid spacing $\Delta x$ must satisfy the CFL condition $\Delta t \leq \Delta x /c_e$ for numerical stability, where $c_e$ is an elastic wave speed. In materials such as metals and metallic glasses, elastic waves travel on the order of kilometers per second. Spatial discretizations capable of resolving fine-scale features of interesting physical phenomena in these materials can be as small as micrometers. For $\dx = 1 \text{~\textmu{}m}$ and $c_e = 1\text{~km/s}$, the CFL condition requires $\dt \leq 1\text{~ns}$, an extreme restriction for phenomena that occur on realistic timescales of hours or days.

\subsection{Quasi-static hypo-elastoplasticity}
We consider a scenario in which plastic deformation occurs on a timescale much greater than the time for waves to propagate through the material. In this setting, macroscopic plastic deformation takes place due to the accumulation of small velocity gradients over long times. The details of a limiting procedure describing this physical regime were performed in previous two-dimensional work~\cite{rycroft15} and will not be reproduced here.

In this quasi-static limit, the equation for the velocity in Eq.~\ref{eqn:nwtn_law} can be approximately replaced by a constraint on the stress
\begin{equation}
    \nabla \cdot \bsig \approx \mathbf{0}.
    \label{eqn:qs}
\end{equation}
Equation~\ref{eqn:qs} dictates that each infinitesimal material element remains in approximate quasi-static equilibrium, and is thus referred to as the quasi-static constraint. The evolution equation for the stress in Eq.~\ref{eqn:lnr_elas} is unaffected by the limiting procedure, and hence Eq.~\ref{eqn:lnr_elas} must be solved subject to the global constraint Eq.~\ref{eqn:qs} to obtain solutions valid in this limit.

At this stage, it is unclear how to do so. The velocity $\vv$ appears in Eq.~\ref{eqn:lnr_elas} through $\tD$, but there is no longer an equation that can be integrated explicitly to solve for it. It is also not guaranteed that solutions of Eq.~\ref{eqn:lnr_elas} subject to the constraint in Eq.~\ref{eqn:qs} will agree with solutions of Eq.~\ref{eqn:lnr_elas} and Eq.~\ref{eqn:nwtn_law}.

\subsection{Incompressible fluid dynamics}
We now demonstrate an analogy between the computational issues presented in the previous section and those encountered in incompressible fluid dynamics. Consider a fluid with velocity $\vv$, pressure $p$, and density $\rho$. The fluid velocity field obeys the Navier--Stokes equation,
\begin{equation}
    \drt{\vv} = -\nabla p + \nu \nabla^2 \vv.
    \label{eqn:navier}
\end{equation}
If the fluid is compressible, then its density satisfies
\begin{equation}
    \drt{\rho} = -\rho \left(\nabla \cdot \vv\right),
    \label{eqn:rho}
\end{equation}
along with an equation of state linking the fluid density to the fluid pressure. Using an explicit scheme to solve the hyperbolic system in Eqs.~\ref{eqn:navier} \& \ref{eqn:rho} will resolve sound waves in the fluid, which leads to timestep restrictions from the CFL condition. In the long-time limit, Eq.~\ref{eqn:rho} is traded for the \textit{incompressibility constraint} on the velocity field,
\begin{equation}
    \nabla \cdot \vv = 0.
    \label{eqn:incomp}
\end{equation}
This limit reduces the coupled partial differential equations for the pressure and velocity to a single constrained equation for the velocity. The pressure is present in the equation for the velocity, though its evolution equation has been exchanged for the incompressibility constraint; this is much like the quasi-static limit of hypo-elastoplasticity described in the preceding section.

Chorin~\cite{chorin67,chorin68} developed a numerical method for this system of equations that involves the use of an orthogonal projection, spurring significant research into related algorithms~\cite{brown01, min06}. Such projection methods proceed via a two-step procedure, where an intermediate velocity $\vv^*$ is first computed which does not obey the incompressibility constraint. $\vv^*$ is then orthogonally projected onto the manifold of divergence-free solutions through the solution of an elliptic problem for an auxiliary field related to the pressure. The process of projection simultaneously enforces the constraint and enables computation of the pressure field.

\begin{figure}
    \begin{overpic}[width=\textwidth]{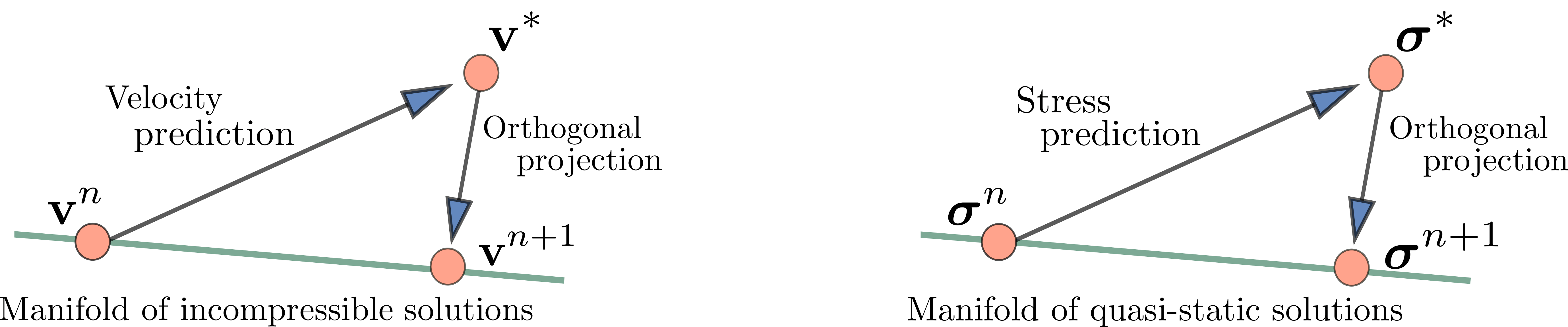}
    \put(0, 20){(a)}
    \put(58, 20){(b)}
    \end{overpic}
    \vspace{5mm}
    \caption{The projection-based timestepping scheme for (a) the velocity field in incompressible fluid dynamics and (b) the stress tensor in quasi-static hypo-elastoplasticity. In both cases, an intermediate field value (denoted with a superscript $*$) is first computed which does not obey the divergence-free constraint. This intermediate field value is then projected back onto the manifold of divergence-free solutions to compute the field at the next timestep.\label{fig:proj}}
\end{figure}

One typical approach is to employ a Hodge decomposition~\cite{brown01, min06},
\begin{equation}
    \vv^* = \vv + \nabla \phi,
    \label{eqn:v_star}
\end{equation}
where $\vv$ is the desired divergence-free velocity field and $\phi$ is an auxiliary field. One then updates $\vv^*$ for a fixed interval of time via the equation
\begin{equation}
    \vv^*_t + \left(\vv \cdot \nabla\right)\vv + \nabla q = \nu \nabla^2 \vv^*,
    \label{eqn:v_star_evol}
\end{equation}
where $\nabla q$ is an approximation to the pressure gradient. Substituting Eq.~\ref{eqn:v_star} into Eq.~\ref{eqn:v_star_evol} leads to a formula for the pressure
\begin{equation}
    \nabla p = \nabla \left(q + \phi_t\right) - \nu \nabla^2 \nabla \phi,
    \label{eqn:p_comp}
\end{equation}
from which $p$ can be computed. The divergence of Eq.~\ref{eqn:v_star} implies that $\nabla \cdot \vv = 0$ if $\phi$ is such that
\begin{equation}
    \nabla^2 \phi = \nabla \cdot \vv^*.
    \label{eqn:phi}
\end{equation}
The Poisson problem in Eq.~\ref{eqn:phi} can be solved for $\phi$ using standard techniques of numerical linear algebra, and the projection can be completed by computing
\begin{equation}
    \vv = \vv^* - \nabla \phi.
    \label{eqn:v_proj}
\end{equation}
Boundary conditions on $\phi$ in Eq.~\ref{eqn:phi} depend on the physical scenario of interest, and are critical for obtaining higher-order methods~\cite{brown01, min06}. The algorithm typically proceeds by advancing Eq.~\ref{eqn:v_star_evol} for a single timestep, computing $\vv$ according to Eq.~\ref{eqn:v_proj}, and then setting $\vv^* = \vv$ for the beginning of the next timestep~\cite{brown_acc_proj}. This procedure is schematically represented in discrete-time in Fig.~\ref{fig:proj}(a). Projection methods avoid the CFL condition associated with compressive waves in the fluid, and hence can use significantly larger timesteps than explicit methods. An extension of projection methods known as gauge methods do not reset $\vv^* = \vv$ at the end of timestep, and instead allow it to continue to evolve during the computation~\cite{saye_dg1, saye_dg2, Summers1881, Cortez1996, Buttke1993}

It is possible to demonstrate that solving Eq.~\ref{eqn:phi} represents an orthogonal projection. We define the inner product between two vector-valued fields,
\begin{equation}
    \langle \mathbf{v}, \mathbf{u} \rangle = \int_\Omega \mathbf{v} \cdot \mathbf{u} \,d^3\bx,
    \label{eqn:inner_v}
\end{equation}
where $\Omega$ is the simulation domain. Using this inner product, we can compute
\begin{align}
    \langle \vv^{n+1} - \vv^n, \vv^{n+1} - \vv^*\rangle &= -\int_\Omega \left(\vv^{n+1} - \vv^n\right)\cdot\nabla \phi(\bx) d^3\bx \nonumber \\
    &= \int_\Omega \left(\nabla \cdot \vv^{n+1} - \nabla \cdot \vv^n\right)\phi(\bx) d^3\bx = 0,
    \label{eqn:orthog_v}
\end{align}
thereby establishing that the projection $\vv^{n+1} - \vv^*$ is orthogonal to
the difference between the two velocity fields, $\vv^{n+1}-\vv^n$.

\subsection{A family of projection methods for hypo-elastoplasticity}
\label{sec:qs_alg_gen}
We now formulate a three-dimensional projection method for solving Eq.~\ref{eqn:lnr_elas} subject to the quasi-static constraint Eq.~\ref{eqn:qs}. We define an intermediate stress
\begin{equation}
    \bsig^* = \bsig + \tC : \nabla \bPhi,
    \label{eqn:sig_star}
\end{equation}
where $\bPhi(\bx, t)$ is an auxiliary vector field. We can solve for $\bsig^*$ by dropping the $\tC : \tD$ term in Eq.~\ref{eqn:lnr_elas},
\begin{equation}
    \bsig^*_t + \left(\vv \cdot \nabla\right)\bsig = \bL \bsig + \bsig \bL^\Trans - \Tr(\bL)\bsig + \tC : \left(\nabla \mathbf{q} - \Dpl\right).
    \label{eqn:sig_star_evol}
\end{equation}
In Eq.~\ref{eqn:sig_star_evol}, $\mathbf{q}$ represents an approximation to the velocity $\vv$. Substituting Eq.~\ref{eqn:sig_star} into Eq.~\ref{eqn:sig_star_evol}, we find
\begin{equation}
    \tC : \tD = \tC : \nabla \left(\mathbf{q} - \bPhi_t \right),
    \label{eqn:D_recov}
\end{equation}
from which $\tD$ can be computed. Taking the divergence of Eq.~\ref{eqn:sig_star} and requiring $\nabla \cdot \bsig = \vec{0}$, $\bPhi$ must satisfy the equation
\begin{equation}
    \nabla \cdot \left(\tC : \nabla \bPhi\right) = \nabla \cdot \bsig^*.
    \label{eqn:Phi}
\end{equation}
Equation~\ref{eqn:Phi} is a linear system with source term $\nabla \cdot \bsig^*$ that can be solved for $\bPhi$. Hence, we can devise a projection method for quasi-static hypo-elastoplasticity by evolving Eq.~\ref{eqn:sig_star_evol} for a single timestep, finding $\bPhi$ according to Eq.~\ref{eqn:Phi}, and projecting $\bsig^*$ onto the manifold of divergence-free solutions by computing $\bsig = \bsig^* - \tC : \nabla \bPhi$. The algorithm may then proceed by setting $\bsig^* = \bsig$ for the start of the next timestep. We represent this algorithm schematically in Fig.~\ref{fig:proj}(b). A projection method is defined by the choice of the approximate velocity field $\mathbf{q}$, the auxiliary vector field $\bPhi$, and the integration method for Eq.~\ref{eqn:sig_star_evol}. By instead allowing $\bsig^*$ to evolve over the course of the computation rather than setting $\bsig^* = \bsig$ after each timestep, we expect it to be possible to develop gauge methods for quasi-static hypo-elastoplasticity in an analogous manner to the case of fluid dynamics.

As in the case of fluid dynamics, we can show that the projection is orthogonal in a suitable inner product. To do so, we define an inner product between two stress tensors as in~\cite{rycroft15},
\begin{equation}
    \langle \bsig, \bsig' \rangle = \int_\Omega \bsig : \mathbf{S} : \bsig' \, d^3\bx,
    \label{eqn:inner_stress}
\end{equation}
where $\mathbf{S} = \mathbf{C}^{-1}$ is the stiffness tensor. Equation~\ref{eqn:inner_stress} computes the elastic strain energy of a material with stress field $\bsig$ and strain field $\mathbf{S} : \bsig'$, or vice-versa by symmetry. Because $\mathbf{S}$ is a symmetric positive definite tensor for physically realistic Lam\'e parameters, this definition is an inner product. By explicit computation,
\begin{align}
    \langle \bsig^{n+1} - \bsig^{n}, \bsig^{n+1} - \bsig^{*} \rangle &= \int \left(\bsig^{n+1} - \bsig^{n}\right) : \mathbf{S} : \tC : \nabla \bPhi\, d^3\bx \nonumber\\
    &= \int \left(\bsig^{n+1} - \bsig^{n}\right) : \nabla \bPhi \, d^3\bx\nonumber \\
    &= -\int \left(\nabla \cdot \bsig^{n+1} - \nabla \cdot \bsig^n\right)\cdot\bPhi \, d^3\bx = 0.
\end{align}

\section{Numerical implementation}
\label{sec:numerics}
\begin{table}
    \caption{Material parameters used in this study, for both linear elasticity and the STZ model of amorphous plasticity. The Boltzmann constant $k_B$ is used to convert energetic values to temperatures.}
    \vspace{5mm}
    \label{table:params}
    \centering
    \begin{tabular}{l|l}
        Parameter & Value \\ \hline
        Young's modulus $E$ & 101~GPa\\
        Poisson ratio $\nu$ &  0.35\\
        Bulk modulus $K$ & 122~GPa\\
        Shear modulus $\mu$ & 37.4~GPa\\
        Density $\rho_0$ & $6125\text{~kg~m}^{-3}$\\
        Shear wave speed $c_s$ & $2.47\text{~km~s}^{-1}$\\
        Yield stress $s_Y$ & 0.85~GPa\\
        Molecular vibration timescale $\tau_0$ & $10^{-13}$~s\\
        Typical local strain $\epsilon_0$ & 0.3\\
        Effective heat capacity $c_0$ & 0.4\\
        Typical activation barrier $\Delta/k_B$ & 8000~K\\
        Typical activation volume $\Omega$ & 300~\AA${}^3$\\
        Thermodynamic bath temperature $T$ & 400~K\\
        Steady state effective temperature  $\chi_\infty$ & 900~K\\
        STZ formation energy $e_z/k_B$ & 21000~K
    \end{tabular}
\end{table}

In this section, we describe an implementation of an explicit forward Euler method to solve Eqs.~\ref{eqn:lnr_elas} \& \ref{eqn:nwtn_law}, as well as a specific instance of the quasi-static projection method in Eqs.~\ref{eqn:sig_star_evol}~\&~\ref{eqn:Phi}. We model elastoplastic deformation in a BMG using an athermal variant of the STZ theory.

\subsection{Plasticity model}
\label{subsec:plasticity}
As a plasticity model for a metallic glass, we use an athermal form of the STZ theory suitable for studying diverse materials including BMGs below the glass transition temperature, dense granular materials, and soft materials such as foams or colloidal glasses~\cite{bouchbinder07,langer08}. Within the STZ theory, irreversible molecular rearrangements are assumed to occur sporadically throughout an otherwise elastic material, and each rearrangement induces a small increment of strain. The accumulation of many such events leads to macroscopic plastic deformation. It is assumed that when local stresses surpass the material yield stress $s_y$, these rearrangements occur at rare and localized sites known as STZs. Thermal fluctuations of the atomic configuration are neglected in the athermal formulation: molecular rearrangements are entirely driven by external mechanical forces. Thermal theories introduce additional coupling between a \textit{configurational} subsystem governing the rearrangements that occur at the STZs, and a \textit{kinetic/vibrational} subsystem governing the thermal vibrations of atoms in their cage of nearest neighbors~\cite{kamrin14a}.

STZs may be conceptualized as clusters of atoms predisposed to configurational rearrangements when subjected to external shear~\cite{bouchbinder07}. Each rearrangement corresponds to a transition in the configurational energy landscape; these transitions are usually towards a lower-energy configuration, but there is a small probability for a reverse transition. Before the application of external shear, the material sample is at a local minimum. External shear alters the shape of the energy landscape, and can make transitions to other states considerably more likely.

The density of STZs in space follows a Boltzmann distribution in an effective disorder temperature denoted by $\chi$~\cite{bouch_linear, bouch_eff_dyn, cugli1, cugli2}. $\chi$ governs the out-of-equilibrium configurational degrees of freedom of the material and has many properties of the usual temperature: it is measured in Kelvin, and can be obtained as the derivative of a configurational energy with respect to a configurational entropy~\cite{falk_langer_rev}. $\chi$ is distinct from the thermodynamic temperature $T$, though it plays the same role for the configurational subsystem as $T$ does for the kinetic/vibrational subsystem.

The plastic rate of deformation tensor is proportional to the deviatoric part of the stress tensor \smash{$\bsig_0 = \bsig - \frac{1}{3}\mathbf{I}\tr(\bsig)$}, so that \smash{$\Dpl = D^{\text{pl}}\frac{\bsig_0}{\bar{s}}$}. $\bar{s}$ is a local stress measure essentially given by the Frobenius norm of the deviatoric stress tensor, \smash{$\bar{s}^2 = \frac{1}{2} \sum_{ij} \sigma_{0, ij}^2$}. The magnitude of the plastic rate of deformation is given by
\begin{equation}
    \tau_0 D^{\text{pl}} = e^{-e_z/k_B \chi} \sC(\bar{s}, T)\left(1 - \frac{s_Y}{\bar{s}}\right),
    \label{eqn:stz_dpl}
\end{equation}
where $\tau_0$ is a molecular vibration timescale, $e_z$ is a typical STZ formation energy, and $k_B$ is the Boltzmann constant. $\sC(\bar{s}, T)$ represents the total STZ transition rate. With $\sR(\pm\bar{s}, T)$ denoting the forward and reverse rates between two configurational states, the total transition rate is $\sC(\bar{s}, T) = \frac{1}{2}\left(\sR(\bar{s}, T) + \sR(-\bar{s}, T)\right)$. The transitions follow a linearly stress-biased thermal activation process,
\begin{equation}
    \sR(\pm \bar{s}, T) = \exp\left(-\frac{\Delta \mp \Omega \epsilon_0 \bar{s}}{k_B T}\right).
    \label{eqn:stz_R}
\end{equation}
$\Delta$ is a typical energetic barrier for a transition, $\Omega$ is a typical STZ volume, and $\epsilon_0$ is a typical local strain due to an STZ transition. While thermal fluctuations are neglected in the athermal model, the thermodynamic temperature still sets the magnitude of transition rates in the system. Using the form Eq.~\ref{eqn:stz_R} yields the overall transition rate
\begin{equation}
    \sC(\bar{s}, T) = e^{-\Delta/k_B T}\cosh\left(\frac{\Omega \epsilon_0 \bar{s}}{k_B T}\right).
    \label{eqn:stz_C_form}
\end{equation}
The effective temperature satisfies~\cite{bouchbinder07, bouchbinder07b, manning07, manning09}
\begin{equation}
    c_0\frac{d \chi}{dt} = \frac{\left(\Dpl : \bsig_0\right)}{s_Y}\left(\chi_\infty - \chi\right) + l^2 \nabla \cdot \left(D^{\text{pl}}\nabla\chi\right).
    \label{eqn:chi_evo}
\end{equation}
Equation~\ref{eqn:chi_evo} consists of a term causing growth to an asymptotic value $\chi_\infty$ and a diffusive term with length scale $l$. Both saturation to $\chi_\infty$ and diffusion occur in response to plastic deformation. The term $\Dpl : \bsig_0$ is the energy dissipation rate due to externally applied mechanical work -- STZs are created and annihilated proportionally -- and $c_0$ is an effective heat capacity. Eq.~\ref{eqn:chi_evo} is thus essentially a heat equation, representing the first law of thermodynamics for the configurational subsystem~\cite{bouchbinder07}. The interdependence of Eqs.~\ref{eqn:stz_dpl}~\&~\ref{eqn:chi_evo} enables the development of shear bands via positive feedback, as increasing $\chi$ also increases $D^{\text{pl}}$~\cite{manning07, manning09}.

\subsection{Explicit method}
For the explicit method, we discretize Eqs.~\ref{eqn:lnr_elas}, \ref{eqn:nwtn_law}, and \ref{eqn:chi_evo} with a forward Euler step. This leads to the coupled set of discrete-time equations
\begin{align}
    \rho \frac{\vv^{n+1} - \vv^n}{\Delta t} &= -\left(\vv^n \cdot \nabla\right)\vv^n + \nabla \cdot \bsig^n + \kappa \nabla^2 \vv^n
    \label{eqn:vel_evo_exp},\\
    \frac{\bsig^{n+1} - \bsig^n}{\Delta t} &= -\left(\vv^n \cdot \nabla \right)\bsig^n + \bL^n\bsig^n + \bsig^n \left(\bL^\Trans\right)^n \nonumber\\&\phantom{=} \ \ \ \ \ \ \ \ \ \ \ + \Tr(\bL^n)\bsig^n + \tC : \left(\tD^n - \frac{\left(D^{\text{pl}}\right)^n}{\bar{s}^n}\bsig_0^n\right),
    \label{eqn:stress_evo_exp}\\
    c_0\frac{\chi^{n+1} - \chi^n}{\Delta t} &= \frac{\left((\Dpl)^n : \bsig_0^n\right)}{s_Y}\left(\chi_\infty - \chi^n\right) + l^2 \nabla \cdot \left(\left(D^{\text{pl}}\right)^n\nabla\chi^n\right).
    \label{eqn:chi_evo_exp}
\end{align}
The small viscous stress term $\kappa\nabla^2\vv$ in Eq.~\ref{eqn:vel_evo_exp} is artificially imposed for numerical stability, but it is not needed in the quasi-static method. This term is physically reasonable, and could represent damping within the material. The requirement of damping in the explicit scheme for stability should be expected. If one focuses on elastic waves in the linearized regime then Eq.~\ref{eqn:chi_evo_exp} can be neglected, and the leading-order terms in Eqs.~\ref{eqn:vel_evo_exp} \& \ref{eqn:stress_evo_exp} are $\nabla \cdot \bsig^n$ and $\tC : \tD^n$, respectively. Hence Eqs.~\ref{eqn:vel_evo_exp} \& \ref{eqn:stress_evo_exp} form a first-order hyperbolic system for stress and velocity~\cite{leveque_fv,rycroft12}. We implement the explicit method using centered differencing for the spatial derivatives (Sec.~\ref{subsec:discr}), and it is well-known that such discretizations are unstable in the absence of any damping~\cite{leveque_fd}. The scheme can be stabilized with a damping $\kappa= \kappa' h$ where $\kappa'$ is a constant that we determine empirically.

In three dimensions, the damping term induces a restriction on the timestep \smash{$\Delta t \leq \frac{h^2}{6\kappa}$}. Hence, if $\kappa$ is viewed as a physical constant, this condition is more restrictive than the CFL condition. Here, we choose $\kappa$ to scale with the grid spacing as required for stability, and hence the timestep restriction scales in the same way as the CFL restriction.

\subsection{Quasi-static method}
We now formulate a specific three-dimensional projection method for solving Eq.~\ref{eqn:lnr_elas} subject to the quasi-static constraint Eq.~\ref{eqn:qs}. We first neglect the $\tC:\tD$ term in Eq.~\ref{eqn:lnr_elas} and compute an intermediate stress $\bsig^*$,
\begin{equation}
    \bsig^* = \bsig^n + \Delta t \left(\bL^n\bsig^n + \bsig^n \left(\bL^\Trans\right)^n - \Tr(\bL^n)\bsig^n - \tC:\left(\frac{D^{\text{pl}}}{\bar{s}^n}\bsig_0^n\right)\right).
    \label{eqn:sig_star_qs}
\end{equation}
If the velocity at the next timestep $\vv^{n+1}$ were known, we could compute
\begin{equation}
    \tD^{n+1} = \frac{1}{2}\left(\left(\nabla \vv^{n+1}\right) + \left(\nabla \vv^{n+1}\right)^\Trans\right)
\end{equation}
and complete the forward Euler step in Eq.~\ref{eqn:sig_star_qs} as
\begin{equation}
    \bsig^{n+1} = \bsig^* + \Delta t \left(\tC:\tD^{n+1}\right).
    \label{eqn:sig_n+1}
\end{equation}
Taking the divergence of Eq.~\ref{eqn:sig_n+1}, requiring $\nabla\cdot\bsig^{n+1} = 0$, and rearranging terms leads to the equation
\begin{equation}
    \Delta t\nabla\cdot \left(\tC : \tD^{n+1}\right) = -\nabla\cdot \bsig^*.
    \label{eqn:proj}
\end{equation}
Equation \ref{eqn:proj} is a linear system for the velocity $\vv^{n+1}$ involving mixed spatial derivatives. The source term is given by $-\nabla\cdot\bsig^*$. After solving for $\vv^{n+1}$, it can be used to compute $\bsig^{n+1}$ via Eq.~\ref{eqn:sig_n+1}. Through this process, $\bsig^*$ is projected down to the manifold of divergence-free solutions to arrive at $\bsig^{n+1}$.

The mixed derivatives in Eq.~\ref{eqn:proj} increase the complexity of the projection for hypo-elastoplasticity when compared to the Poisson problem in fluid dynamics, but Eq.~\ref{eqn:proj} can nevertheless be solved rapidly via standard techniques of numerical linear algebra such as the multigrid method. The multigrid method relies on the Gauss--Seidel method for iterative smoothing of the solution, and Gauss--Seidel smoothing is guaranteed to converge if either the linear system is (A) weakly diagonally dominant, or (B) symmetric positive definite. In general the linear system in Eq.~\ref{eqn:proj} will not satisfy condition A, but will satisfy condition B. Hence Gauss--Seidel smoothing is guaranteed to converge, which we use as a component in a multigrid method---details of this multigrid solver are presented later. A connection to the general continuous-time framework presented in Sec.~\ref{sec:qs_alg_gen} is provided in \ref{app:alg_conn}.


\subsection{Discretization and finite difference stencils}
\label{subsec:discr}
The evolution equation for the stress, Eq.~\ref{eqn:lnr_elas}, depends on spatial derivatives of the velocity, while the equation satisfied by the velocity, Eq.~\ref{eqn:nwtn_law}, depends on spatial derivatives of the stress. We exploit this structure through a staggered grid with uniform spacing $\Delta x = \Delta y = \Delta z = h$. The stress tensor $\bsig$ and effective temperature $\chi$ are stored at cell centers and indexed by half-integers, while the velocity $\vv$ is stored at cell corners and indexed by integers, as shown in Fig.~\ref{fig:grid_cell_ghost}(a).

Let $(\p f /\p x)_{i, j, k}$ denote the partial derivative of a field $f$ with respect to $x$ evaluated at grid point $(i, j, k)$. The staggered centered difference is
\begin{align}
  \left(\prx{f}\right)_{i+\half, j+\half, k+\half} &= \frac{1}{4h} \Big(f_{i+1, j, k} - f_{i, j, k} + f_{i+1, j+1, k} - f_{i, j+1, k} \nonumber \\
  &\phantom{=} + f_{i+1, j, k+1} - f_{i, j, k+1} + f_{i+1, j+1, k+1} - f_{i, j+1, k+1}\Big).
    \label{eqn:stag_stenc}
\end{align}
Equation \ref{eqn:stag_stenc} averages four edge-centered centered differences surrounding the cell center and has a discretization error of size $\bigO(h^2)$. The derivative at a cell corner is obtained by the replacement \smash{$(i, j, k) \rightarrow (i - \frac{1}{2}, j - \frac{1}{2}, k - \frac{1}{2})$}. The diffusive term appearing in the velocity update in Eq.~\ref{eqn:vel_evo_exp} is computed via the standard centered difference formula,
\begin{equation}
    \left(\frac{\p^2 f}{\p x^2}\right)_{i, j, k} = \frac{f_{i+1, j, k} - 2f_{i, j, k} + f_{i-1, j, k}}{h^2}.
    \label{eqn:col_snd_deriv}
\end{equation}
The advective derivatives in Eqs.~\ref{eqn:vel_evo_exp} \& \ref{eqn:stress_evo_exp} must be upwinded for stability; we use the second-order essentially non-oscillatory (ENO) scheme~\cite{shu88}. With $\left[f_{xx}\right]_{i, j, k}$ denoting the second derivative with respect to $x$ of the field $f$ at grid point $(i, j, k)$ computed using Eq.~\ref{eqn:col_snd_deriv}, the ENO derivative is defined in the $x$ direction as
\begin{equation}
    \left(\prx{f}\right)_{i, j, k} = \frac{1}{2h}
    \begin{cases}
        -f_{i+2, j, k} + 4f_{i+1, j, k} - 3f_{i, j, k} &\text{if $u_{i, j, k} < 0$ and $\left|\left[f_{xx}\right]_{i, j, k}\right| > \left|\left[f_{xx}\right]_{i+1, j, k}\right|$,}\\
        3f_{i, j, k} - 4f_{i-1, j, k} + f_{i-2, j, k} &\text{if $u_{i, j, k} > 0$ and $\left|\left[f_{xx}\right]_{i, j, k}\right| > \left|\left[f_{xx}\right]_{i-1, j, k}\right|$,}\\
        f_{i+1, j, k} - f_{i-1, j, k} &\text{otherwise}.
    \end{cases}
    \label{eqn:ENO}
\end{equation}
Equation \ref{eqn:ENO} uses the curvature of $f$ to switch between an upwinded three-point derivative and a centered difference. Versions of Eqs.~\ref{eqn:stag_stenc}, \ref{eqn:col_snd_deriv}, \& \ref{eqn:ENO} in the $y$ and $z$ coordinates are obtained analogously.

To solve Eq.~\ref{eqn:proj} via numerical linear algebra, the spatial derivatives must first be discretized using finite differences. In addition to the finite differences discussed above, Eq.~\ref{eqn:proj} also contains mixed partial derivatives. The $xy$-derivative is computed numerically as
\begin{equation}
    \left(\frac{\p^2 f}{\p x \p y}\right)_{i,j,k} = \frac{f_{i+1, j+1, k} - f_{i+1, j-1, k} - f_{i-1, j+1, k} + f_{i-1, j-1, k}}{4h^2},
\end{equation}
with analogous expressions for other mixed partials.

\subsection{Parallelization via MPI and domain decomposition}
\begin{figure}
    \begin{tabular}{cc}
        \begin{overpic}[width=\subpanelwid]{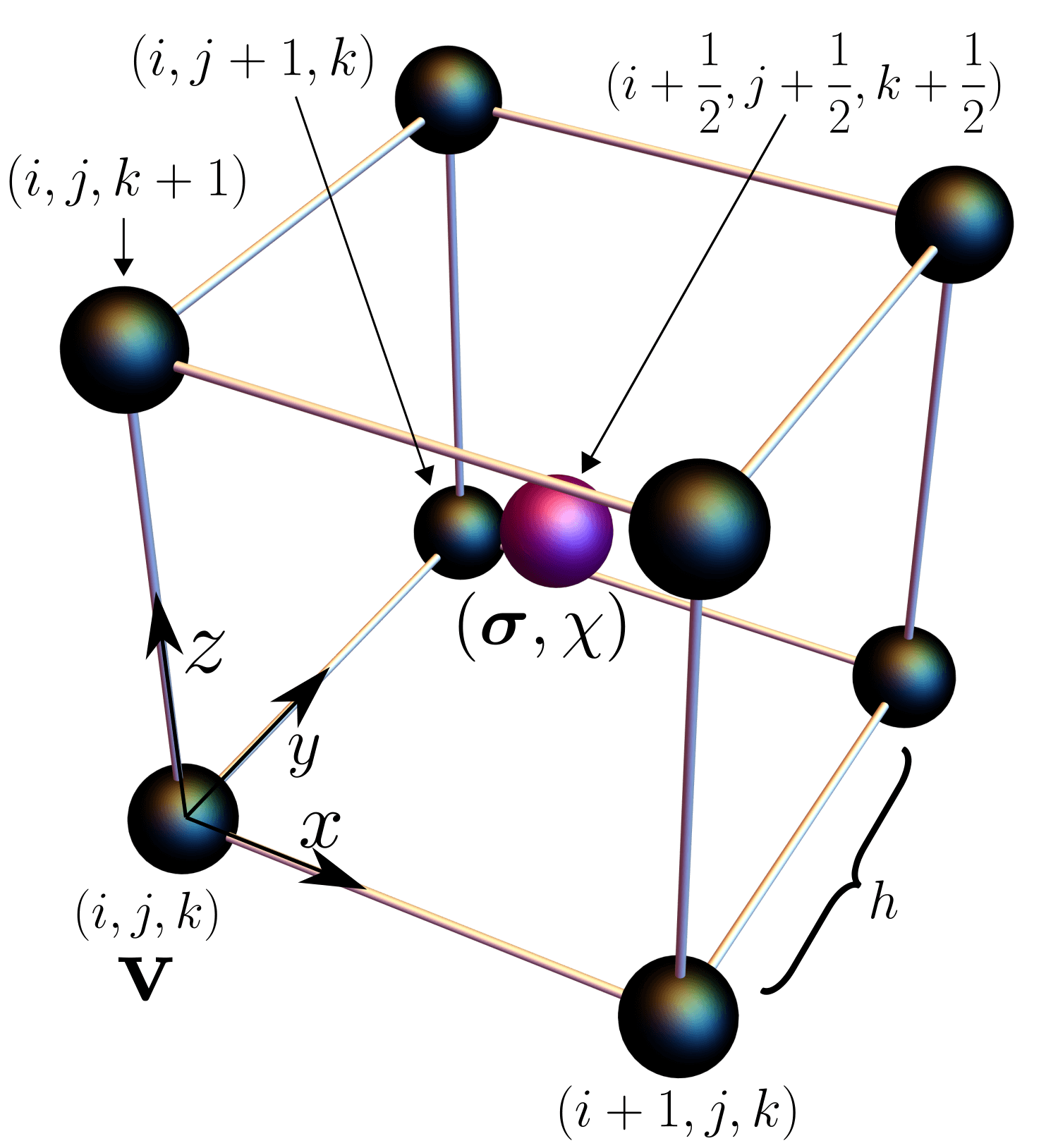}
        \put(0, 93){(a)}
        \end{overpic} &
        \begin{overpic}[width=\subpanelwid]{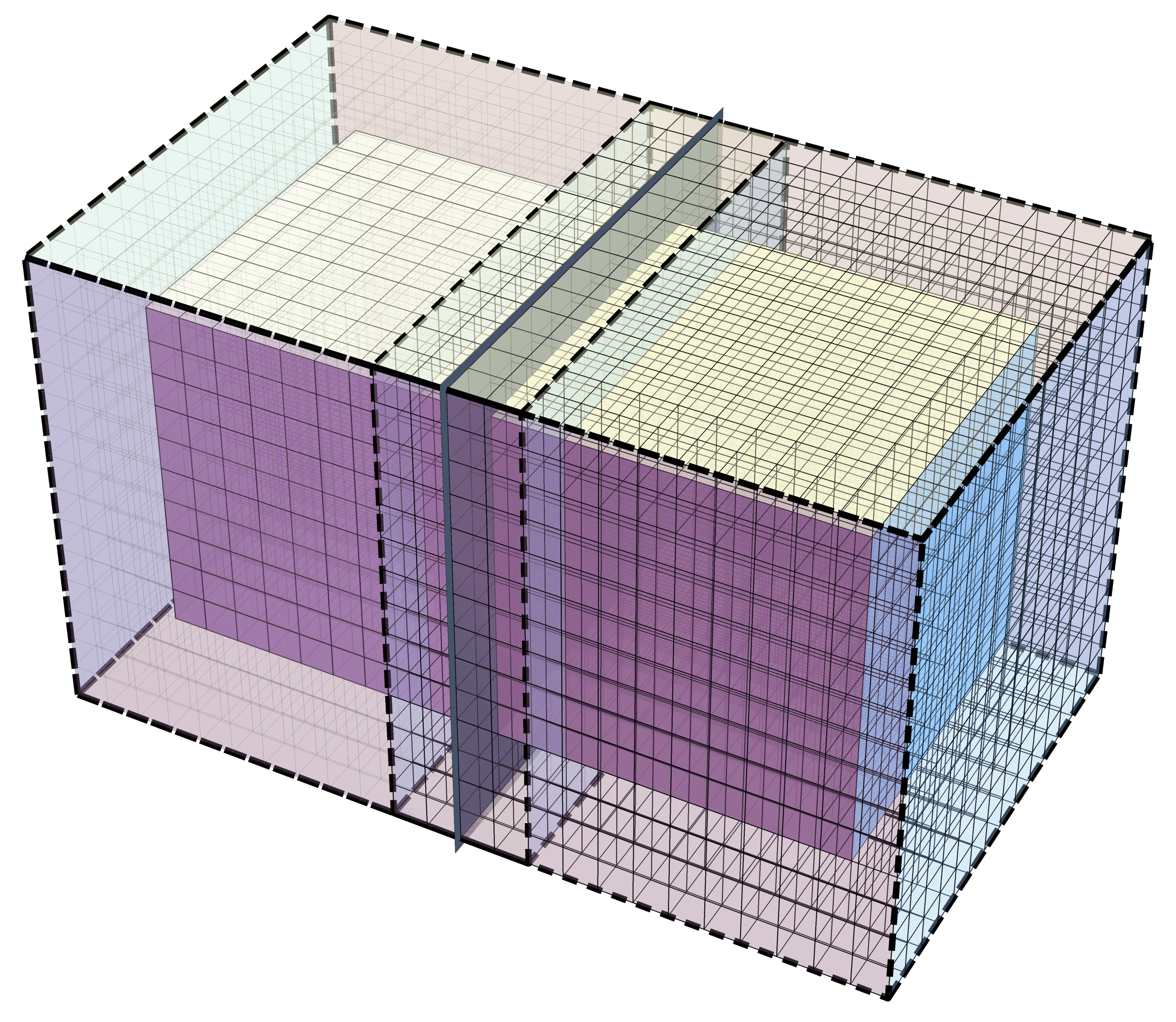}
        \put(0, 100){(b)}
        \end{overpic}
    \end{tabular}
    \vspace{2mm}
    \caption{(a) The arrangement of discretized simulation fields. The cube here corresponds to a single grid cell of side length $h$. Velocities $\vv$ are stored at cell corners denoted by black spheres and indexed by integers. Stresses $\bsig$ and effective temperatures $\chi$ are stored at cell centers indicated by the purple sphere and indexed by half integers. (b) Processor ghost regions. The two solid cubes represent two adjacent processor subdomains; the boundary between them is indicated by a gray plane in the center of the figure. Surrounding each processor subdomain is a transparent two-grid-point ghost region bounded by black dashed lines. For clarity, the ghost region grid in the left processor subdomain has been drawn in a thin gray, while the ghost region grid in the right processor subdomain has been drawn in black. These two processors communicate the overlapping rectangular strip surrounding the separating plane in the center of the figure.}
    \label{fig:grid_cell_ghost}
\end{figure}

We solve Eqs.~\ref{eqn:vel_evo_exp} \& \ref{eqn:stress_evo_exp} in parallel with a custom C++ code that uses the MPI library for parallelization~\cite{openmpi}. We use a global grid comprised of $Q \times M \times N$ grid cells. Each grid cell is labeled by its lower corner index $(i,j,k)$ as shown in Fig.~\ref{fig:grid_cell_ghost}(a), with indices in the ranges $i\in\{0,1,\ldots,Q-1\}$, $j\in\{0,1,\ldots,M-1\}$, $k\in\{0,1,\ldots,N-1\}$.
The global grid is split across processors using a standard Cartesian decomposition into $P_x \times P_y \times P_z$ subdomains. Specifically, each processor is indexed as $(I,J,K)$ with $I\in\{0,\ldots,P_x-1\}$, $J\in\{0,\ldots,P_y-1\}$, and $K\in\{0,\ldots,P_z-1\}$. Processor $(I,J,K)$ is responsible for the grid cells
\begin{align}
  i &\in \{\lfloor I Q / P_x \rfloor, \ldots, \lfloor (I+1) Q / P_x \rfloor -1 \}, \nonumber \\
  j &\in \{\lfloor J M / P_y \rfloor, \ldots, \lfloor (J+1) M / P_y \rfloor -1 \}, \nonumber \\
  k &\in \{\lfloor K N / P_z \rfloor, \ldots, \lfloor (K+1) N / P_z \rfloor -1 \}.
\end{align}
Hence the subdomain assigned to each processor is identical up to one grid point in each direction.

The finite difference stencils in Eqs.~\ref{eqn:stag_stenc} \& \ref{eqn:col_snd_deriv} require data from adjacent gridpoints, and the ENO derivative in Eq.~\ref{eqn:ENO} requires data from at most two grid points away. On grid points within two points of a subdomain boundary, the derivative calculation can therefore require inaccessible data in a distributed memory setting. To handle this, we pad each processor subdomain with ghost regions of width two. A ghost region is a cubical shell of non-physical grid points whose field values are filled with data from adjacent processors, so that each subdomain can locally access values computed and stored in adjacent subdomains (Fig.~\ref{fig:grid_cell_ghost}(b)). At the simulation boundaries, ghost regions are used to enforce boundary conditions. For periodic boundary conditions, the ghost regions wrap around to processor subdomains on the other side of the simulation. For non-periodic conditions, the ghost grid points store values necessary to enforce the given boundary condition: for a free boundary, the fields are linearly extrapolated into the ghost region. For a Dirichlet condition, the physical field values are reflected over the boundary into the ghost region.

At the start of each timestep, each processor communicates with the 26 nearby processors sharing six faces, twelve edges, and eight corner regions. Each processor sends data to nearby processors and receives the data it requires from the same nearby processors via non-blocking communication. Each processor waits to ensure it has received all face data before updating faces, edge data before updating edges, and corner data before updating corners. This ensures that components of each ghost region can be updated when the data is available without waiting for all data to arrive. The total cost of parallel communication scales with the surface area of a processor subdomain. Usually, the total number of processors $P_\textrm{tot}$ is chosen to be power of two, and the precise subdomain decomposition is determined by considering all triplets $(P_x,P_y,P_z)$ that satisfy $P_xP_yP_z = P_\textrm{tot}$, and selecting the one that minimizes the surface area between the subdomains.

\subsection{Performing the projection step}
We solve Eq.~\ref{eqn:proj} for the velocity using a custom parallel implementation of the geometric multigrid method, a linear system solver that is particularly suited to elliptic problems that take place on a physical grid~\cite{Briggs2000}. Let $G_0$ be the original grid, and let $\mathbf{A}_0\mathbf{x}_0 = \mathbf{b}_0$ be the linear system to solve on this grid. In the multigrid method, a hierarchy of progressively coarser grids $G_1,G_2, \ldots, G_g$ is introduced. In our implementation, if $G_k$ has resolution $Q_k \times M_k \times N_k$, then $G_{k+1}$ has resolution \smash{$\lceil Q_k/2\rceil,\lceil M_k/2\rceil,\lceil N_k/2\rceil$}. Interpolation operators $\mathbf{T}_k: G_k \to G_{k-1}$ are introduced based on linear interpolation, and restriction operators $\mathbf{R}_k : G_k \to G_{k+1}$ are introduced based on local averaging. Both $\mathbf{T}_k$ and $\mathbf{R}_k$ can be represented as rectangular matrices, and in our implementation $\mathbf{R}_{k-1} = \mathbf{T}_k^\Trans$---this condition is not necessary for a practical implementation, but is useful in some convergence proofs~\cite{demmel}.

Our multigrid implementation uses the standard V-cycle~\cite{Briggs2000,demmel} with two pre-smoothing steps and two post-smoothing steps. On $G_0$ the grid is decomposed among the processors in the same way as the simulation fields. The smoothing steps are performed using the Gauss--Seidel method on each processor, with the ghost regions being synchronized after each step. This requires building a representation of the linear system on each grid, which we do via recursive matrix multiplication~\cite{xu92,rycroft13},
\begin{equation}
  \mathbf{A}_k = \mathbf{R}_{k-1} \mathbf{A}_{k-1} \mathbf{T}_k.
\end{equation}
The implementation works with periodic and non-periodic boundary conditions, and arbitrary grid dimensions. As the grids are coarsened, the amount of work on each grid is rapidly reduced, to the point where it is no longer effective for all processors to share the work. The implementation therefore has the ability to amalgamate the coarser problem onto a smaller set of processors, with the rest remaining idle.

The multigrid implementation uses C++ templates, so that the linear system can be compiled to work with an arbitrary data type. For the current problem, $\mathbf{b}_0$ is given by the source term $-\Delta t \nabla \cdot \bsig$ and $\mathbf{x}_0$ contains values of $\vv^{n+1}$ across the entire grid. Hence, we compile the multigrid library where the elements of $\mathbf{b}_0$ and $\mathbf{x}_0$ are 3-vectors, and the block elements of $\mathbf{A}_0$ are $3\times 3$ symmetric matrices. The matrix $\mathbf{A}_0$ is sparse, and a grid point $(i, j, k)$ is only coupled to the 27 grid points in the $3\times 3\times 3$ surrounding cube of grid points given by coordinates $(i + \{-1, 0, 1\}, j + \{-1, 0, 1\}, k + \{-1, 0, 1\})$ in our discretization scheme.

\section{Shearing between two parallel plates}
\label{sec:examples}
In the following sections, we consider several material samples being sheared between two parallel plates. This example is experimentally relevant, has simple boundary conditions, demonstrates complex shear banding dynamics~\cite{maass15, greer-2013, zhang-2006, bing-2005, maas-2014, antonaglia-2014, sun-2013, schuh-2007}, and has been studied previously in two dimensions~\cite{rycroft15}. It represents a natural physical scenario to compare three-dimensional results to two-dimensional results, compare simulation data to experiments, and to quantitatively compare the explicit and quasi-static methods.

In all simulations, we consider a domain periodic in the $x$ and $y$ directions with shear velocity applied on the top and bottom boundaries in $z$. The domain occupies $-L \leq x < L, -L \leq y < L$, and $-\gamma L \leq z \leq \gamma L$ with \smash{$\gamma = \frac{1}{2}$} and $L = 1\text{~cm}$. A natural unit of time is given as $t_s = L/c_s$ where $c_s = \sqrt{\mu/\rho}$ is the material shear wave speed, and we measure time in this scale. The boundary conditions are given by
\begin{equation}
    \vv(x, y, \pm \gamma L, t) = (\pm U(t), 0, 0),
    \label{eqn:u_bc}
\end{equation}
where the function $U(t)$ is given by
\begin{equation}
    U(t) = \begin{cases}
        \frac{U_B t}{t_s} &\qquad \text{if $t < t_s$,}\\
        U_B &\qquad \text{otherwise.}
    \end{cases}
\end{equation}
The ramp-up in the function $U(t)$ prevents a large deformation rate near the boundary that would be present with $U(t) = U_B$ immediately at $t=0$. The elasticity and plasticity parameters are defined in Table \ref{table:params}. From these values, the natural timescale is $t_s = 4.05~\text{\textmu{}s}$.

\begin{figure}
    \centering
    \includegraphics[width=0.8\textwidth]{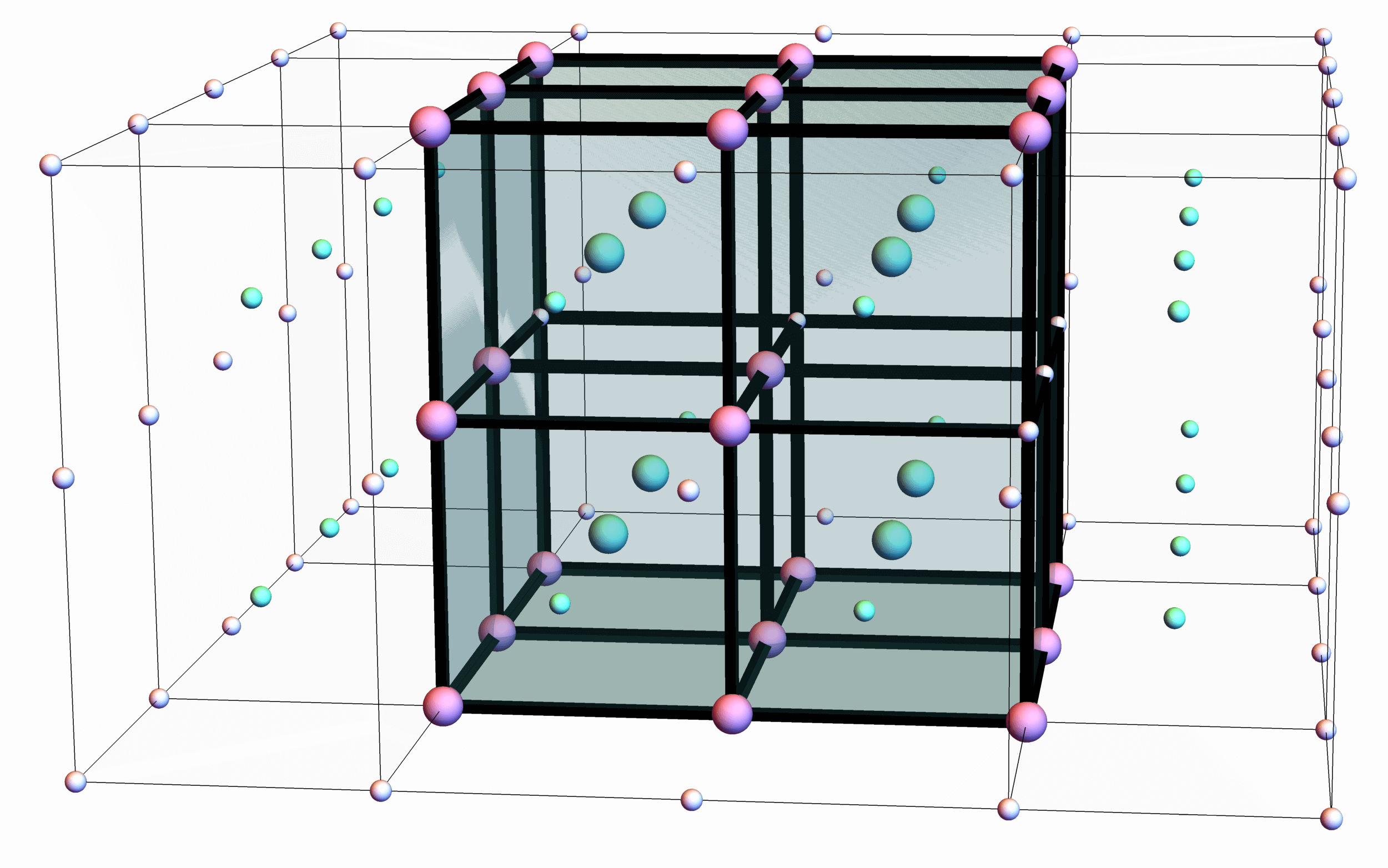}
    \caption{A diagram of the simulation grid layout for the simplified case of $(Q,M,N)=(2,2,2)$. Corner-centered grid points are shown in pink and cell-centered grid points are shown in green. Ghost grid points used for enforcing boundary conditions are shown adjacent to the $\pm x$ and $\pm y$ faces and are surrounded by translucent rectangular prisms. For clarity, these are omitted from the $\pm z$ faces, and only one cell-centered ghost point is shown. Corner-centered ghost points and cell-centered ghost points are smaller and are shown in lighter pink and green than their physical counterparts. The ghost points adjacent to the $\pm x$ and $\pm y$ faces wrap around, and are used to enforce periodic boundary conditions. In the $z$ direction, the ghost grid points are used to linearly extrapolate the $\bsig$ and $\chi$ values, leaving both fields free on the boundary. In the $z$ direction, there is one extra corner-centered grid point, giving the appearance of a grid of size $2\times 2\times 3$. This grid point is used to enforce shear boundary conditions on the velocity field, but the equivalent cell-centered grid points are used to store ghost $\bsig$ and $\chi$ values.}
    \label{fig:grid_setup}
\end{figure}

A diagram of the global three-dimensional grid and the ghost regions at simulation boundaries used for implementing the boundary conditions is shown in Fig.~\ref{fig:grid_setup}. The cell-cornered grid points run according to \smash{$i \in \{0, \hdots, Q-1\}$, $j \in \{ 0, \hdots, M-1\}$, and $k \in \{ 0, \hdots, N\}$}; because the grid is non-periodic in $z$ there is an extra grid point in this direction. The velocities at grid points with $k=0$ and $k=N$ are fixed according to the boundary velocity in Eq.~\ref{eqn:u_bc}. The cell-centered grid points run according to \smash{$i \in \{ \frac{1}{2}, \frac{3}{2}, \hdots Q - \frac{1}{2}\}$, $j \in \{ \frac{1}{2}, \frac{3}{2}, \hdots M - \frac{1}{2}\}$}, and \smash{$k \in \{ \frac{1}{2}, \frac{3}{2}, \hdots N - \frac{1}{2}\}$}. Ghost layers of cell-centered grid points are at \smash{$(i,j,-\frac12)$}, \smash{$(i,j,-\frac{3}{2})$}, \smash{$(i,j,N+\frac12)$}, and \smash{$(i,j,N+\frac32)$} with equivalent expressions in the other dimensions and for cell-cornered grid points. The values of $\bsig$ and $\chi$ in the ghost layers are linearly extrapolated from the two nearest layers to ensure that these fields remain free on the boundary. At the simulation boundaries in the $x$ and $y$ directions, ghost points outside the simulation domain are filled with values that wrap around.

Following the introduction in Sec.~\ref{subsec:plasticity}, the effective temperature $\chi$ is a continuum-scale variable that encodes the density of STZs, and hence the details of the material's microscopic structure. Its initial condition determines the evolution of plastic deformation within the material.

\subsection{Qualitative comparison between explicit and quasi-static methods}
\label{ssec:qual_compare}
\begin{figure}
    \centering
    \fcolorbox{black}{black}{\includegraphics[width=\subpanelwid]{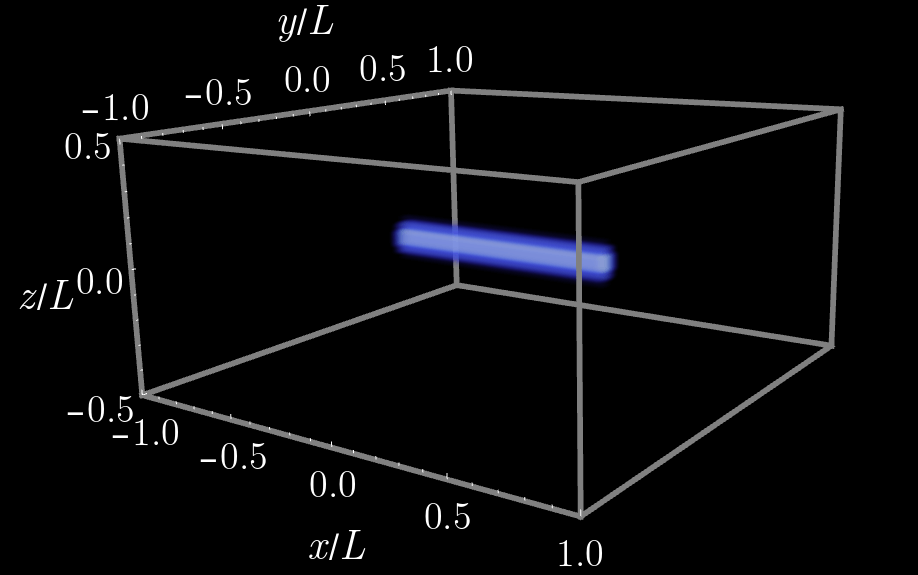}}
    \begin{subfigure}{\textwidth}
        \centering
        \vspace{5mm}
        \includegraphics[width=.75\textwidth]{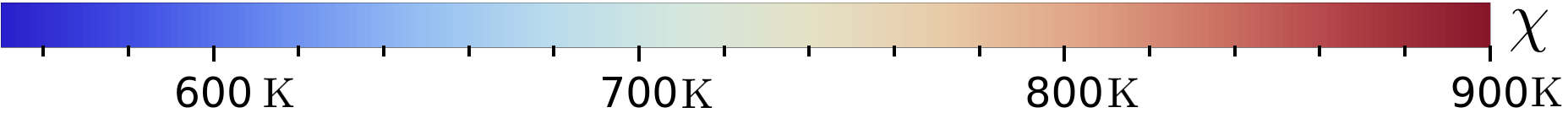}
        \vspace{5mm}
    \end{subfigure}
    \caption{The initial condition for the effective temperature field for the quantitative comparison between the quasi-static and explicit methods. Here, $a = 0.25$ and $\eta = 0.5$ in the opacity function Eq.~\ref{eqn:opac}.}
    \label{fig:qs_d_init}
\end{figure}

We now demonstrate the qualitative equivalence between results computed with the explicit and quasi-static methods. We consider an initial condition corresponding to a finite cylindrical inclusion
\begin{equation}
  \chi(\bx, t=0) = \begin{cases}
      600\text{~K} + (200\text{~K}) e^{-500(z^2 + y^2)/L^2} &\qquad \text{if $|x| < L/2$,}\\
      600\text{~K} &\qquad \text{otherwise.}
  \end{cases}
\end{equation}
Initially the cylindrical inclusion is slightly more amenable to plastic deformation, and hence we expect to see a shear band nucleate from it. To visualize the effective temperature field in three dimensions, we use a custom opacity function defined as
\begin{equation}
    O(\bx) =
    \begin{cases}
        \left(\frac{\chi(\bx) - \chi_{bg}}{\chi_\infty - \chi_{bg}}\right) & \qquad \text{if $\frac{\chi(\bx) - \chi_{bg}}{\chi_\infty - \chi_{bg}} > \tfrac12,$} \\
        \exp\left(-a \left(\frac{\chi_\infty - \chi_{bg}}{\chi(\bx) - \chi_{bg}}\right)^\eta\right) & \qquad \text{otherwise.}
    \end{cases}
    \label{eqn:opac}
\end{equation}
Equation \ref{eqn:opac} sets the opacity of a grid point based on the value of $\chi(\bx)$. The parameters $a$ and $\eta$ are chosen on a case-by-case basis to reveal the most interesting features\footnote{Ideally, we would like to use the same opacity parameters for all plots. However, due to significant variations in the ranges of the $\chi$ fields and in their spatial structures, we found it was necessary to set the parameters on an individual basis. We note, however, that the color scale is absolute across all simulations.}. The initial condition is depicted in Fig.~\ref{fig:qs_d_init}. The grid is of size $64\times 64\times 32$ to accommodate the limitations of the explicit simulation method, corresponding to a grid spacing \smash{$h = \frac{L}{32}$}. A viscous stress constant of \smash{$\kappa' = 4.8 \frac{L}{t_s}$} is used in Eq.~\ref{eqn:vel_evo_exp}. The timestep for the explicit method is \smash{$\Delta t_{\text{e}} = \frac{h^2 t_s}{20L^2}$}.

A typical applied shear velocity that is comparable to a realistic loading rate in a laboratory experiment is $U_b =10^{-7} L/t_s$~\cite{maass15}. With this velocity, running an explicit simulation is prohibitively expensive due to the CFL condition. To ensure that significant plastic deformation occurs on timescales reachable by the explicit method, a scaling parameter $\zeta$ is introduced. The molecular vibration timescale $\tau_0$ is rescaled to $\tau_0\zeta^{-1}$ and the applied shear velocity is inversely rescaled to $U_B = 10^{-7}L/t_s\zeta$. The simulation is conducted until a final time of $t_f = 2\times 10^{6}t_s/\zeta$. As $\zeta$ approaches zero, the quasi-static limit of Eqs.~\ref{eqn:lnr_elas} \& \ref{eqn:nwtn_law} is formally approached. We therefore expect greater agreement for smaller values of $\zeta$. Due to the appearance of \smash{$\frac{1}{\tau_0}$} in Eq.~\ref{eqn:stz_dpl}, the introduction of $\zeta$ has the effect of linearly scaling the magnitude of plastic deformation by a factor of $\zeta$. A quasi-static timestep of $\Delta t_{\text{qs}} = 200 t_s/\zeta$ is used.

\begin{figure}
\fcolorbox{black}{black}{
    \begin{tabular}{cc}
        \begin{subfigure}{\subpanelwid}
            \centering
            \includegraphics[width=\textwidth]{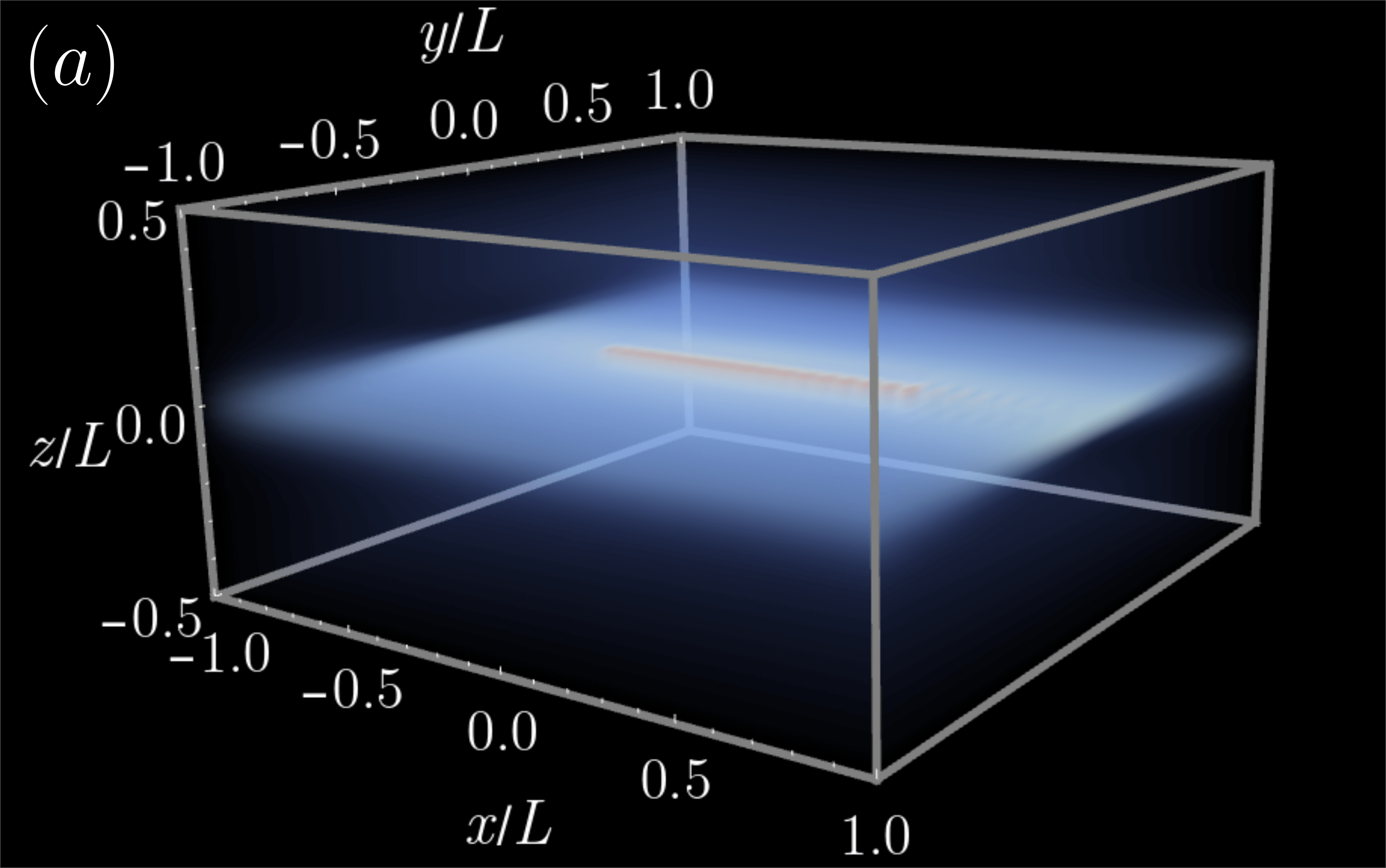}
        \end{subfigure} &

        \begin{subfigure}{\subpanelwid}
            \centering
            \includegraphics[width=\textwidth]{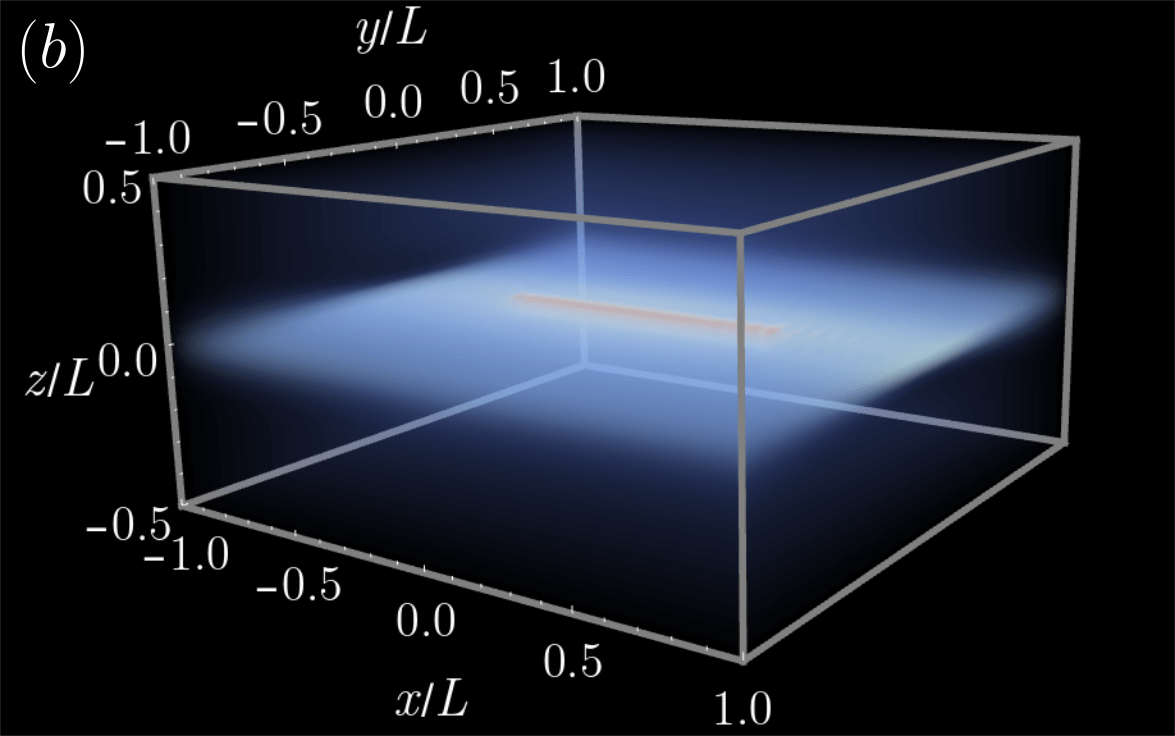}
        \end{subfigure}\\

        \begin{subfigure}{\subpanelwid}
            \centering
            \includegraphics[width=\textwidth]{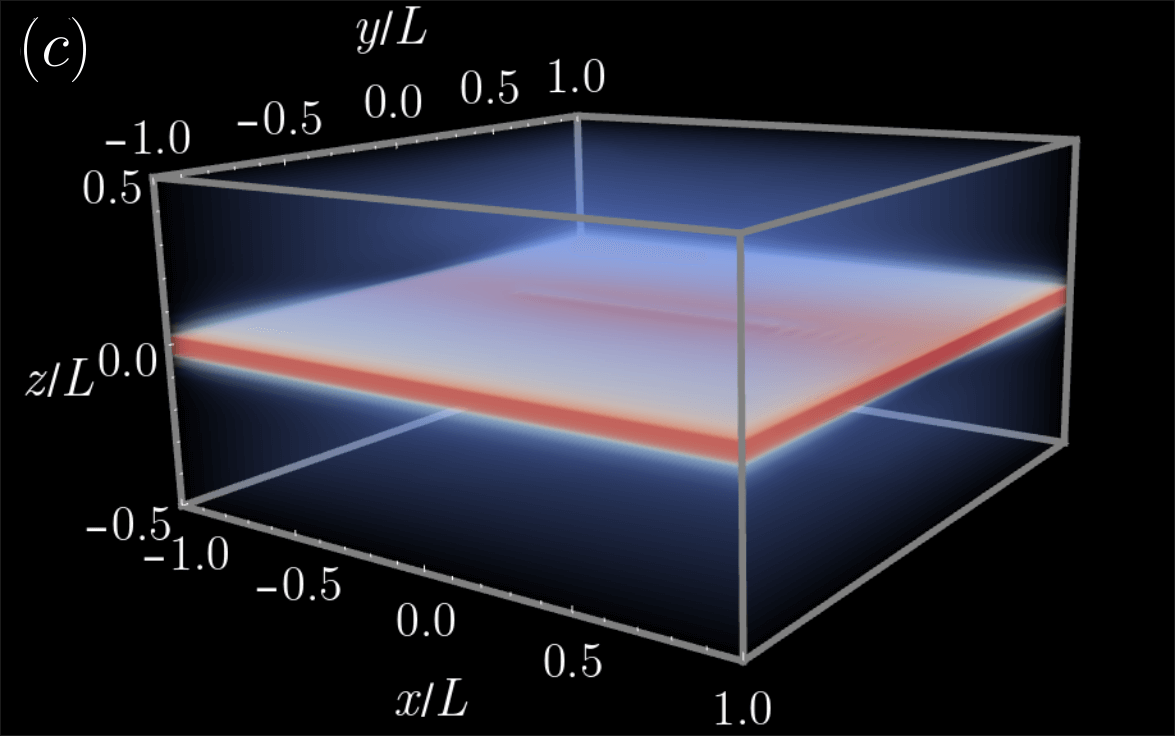}
        \end{subfigure} &

        \begin{subfigure}{\subpanelwid}
            \centering
            \includegraphics[width=\textwidth]{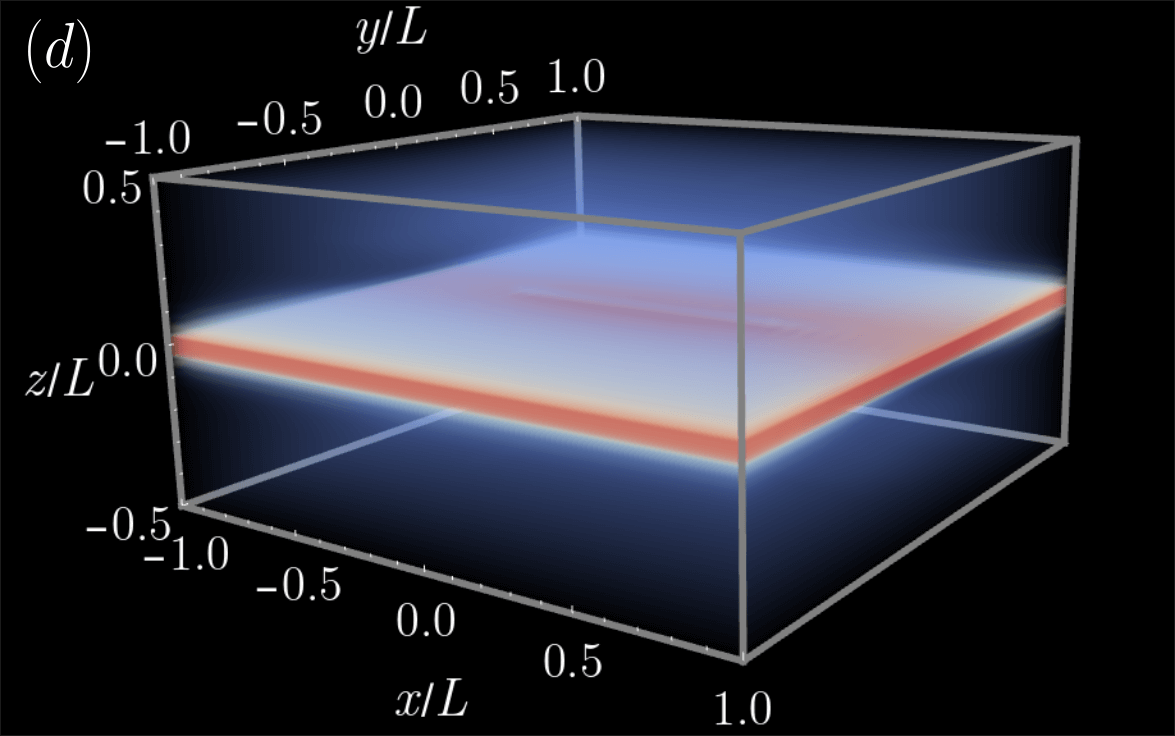}
        \end{subfigure}\\

        \begin{subfigure}{\subpanelwid}
            \centering
            \includegraphics[width=\textwidth]{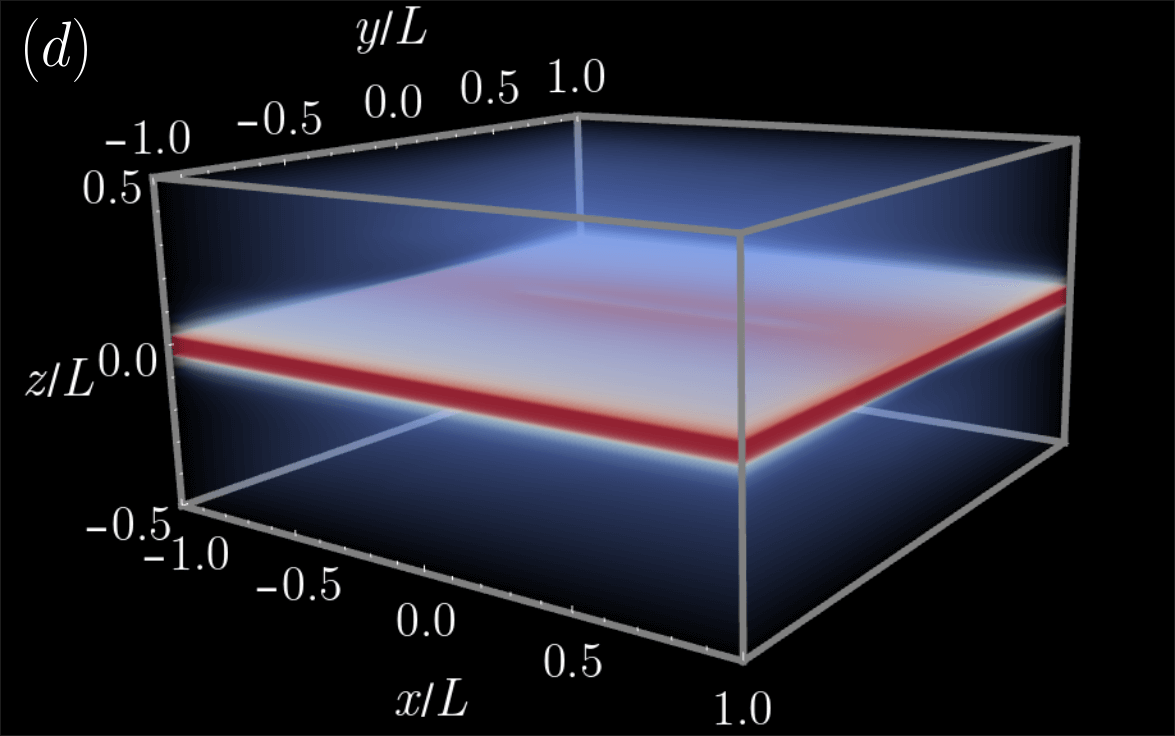}
        \end{subfigure} &

        \begin{subfigure}{\subpanelwid}
            \centering
            \includegraphics[width=\textwidth]{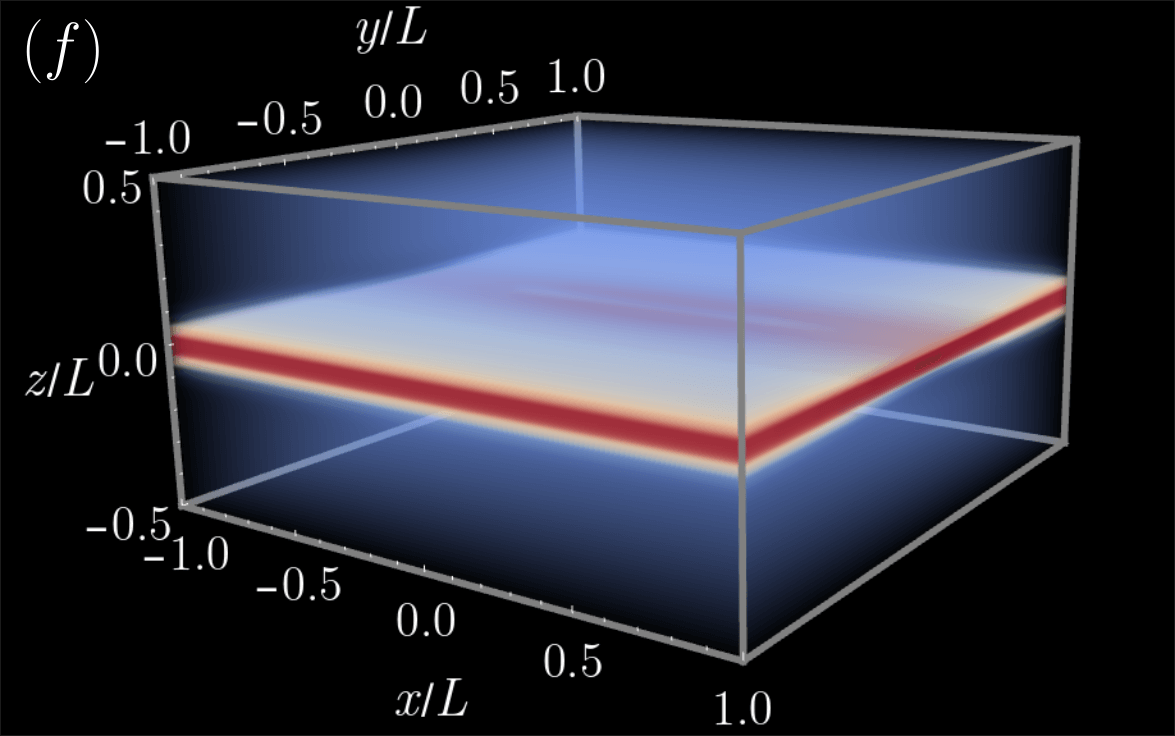}
        \end{subfigure}\\
    \end{tabular}}
    \begin{subfigure}{\textwidth}
        \centering
        \vspace{5mm}
        \includegraphics[width=.75\textwidth]{imgs/rslt_figs/colorbar_2019.png}
        \vspace{5mm}
    \end{subfigure}
    \caption{Snapshots of the effective temperature distribution $\chi(\bx, t)$ for the explicit simulation (left) and quasi-static simulation (right) for $\zeta = 10^4$. The simulation fields are qualitatively similar. In all plots, $a = 0.4$ and $\eta = 1.4$ in the opacity function. (a,b) $t = 50 t_s$. (c,d) $t = 75 t_s$. (e,f) $t = 100 t_s$.}
    \label{fig:qs_d_tem}
\end{figure}

In Fig.~\ref{fig:qs_d_tem}, we show three snapshots of the effective temperature field with $\zeta = 10^4$ from each of the two simulation methods, at $t=50t_s$, $t=75 t_s$, and $t=100t_s$ respectively. The explicit simulation is shown on the left and the quasi-static simulation is shown on the right. The results are qualitatively similar in all three snapshots. At $t=50t_s$, a shear band begins to emerge, nucleating outwards from the center of the simulation. A thin region of higher $\chi$ is visible in the center of the band. By $t = 75 t_s$, the shear band has fully formed and spans the system. At $t=100t_s$, the band grows stronger and $\chi$ continues to increase.

\begin{figure}
    \begin{tabular}{cc}
        \begin{overpic}[width=\subpanelwid]{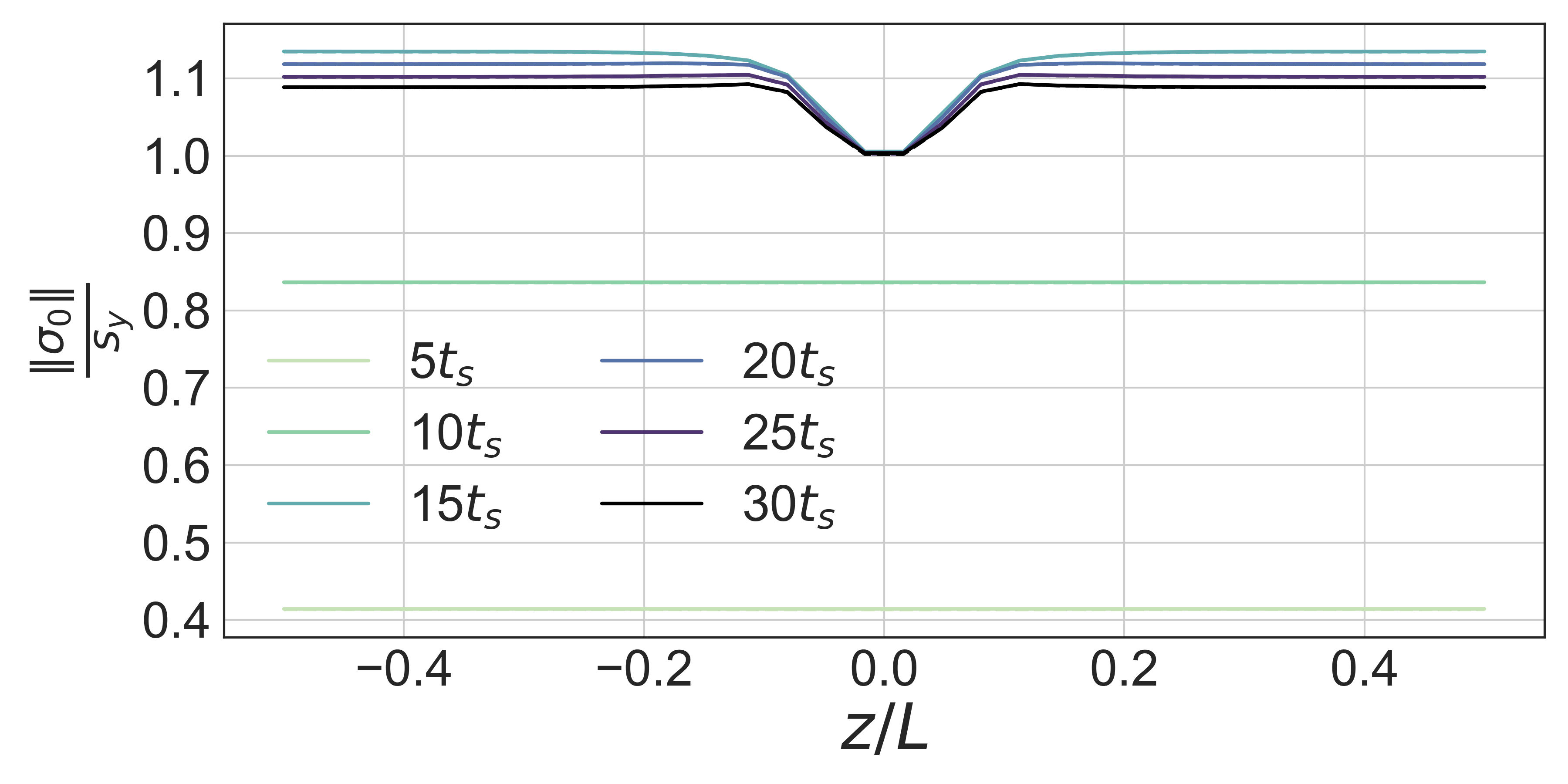}
        \put(0, 52.5){(a)}
        \end{overpic}&
        \begin{overpic}[width=\subpanelwid]{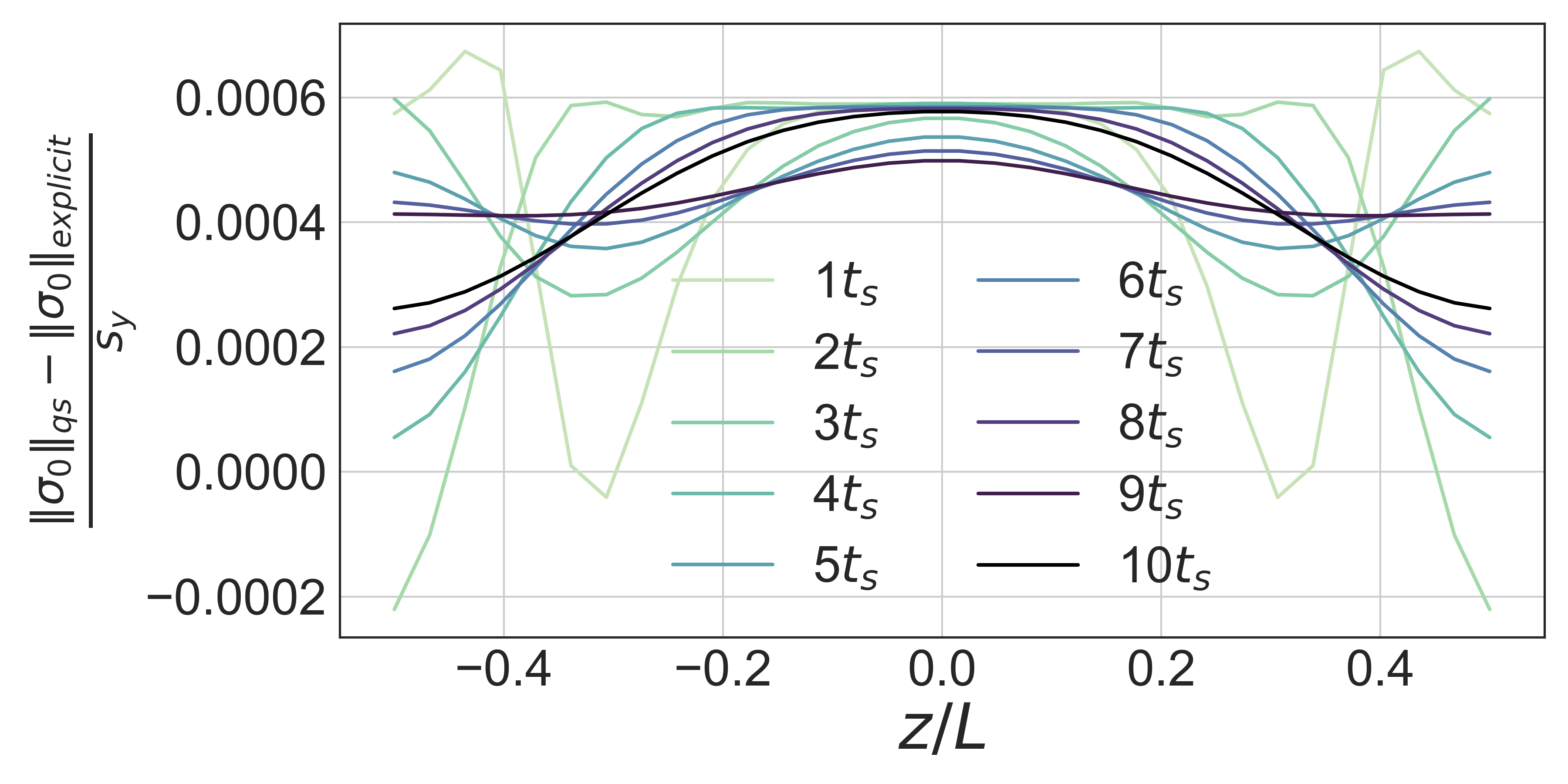}
        \put(0, 52.5){(b)}
        \end{overpic}
    \end{tabular}
    \vspace{5mm}
    \caption{(a) The magnitude of the deviatoric stress tensor $\|\bsig_0\|$ for the explicit and quasi-static simulation methods along a cross section in $z$ for $x=0$ and $y=0$. Results for the explicit and quasi-static simulation methods are shown in dashed and solid lines respectively. Oscillations at $t=5t_s$ and $t=10t_s$ are due to elastic waves propagating through the medium in the explicit simulation, but are difficult to see by eye at this scale, see next pane. As plasticity kicks in past $t=15t_s$, these waves damp out. (b) The difference in the magnitude of the deviatoric stress tensor $\|\bsig_0\|$ for the explicit and quasi-static simulation methods, along a cross section in $z$ for $x=0$ and $y=0$ fixed. The oscillations are due to elastic waves propagating through the medium in the explicit simulation.}
    \label{fig:qs_d_transect}
\end{figure}

Figure \ref{fig:qs_d_transect} shows cross-sections in $z$ for fixed $x=0$ and $y=0$ of $\|\bsig_0\|_{\text{qs}} - \|\bsig_0\|_{\text{e}}$ for several time points before the onset of plastic deformation, highlighting some differences between the two methods. The explicit simulation exhibits oscillations due to elastic waves propagating through the medium. Because the quasi-static method does not resolve these elastic waves, the oscillations are apparent in the deviatoric stress differences. When plastic deformation sets in, plasticity-induced damping removes the elastic waves and the agreement improves.

\subsection{Quantitative comparison between explicit and quasi-static methods}
\label{ssec:quant_compare}
Having demonstrated the qualitative agreement between the two simulation methodologies for $\zeta = 10^4$ in the previous section, we now examine convergence as $\zeta$ is decreased. The same simulation geometry, boundary conditions, and initial conditions in the effective temperature field are used here as in the previous section. To quantitatively compute the agreement between the explicit and quasi-static methods, we define a norm on simulation fields $\mathbf{f}$,
\begin{equation}
    \left\|\mathbf{f}\right\|(t) = \sqrt{\frac{1}{8\gamma L^3}\int_{-\gamma L}^{\gamma L}\int_{-L}^L\int_{-L}^L \left|\mathbf{f}\left(\bx, t\right)\right|^2\, dx \,dy\, dz}.
    \label{eqn:norm}
\end{equation}
The integral in Eq.~\ref{eqn:norm} runs over the entire simulation domain and is computed numerically via the trapezoid rule. The appearance of $|\cdot|$ in Eq.~\ref{eqn:norm} is taken to be the Euclidean norm for vectors, absolute value for scalars, and the Frobenius norm for matrices.

Equation \ref{eqn:norm} is evaluated for $\chi_{\text{qs}} - \chi_{\text{e}}$, $\bsig_{\text{qs}} - \bsig_{\text{e}}$, and $\vv_{\text{qs}} - \vv_{\text{e}}$ at intervals of $0.02t_s$ in pairs of explicit and quasi-static simulations with $\zeta = 10^4, 5\times 10^3, 2.5 \times 10^3$ and $1.25\times 10^3$. In each case, the norm is non-dimensionalized using the quantities $\chi_\infty$, $s_Y$, and $U_B$ to ensure the norm values for quantities of different units are of comparable magnitude. For each simulation, the explicit timestep is \smash{$\Delta t_\text{e} = \frac{t_s h^2}{20 L^2}$} and the quasi-static timestep is \smash{$\Delta t_\text{qs} = \frac{100 t_s}{\zeta}$}.

\begin{figure}
    \centering
    \begin{tabular}{cc}
        \begin{overpic}[width=\subpanelwid]{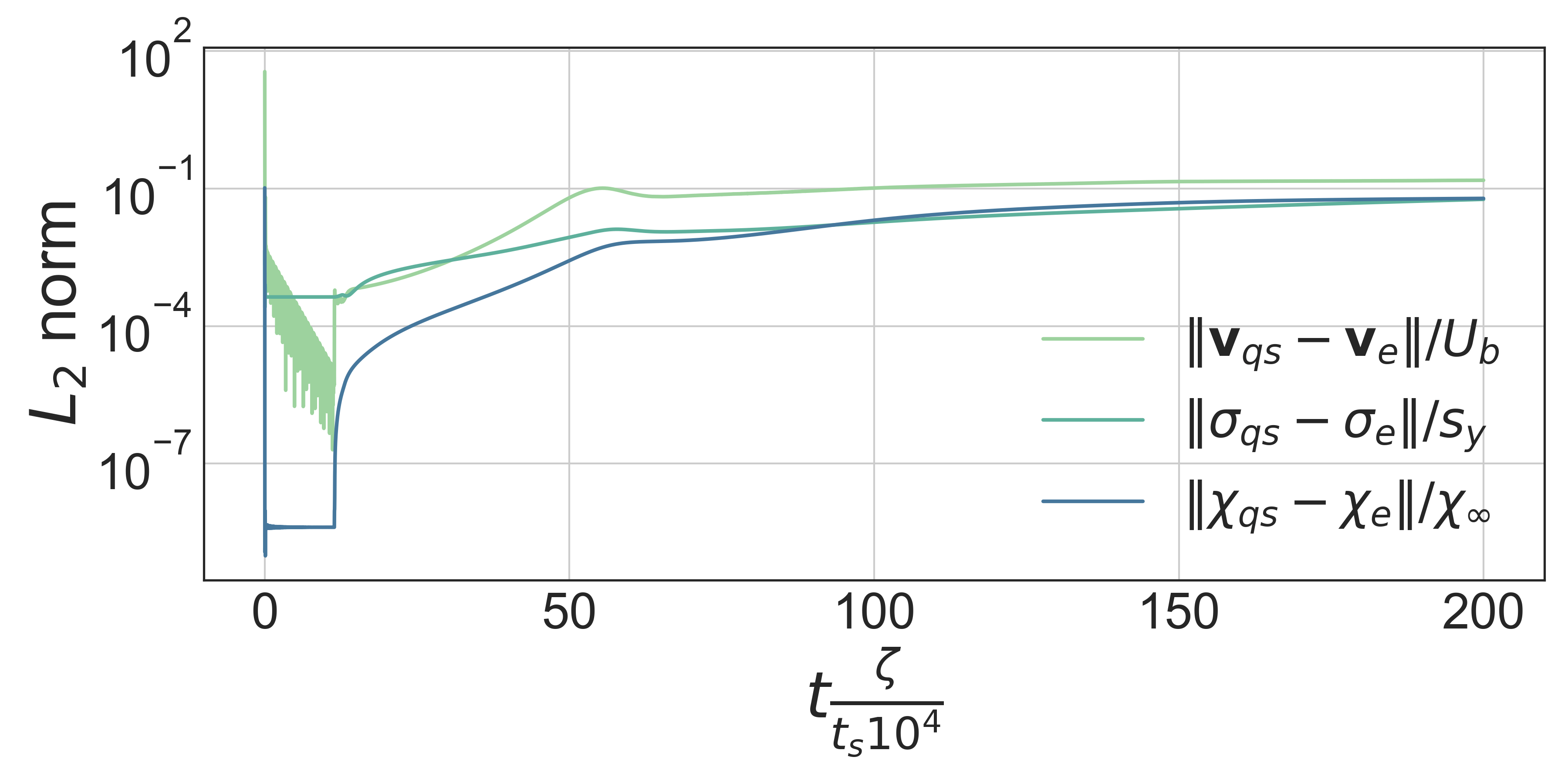}
        \put(0, 52.5){(a)}
        \end{overpic} &
        \begin{overpic}[width=\subpanelwid]{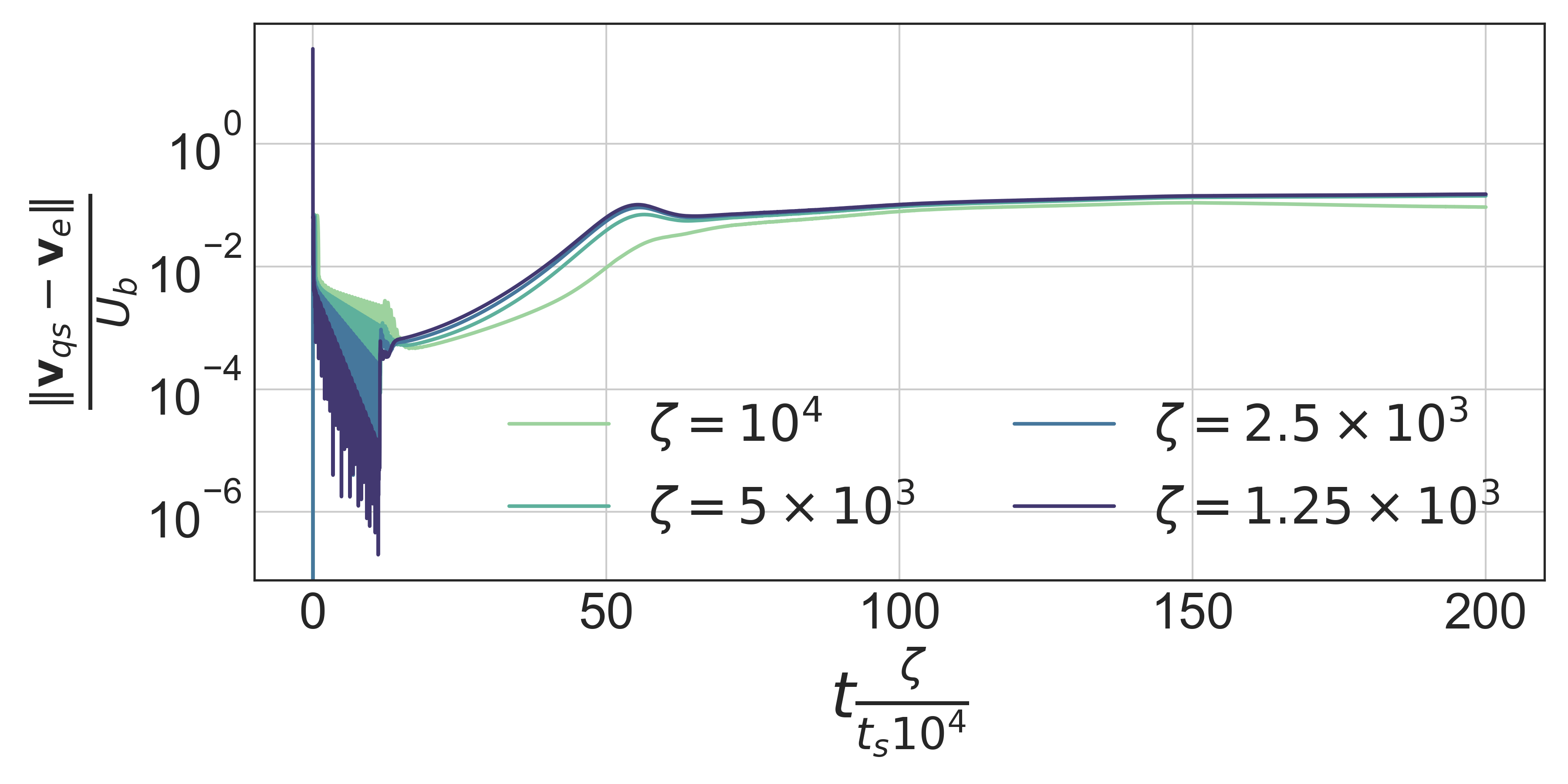}
        \put(0, 52.5){(b)}
        \end{overpic}\\
        \begin{overpic}[width=\subpanelwid]{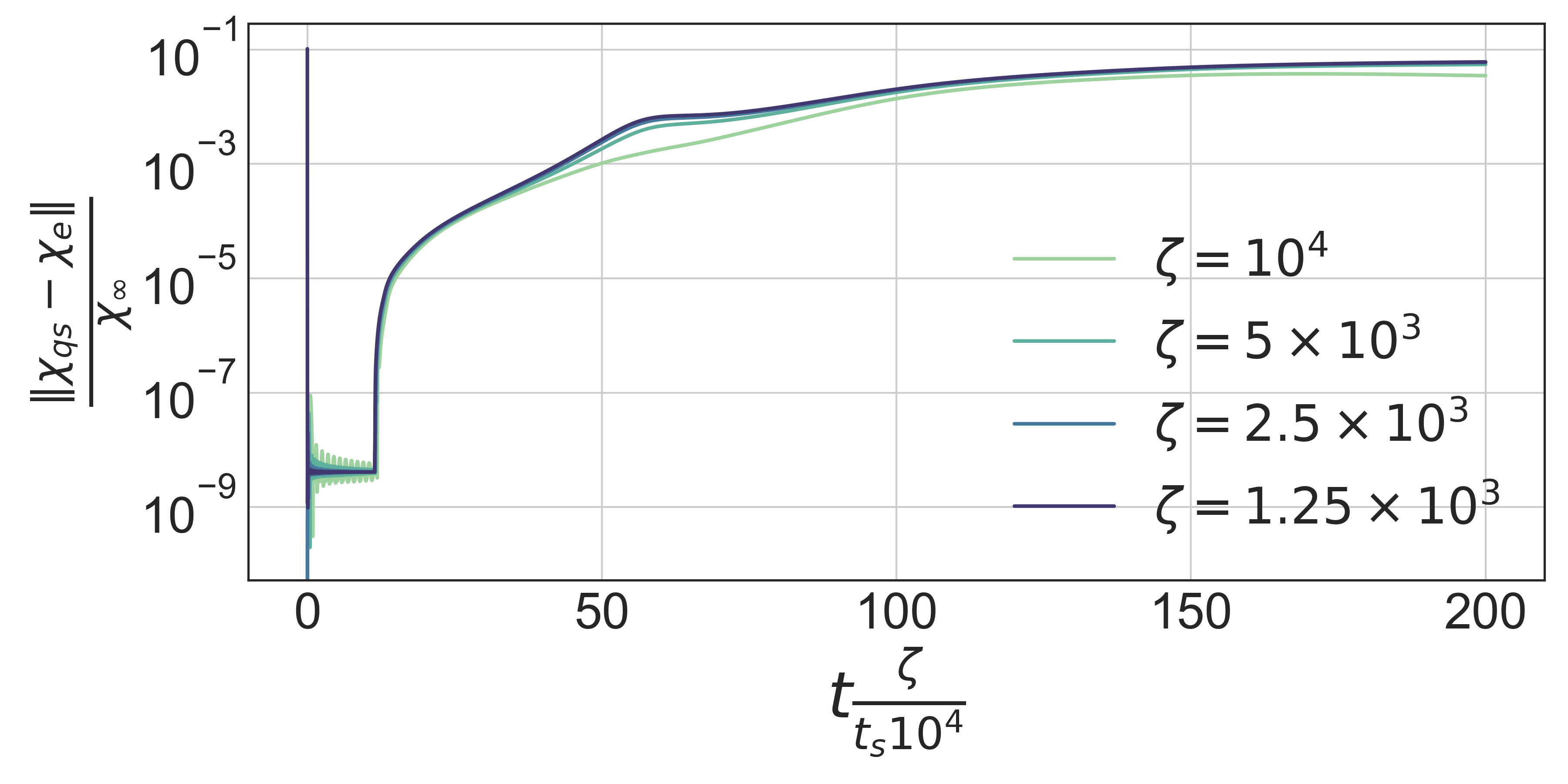}
        \put(0, 52.5){(c)}
        \end{overpic}&
        \begin{overpic}[width=\subpanelwid]{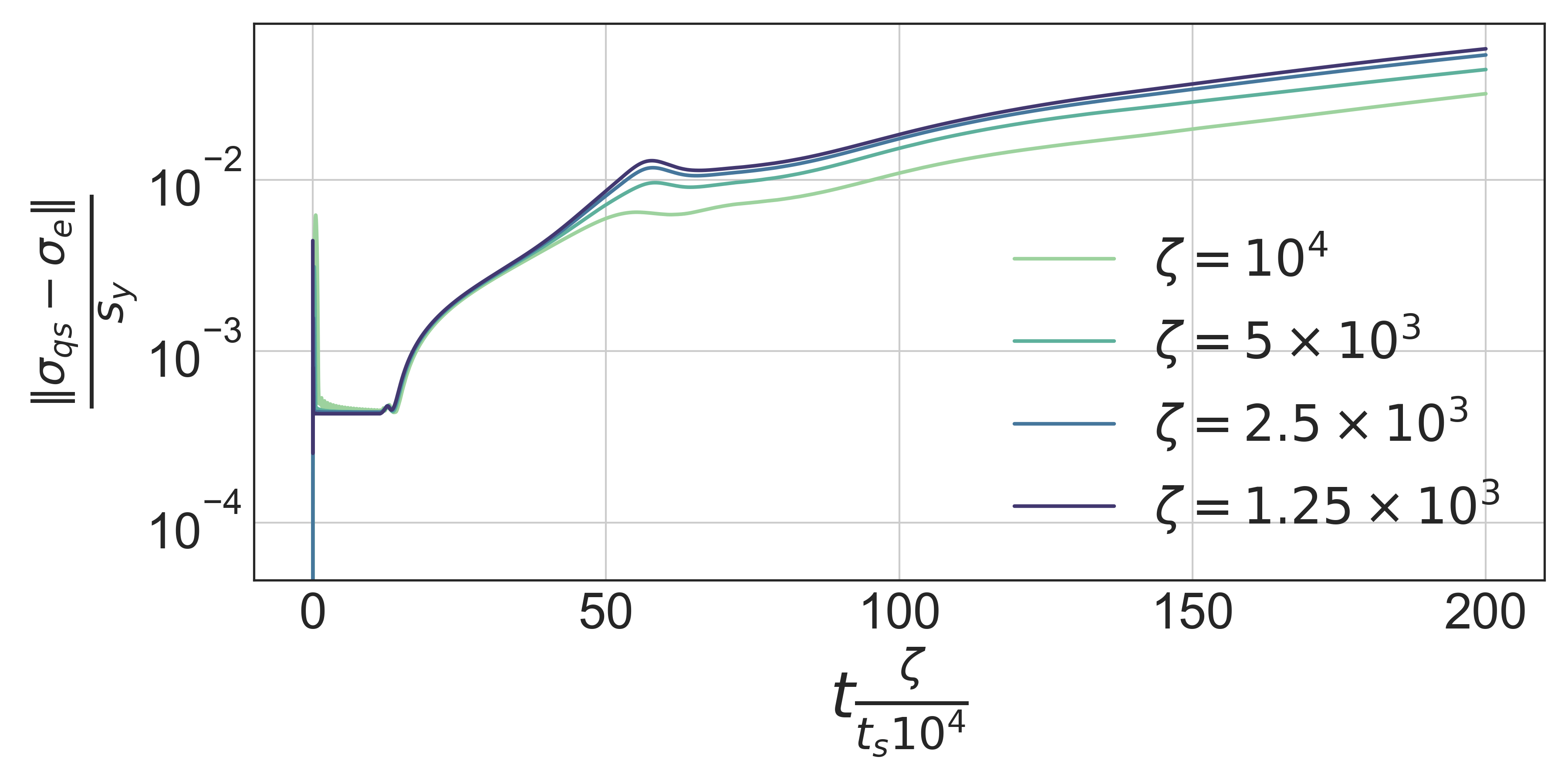}
        \put(0, 52.5){(d)}
        \end{overpic}\\
    \end{tabular}
    \vspace{5mm}
    \caption{$L_2$ norm of the $\chi$, $\vv$, and $\bsig$ simulation field differences between the explicit and quasi-static method computed using Eq.~\ref{eqn:norm} and normalized by the respective characteristic variables. (a) A comparison of the four different field norms, for the value of $\zeta = 10^4$. The remaining three panels show the differences in (b) velocity, (c) effective temperature, and (d) stress, respectively, for a range of values of $\zeta$.}
    \label{fig:qs_d_comp}
\end{figure}

Plots of all three norm values are shown as a function of time in Fig.~\ref{fig:qs_d_comp}(a) for the value of $\zeta = 1.25\times10^3$. Oscillations due to elastic waves are visible in all simulation fields until around $t \approx 12t_s$ when the yield stress is reached. After the onset of plastic deformation, the norm in effective temperature increases steadily, most rapidly during the period of shear band nucleation from $t\approx 12 t_s$ to $t \approx 25 t_s$. The disagreements in $\bsig$ and $\vv$ decrease during the elastic region, and steadily increase after plastic deformation begins.

The remaining three panels in Fig.~\ref{fig:qs_d_comp} show the quantitative comparisons as a function of time for values of $\zeta = 10^4, 5\times 10^3, 2.5\times 10^3$, and $1.25\times 10^3$ for $\vv$, $\chi$, and $\bsig$. In all plots, better agreement with smaller $\zeta$ is observed during the elastic regime and during the onset of plasticity while $t \leq 12t_s$. After shear band nucleation from $12 t_s \leq t \leq 25 t_s$, all values of $\zeta$ have roughly equal error magnitudes, with slightly greater agreement for higher values of $\zeta$. This is consistent with previous comparisons in two dimensions, where the dominant factor governing the disagreement between the two simulation methods was shown to be due to differences in the discretization rather than the value of $\zeta$ itself~\cite{rycroft15}.

\begin{figure}
    \fcolorbox{black}{black}{
    \begin{tabular}{cc}
        \begin{subfigure}{\subpanelwid}
            \centering
            \includegraphics[width=\textwidth]{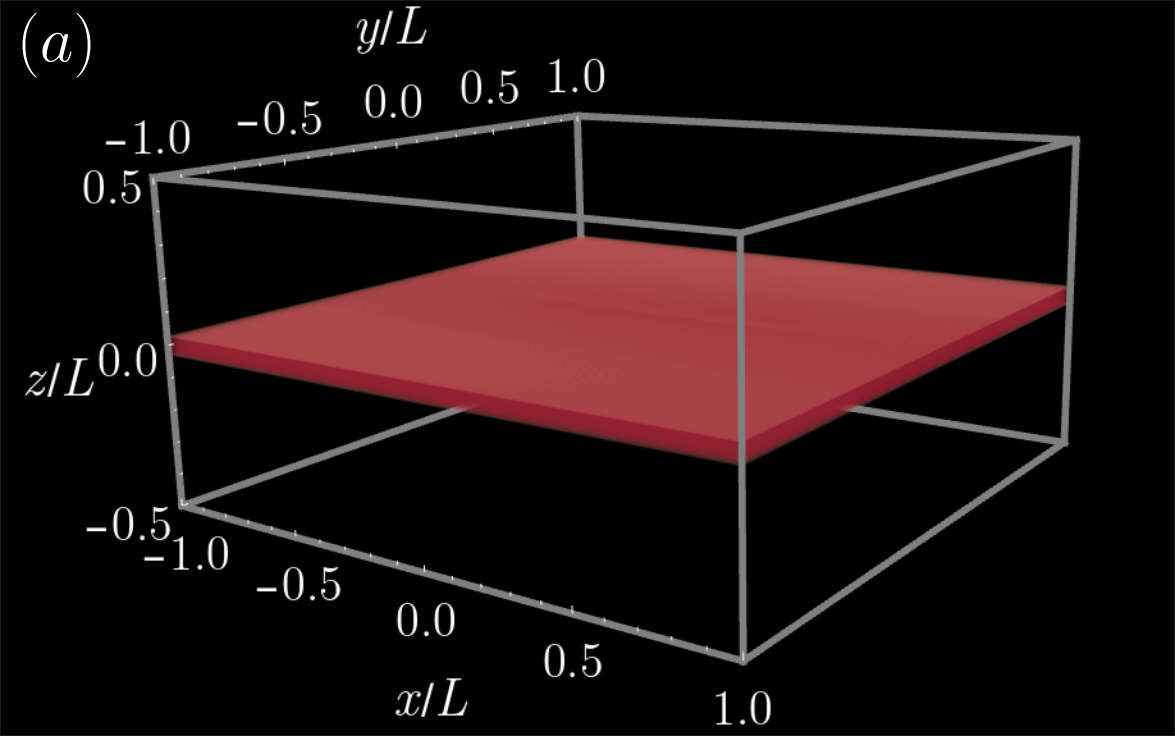}
        \end{subfigure} &
        \begin{subfigure}{\subpanelwid}
            \centering
            \includegraphics[width=\textwidth]{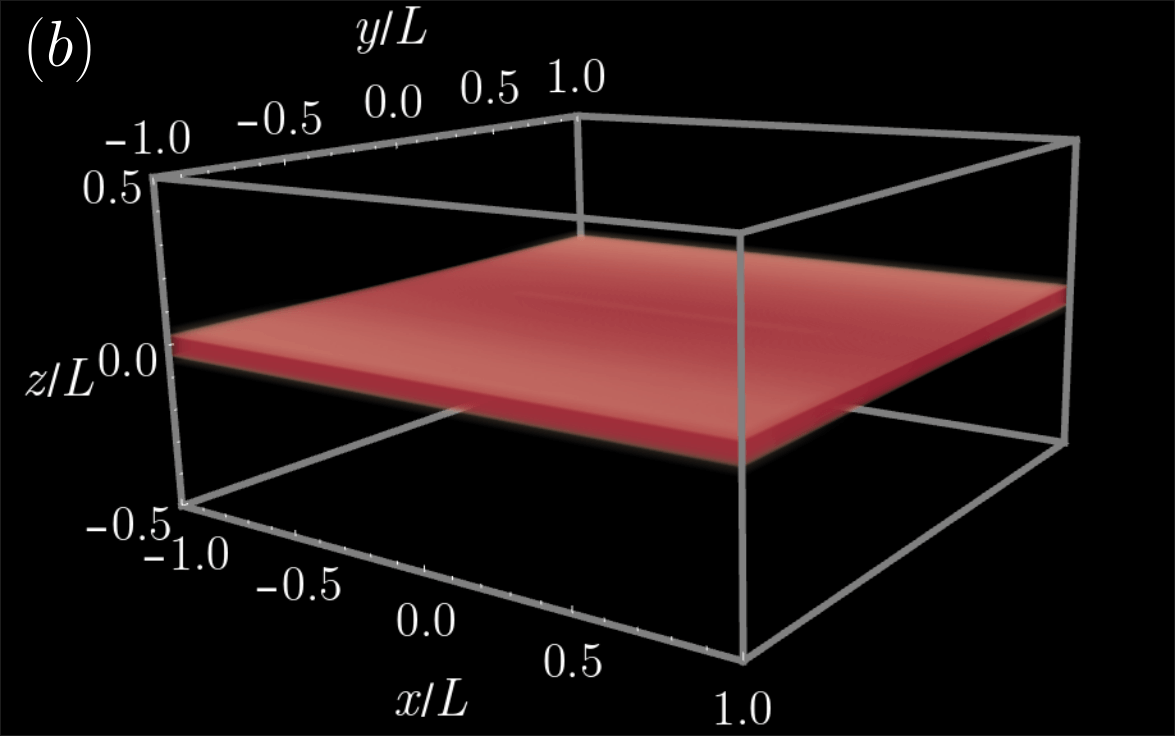}
        \end{subfigure} \\
        \begin{subfigure}{\subpanelwid}
            \centering
            \includegraphics[width=\textwidth]{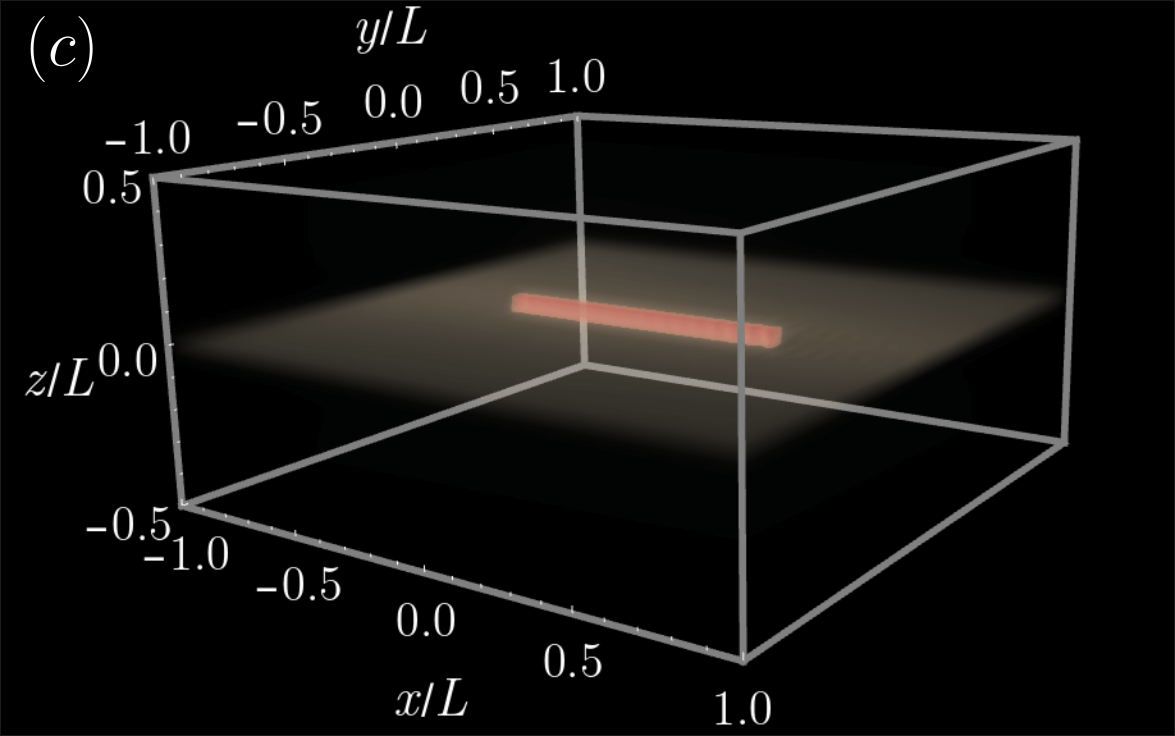}
        \end{subfigure} &
        \begin{subfigure}{\subpanelwid}
            \centering
            \includegraphics[width=\textwidth]{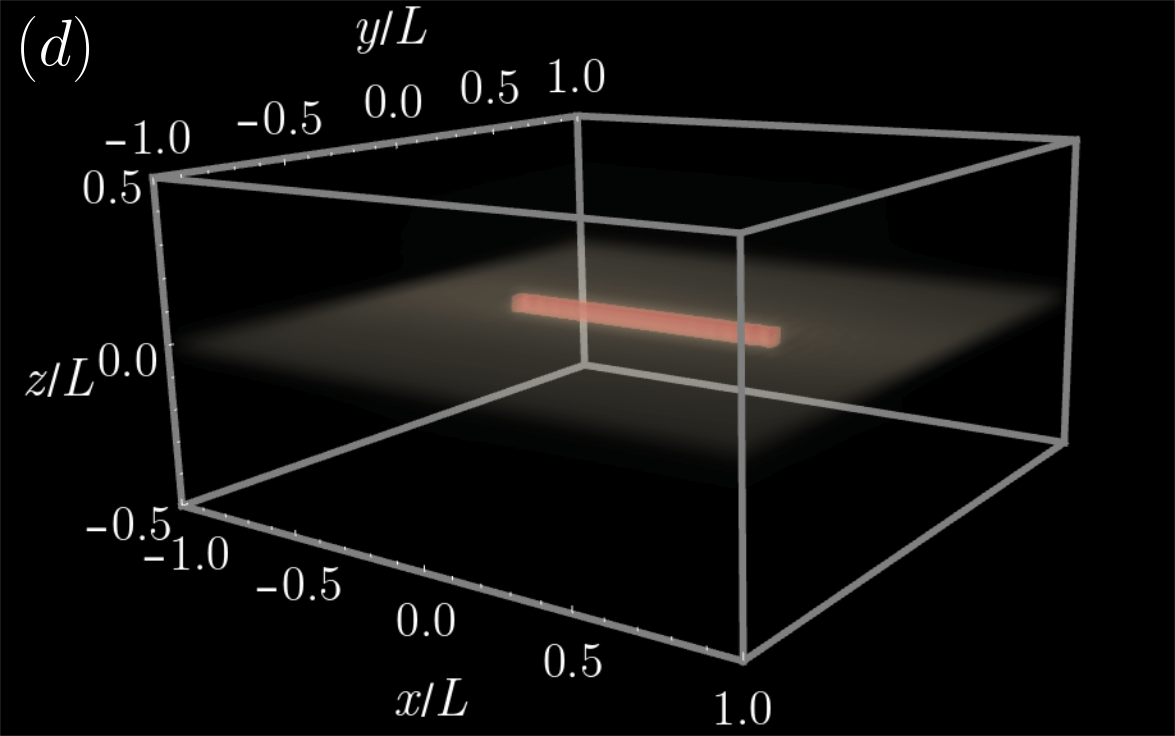}
        \end{subfigure} \\
        \begin{subfigure}{\subpanelwid}
            \centering
            \includegraphics[width=\textwidth]{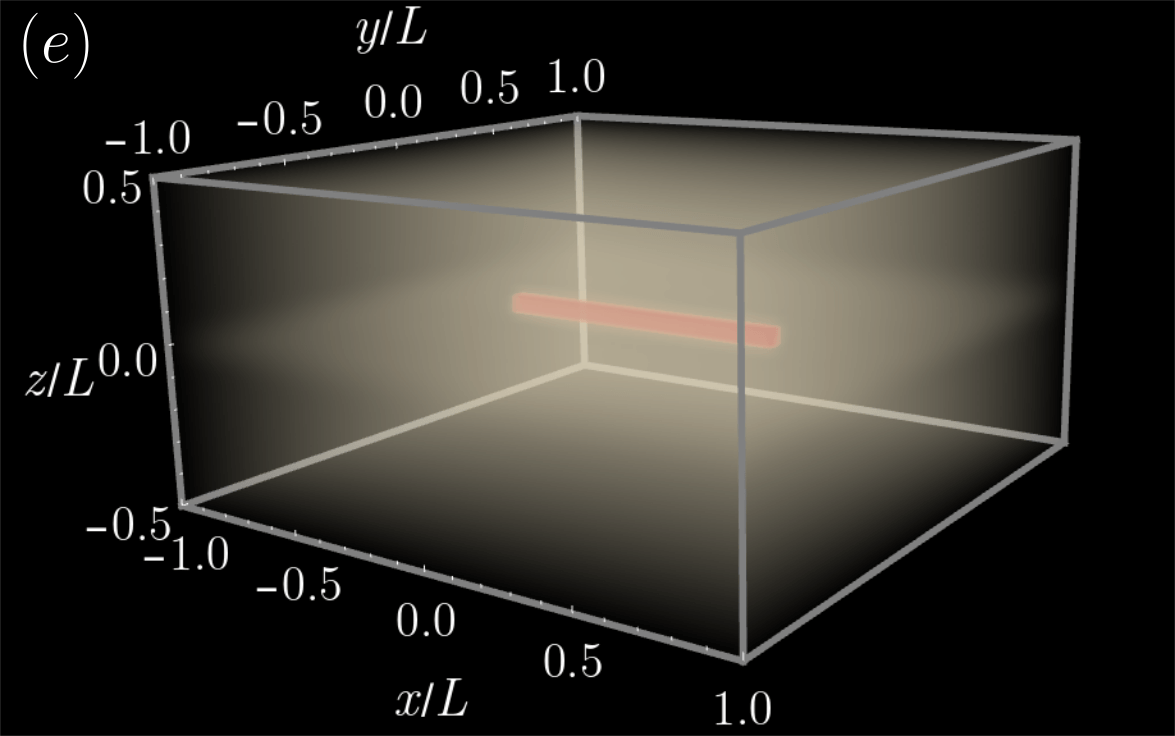}
        \end{subfigure} &
        \begin{subfigure}{\subpanelwid}
            \centering
            \includegraphics[width=\textwidth]{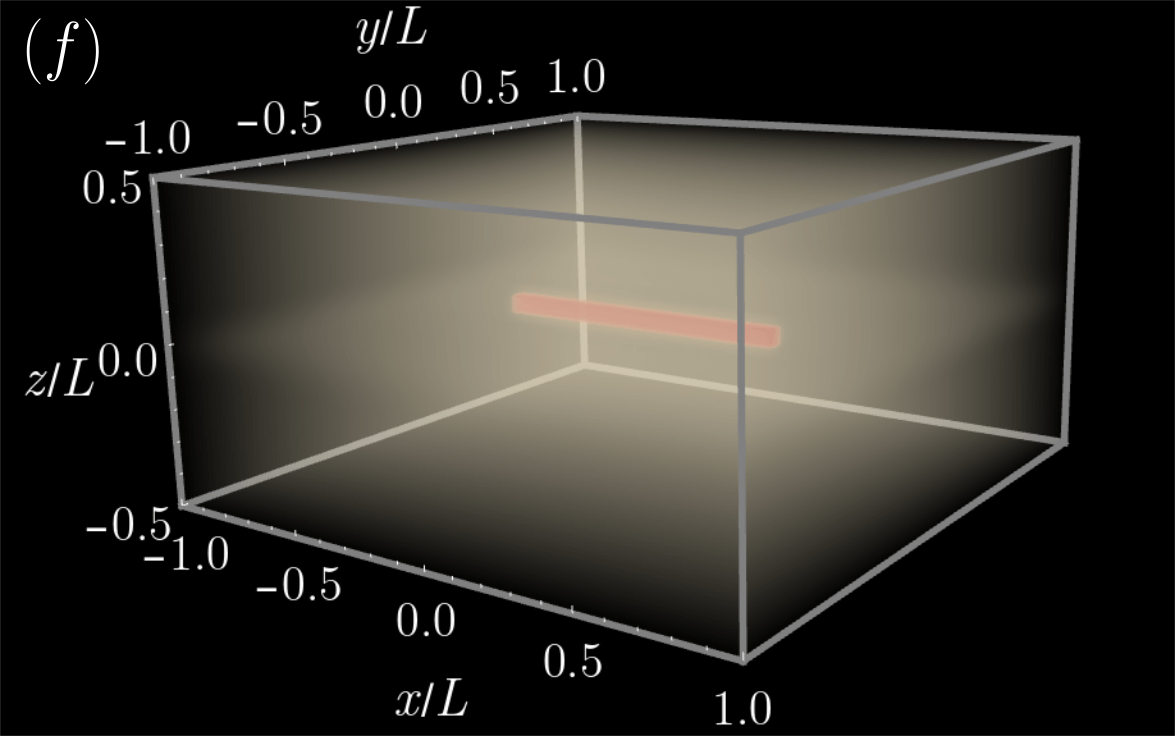}
        \end{subfigure}
    \end{tabular}}
    \begin{subfigure}{\textwidth}
        \centering
        \vspace{5mm}
        \includegraphics[width=.75\textwidth]{imgs/rslt_figs/colorbar_2019.png}
        \vspace{5mm}
    \end{subfigure}
    \caption{Qualitative demonstration of the effect of increasing the background $\chi$ field with $\chi_{\text{bg}}$ set to (a,b) $600\text{~K}$ (c,d) $650\text{~K}$ and (e,f) $700\text{~K}$. All snapshots are displayed at $t = 10^6 t_s$ for a value of $\zeta = 10^4$, with fixed values of $a=0.75$ and $\eta = 3$ in the opacity function. Simulation results are shown for the explicit method on the left and the quasi-static method on the right. For lower $\chi_\text{bg}$, the shear band is more prominent, develops more rapidly, and has more fine-scale features. The presence of fine-scale features and the rapid development of the band respectively cause the differences in spatial and temporal discretizations to become more pronounced.}
    \label{fig:qs_d_bg_chi}
\end{figure}

\begin{figure}
    \begin{tabular}{cc}
        \begin{overpic}[width=\subpanelwid]{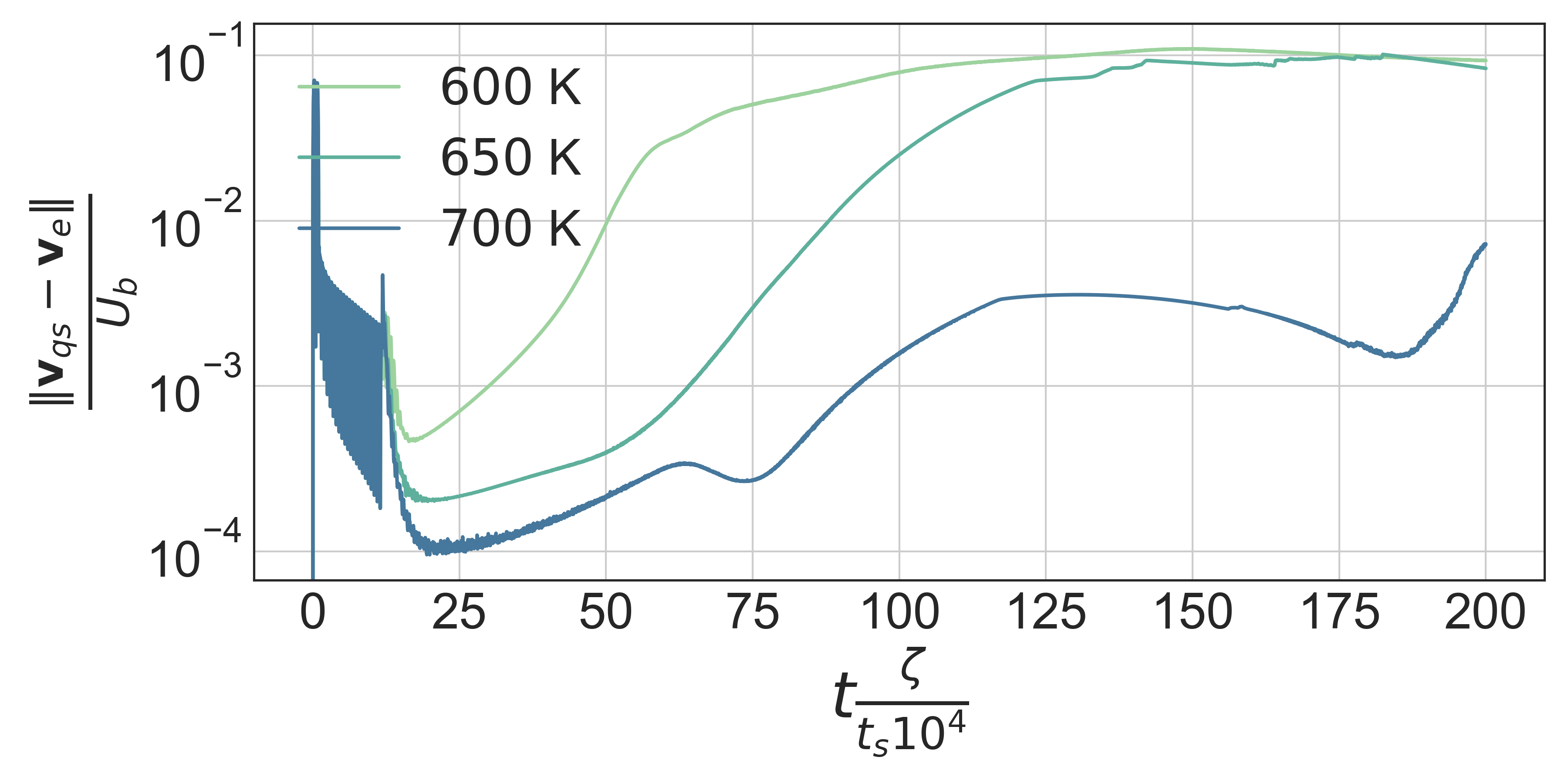}
        \put(0, 52.5){(a)}
        \end{overpic}&
        \begin{overpic}[width=\subpanelwid]{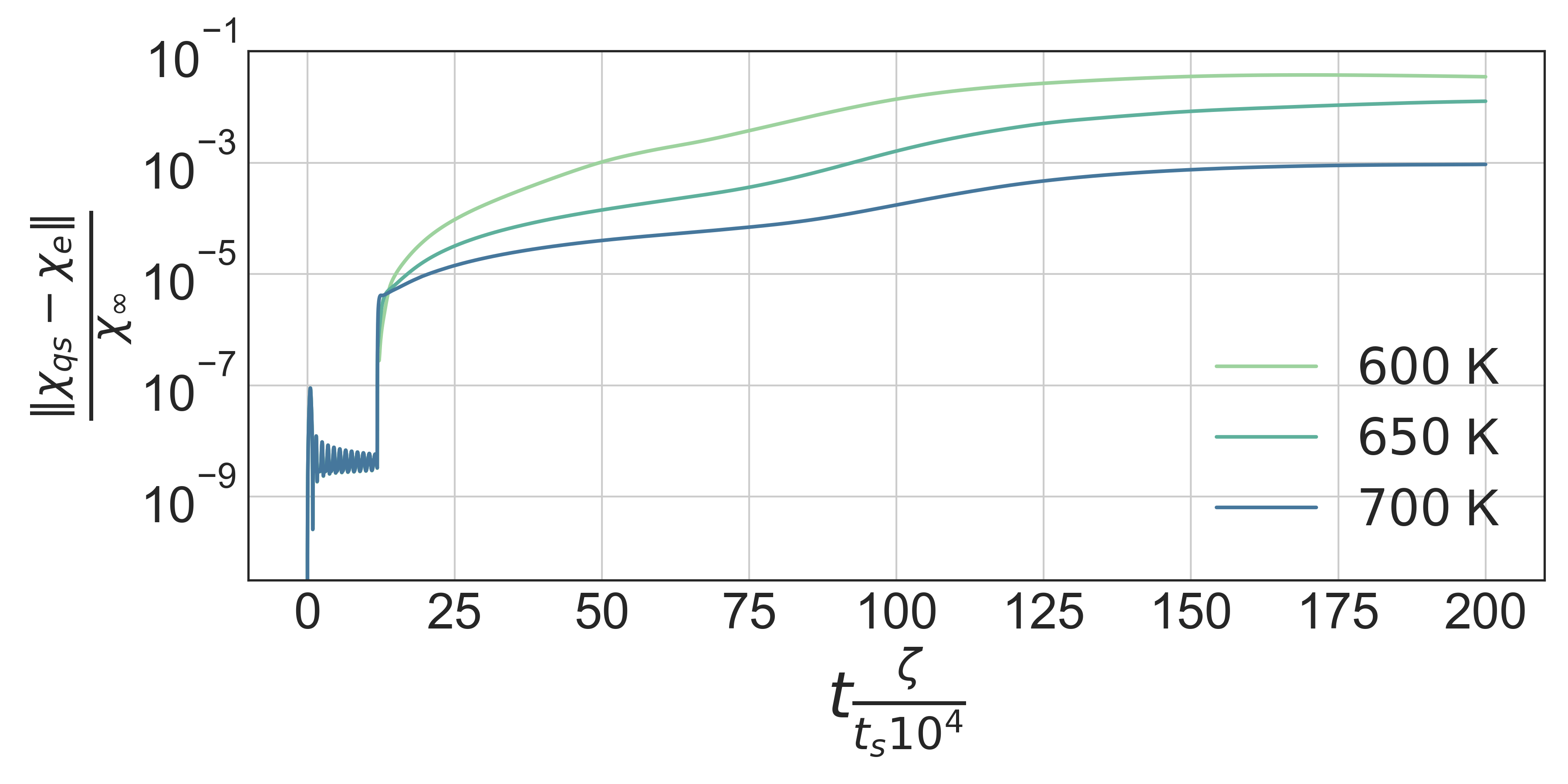}
        \put(0, 52.5){(b)}
        \end{overpic}\\
        \multicolumn{2}{c}{
        \centering
        \begin{overpic}[width=\subpanelwid]{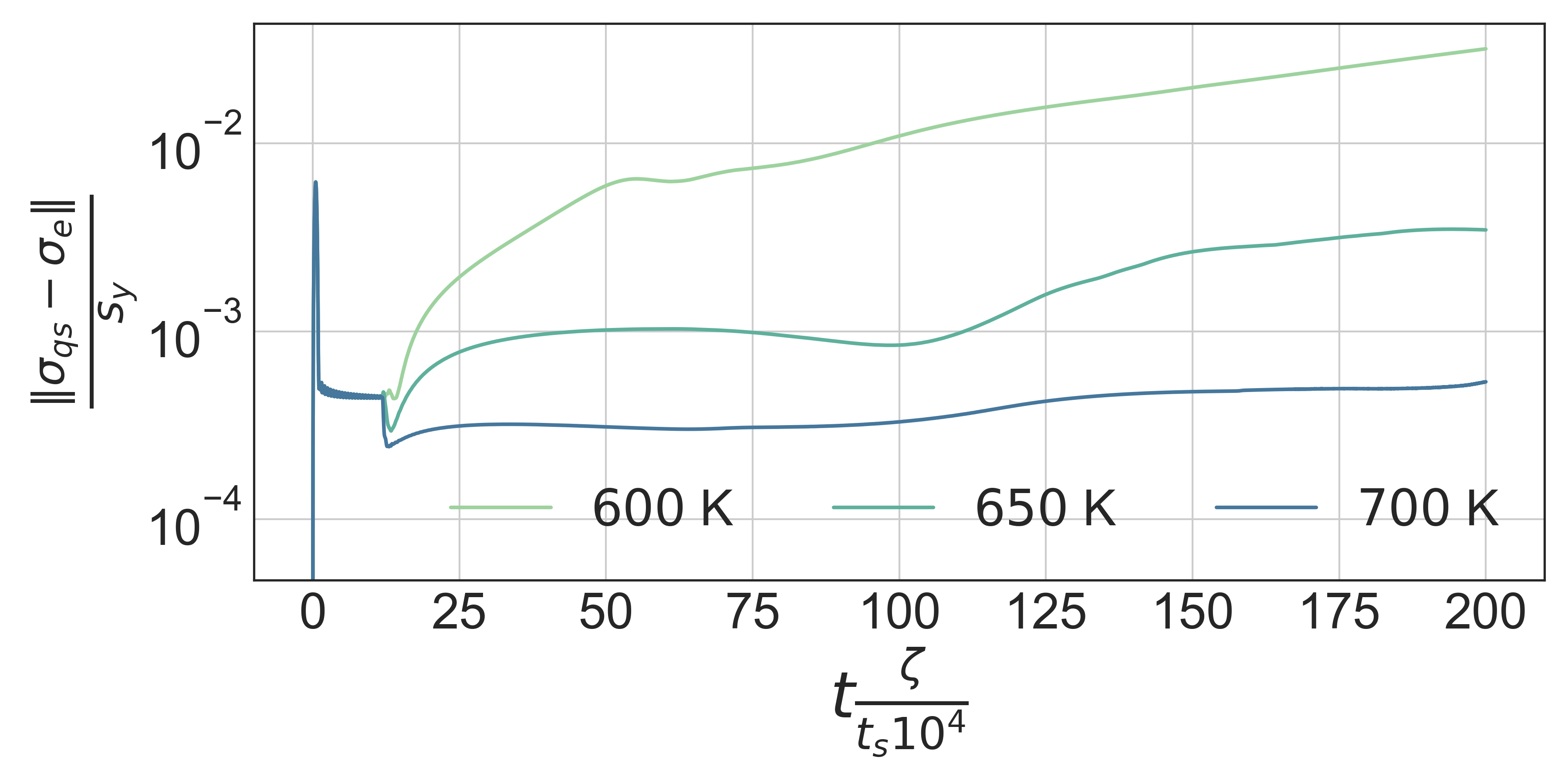}
        \put(0, 45){(c)}
        \end{overpic}}
    \end{tabular}
    \vspace{5mm}
    \caption{Normalized $L_2$ difference in (a) $\vv$ (b) $\chi$ and (c) $\bsig$ between the explicit and quasi-static methods for various choices of background $\chi$ field. Agreement improves as $\chi_{\text{bg}}$ increases. This is due to a reduction in fine-scale features of shear banding with increasing $\chi_{\text{bg}}$, which are resolved non-identically by the different discretizations.}
    \label{fig:qs_d_bg_chi_comp}
\end{figure}

A method to reduce the differences in discretization is to increase the background $\chi$ field. With higher values of background $\chi$, finer-scale features in the shear banding dynamics are less prominent. This ensures that differences in the spatial discretization will be minimized. There is also less rapid development of the shear band, and thus the difference in timestep between the two methods will be less significant. Snapshots of the effective temperature field are shown in Fig.~\ref{fig:qs_d_bg_chi} at $t = 10^6 t_s$ for background $\chi$ field values of $\chi_{\text{bg}} = 600\text{~K}, 650\text{~K}$, and $700\text{~K}$. Figure \ref{fig:qs_d_bg_chi_comp} confirms that the differences between the two types of simulation decreases as $\chi_{\text{bg}}$ increases.

\subsection{Parallel scaling analysis}
\begin{figure}
\fcolorbox{black}{black}{
    \begin{tabular}{cc}

        \begin{subfigure}{\subpanelwid}
            \centering
            \includegraphics[width=\textwidth]{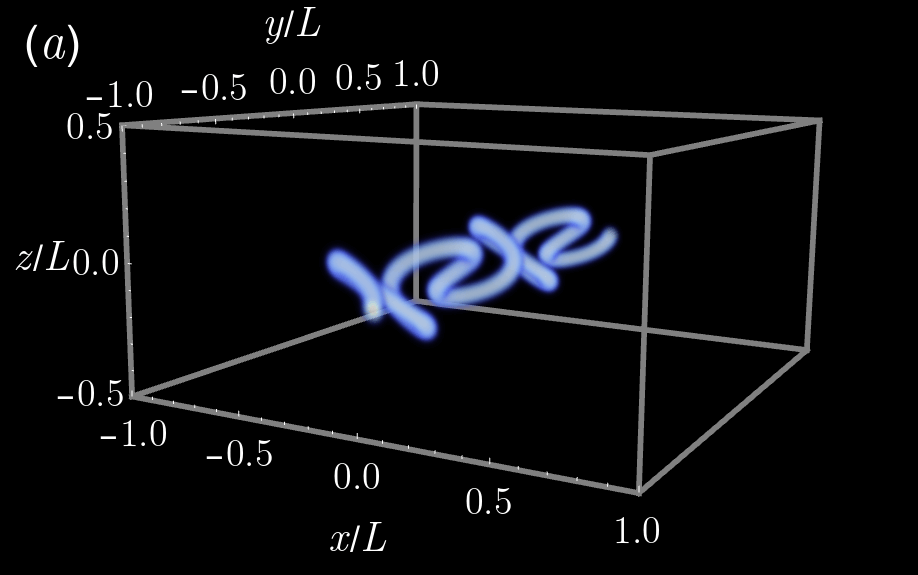}
        \end{subfigure} &

        \begin{subfigure}{\subpanelwid}
            \centering
            \includegraphics[width=\textwidth]{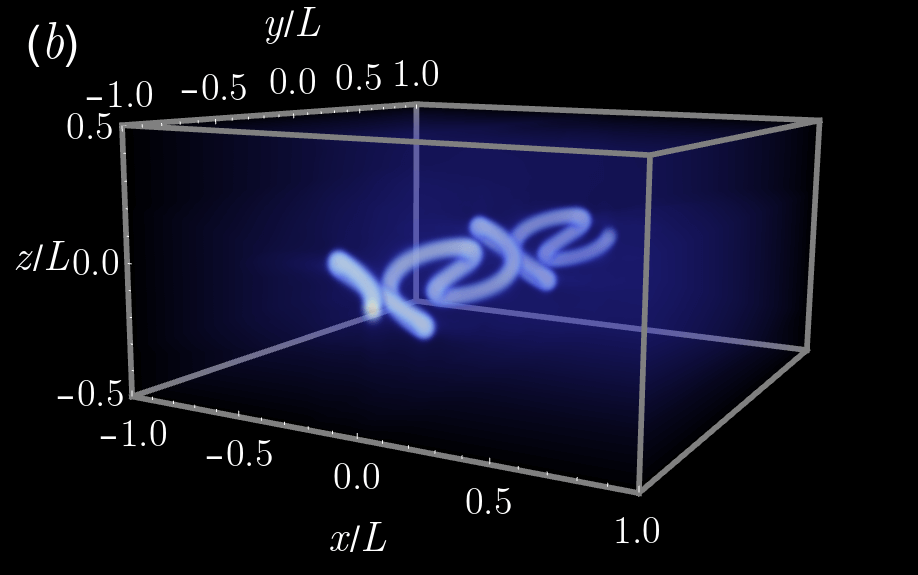}
        \end{subfigure}\\

        \begin{subfigure}{\subpanelwid}
            \centering
            \includegraphics[width=\textwidth]{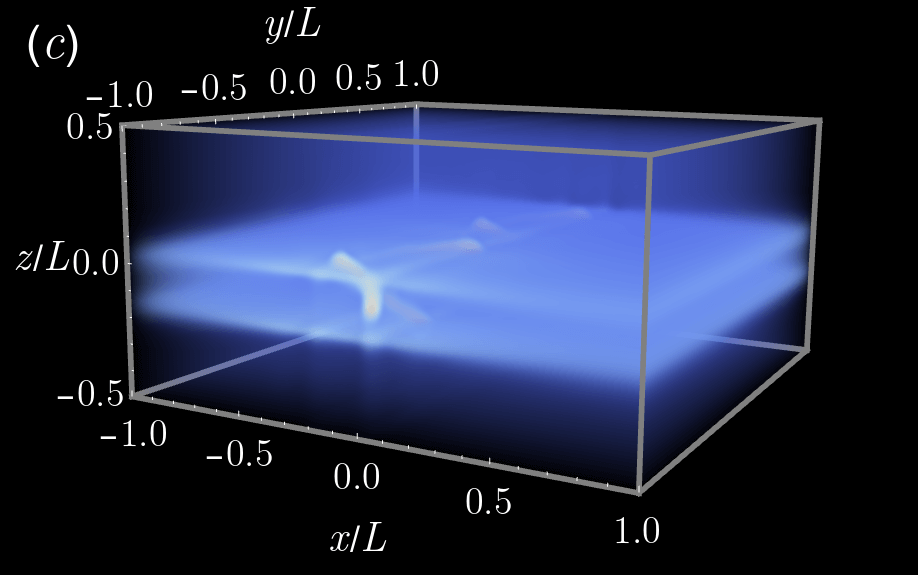}
        \end{subfigure} &

        \begin{subfigure}{\subpanelwid}
            \centering
            \includegraphics[width=\textwidth]{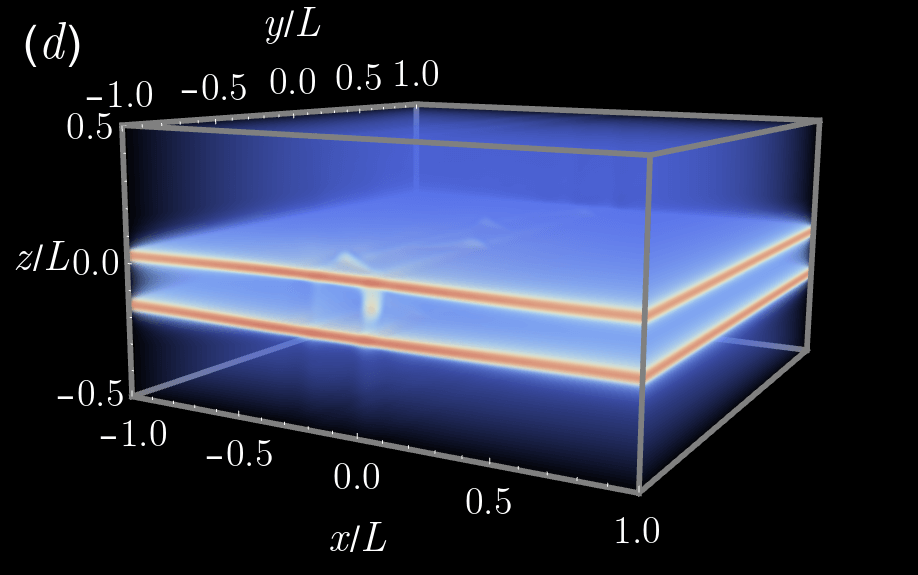}
        \end{subfigure}\\

        \begin{subfigure}{\subpanelwid}
            \centering
            \includegraphics[width=\textwidth]{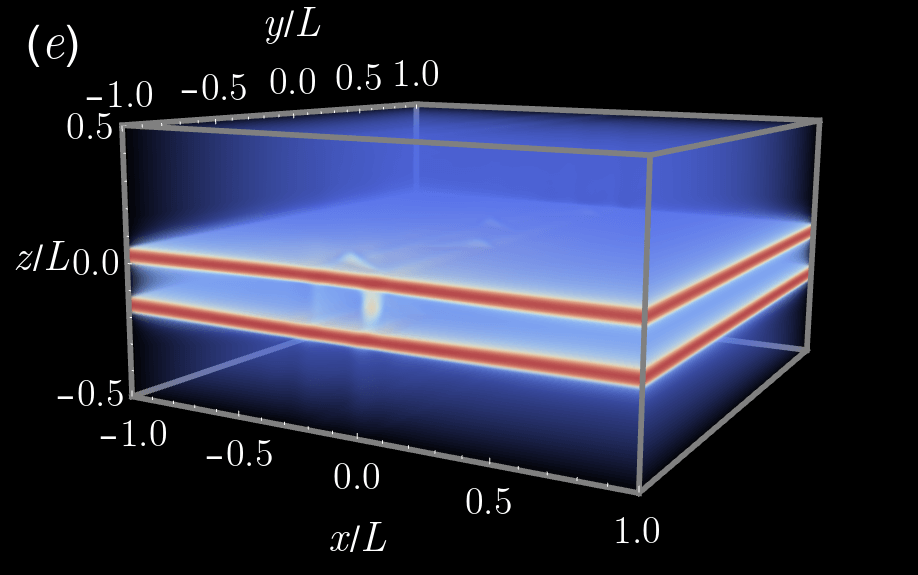}
        \end{subfigure} &

        \begin{subfigure}{\subpanelwid}
            \centering
            \includegraphics[width=\textwidth]{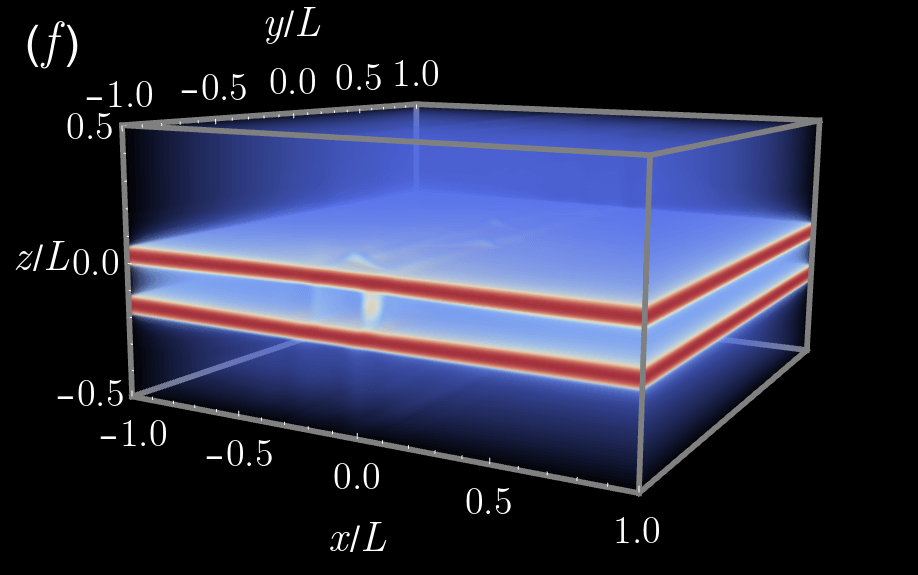}
        \end{subfigure}\\
    \end{tabular}}
    \begin{subfigure}{\textwidth}
        \centering
        \vspace{5mm}
        \includegraphics[width=.75\textwidth]{imgs/rslt_figs/colorbar_2019.png}
        \vspace{5mm}
    \end{subfigure}
    \caption{Snapshots of the effective temperature distribution $\chi(\bx, t)$ for a quasi-static simulation with $\zeta = 1$. The initial condition is given in Eq.~\ref{eqn:case12_chi}, corresponding to a non-commensurate helix. (a) $t = 0 t_s$. $a=0.25$, $\eta = 1.2$. (b) $t = 10^5 t_s$. $a=0.3, \eta=1.3$. (c) $t = 2\times 10^5 t_s$. $a=0.4, \eta=1.5$. (d) $t = 2.5\times 10^5 t_s$. $a=0.4, \eta=1.5$. (e) $t = 3\times 10^5 t_s$. $a=0.4, \eta=1.5$. (f) $t = 4\times 10^5t_s$. $a=0.4, \eta=1.5$.}
    \label{fig:case12}
\end{figure}

In this subsection, we present a parallel scaling analysis of our quasi-static algorithm implementation. All simulations are run on an Ubuntu Linux computer with dual 16-core 2.10~GHz Intel Xeon E5-2683 v4 processors.  We consider two measures of parallel scaling. First, we vary the number of processors $p$ while keeping $N_xN_yN_z/p$ approximately fixed as a test of weak parallel scaling. We denote $N = N_x = N_y = 2N_z$ so that the total number of grid points is $N^3/2$. For this weak scaling experiment, we consider an initial condition in the effective temperature field corresponding to a non-commensurate helix, given by
\begin{align}
    \delta_x &= \frac{x}{L} - \left(\frac{\cos\left(6\pi\left(\frac{y}{L} + 1\right)\right)}{8} - \frac{1}{16}\right),\nonumber\\
    \delta_z &= \frac{z}{L} - \left(\frac{\sin\left(4\pi\left(\frac{y}{L} + 1\right)\right)}{8} - \frac{1}{16}\right),\nonumber\\
    \chi\left(\bx, t=0\right) &= 550\text{~K} + \left(200\text{~K}\right)e^{-750\left(\delta_x^2 + \delta_z^2\right)}.
    \label{eqn:case12_chi}
\end{align}
We consider the values $N = 80, 102, 128, 164, 204, 256$ and matching values $p = 1, 2, 4, 8, 16, 32$. Each experiment uses a boundary shear velocity of $U_b = 10^{-7}\frac{L}{t_s}$, a diffusion lengthscale of $l = \frac{3}{2} h$, a quasi-static timestep of $\Delta t = 200 t_s$, and is simulated to a final time $t_f = 10^6 t_s$. The shear banding dynamics for $N=256$ are shown in Fig.~\ref{fig:case12}. In Fig.~\ref{fig:case12}(a), the initial condition is shown. In Fig.~\ref{fig:case12}(b), the background effective temperature field begins to increase, but shear banding has not yet begun. In Fig.~\ref{fig:case12}(c), clear system-spanning shear bands have begun to nucleate off the top and bottom of the helix. In Figs.~\ref{fig:case12}(d)--(f), the shear bands continue to become more prominent and thicken.

Experimental results are shown in Table \ref{tab:fix_n_over_p}. The weak scaling ratio is computed using the formula \smash{$w = \frac{r t_{1}}{t_p}$} where $t_1$ is the time for the simulation with one processor, $t_p$ is the time for the simulation with $p$ processors, and $r = N^3_p/(p N_1^3)$ is a small adjustment factor for deviations from perfectly equal work. The total time increases with grid size at a significantly reduced rate when compared to serial computation, and the weak scaling ratios for $p\neq 32$ indicate good weak scaling performance. The value of the weak scaling ratios at $p=32$ may indicate an issue such as limitations of communication buffers. These results hold for both the total time and the V-cycle computation, with the V-cycle computations showing mildly worse weak scaling than the overall computation. The number of required V-cycles increases slightly as the grid size is increased.

\begin{table}
  \begin{center}
    \small
    \begin{tabular}{|l|l|l|l|l|l|l|}\hline
                           & \shortstack{\strut$N=256$\\ $p=32$} & \shortstack{\strut$N=204$\\ $p=16$} & \shortstack{\strut$N=164$\\ $p=8$} & \shortstack{\strut$N=128$\\ $p=4$} & \shortstack{\strut$N=102$\\ $p=2$} & \shortstack{\strut $N=80$ \\ $p=1$} \\ \hline
    Total time (hours)     & 3.69         & 2.31         & 2.36        & 1.94        & 1.74        & 1.52   \\\hline
    V-cycle time (hours)   & 2.75         & 1.71         & 1.74        & 1.37        & 1.20        & 1.00   \\\hline
    \# of V-cycles         & 17939           & 17539           & 17171          & 16511          & 15575          & 15073    \\\hline
    Time/V-cycle (seconds) & 0.55        & 0.35        & 0.37       & 0.30       & 0.28       & 0.24 \\\hline
    WS ratio (total time) & 0.42       & 0.68         & 0.69       & 0.80       & 0.90   & 1.0 \\\hline
    WS ratio (V-cycle time) & 0.37 & 0.61         & 0.62       & 0.75       & 0.86   & 1.0 \\\hline
    \end{tabular}
  \end{center}
  \caption{Data describing the total time per simulation, the time spent in all V-cycles per simulation, the total number of V-cycles, the average time per V-cycle, and the weak scaling ratios for total time and V-cycle time. Values of $N$ and $p$ were chosen to keep the number of grid points per processor $N^3/(2p)$ approximately constant. The weak scaling (WS) ratio is computed using the formula $w = r t_1/t_p$ where $t_1$ is the time for the simulation with one processor, $t_p$ is the time for the simulation with $p$ processors, and $r = N^3_p/(p N_1^3)$ is a small adjustment factor for deviations from perfectly equal work.}
    \label{tab:fix_n_over_p}
\end{table}

\begin{table}
  \begin{center}
    \begin{tabular}{|l|l|l|l|l|l|l|}\hline
                           & $N=128$  & $N=256$ & $N=384$  & $N=512$  & $N=768$ \\\hline
    Total time (hours)     & 1.61  & 6.12 & 22.44  & 42.86  & 105.97 \\ \hline
    V-cycle time (hours)   & 1.27   & 3.89 & 14.28  & 30.23  & 75.90 \\ \hline
    \# of V-cycles         & 20480    & 18367   & 24675    & 28175    & 19689   \\ \hline
    Time/V-cycle (seconds) & 0.22  & 0.76 & 2.08  & 3.87  & 13.88 \\ \hline
    \end{tabular}
  \end{center}
  \caption{Data describing the total time per simulation, the time spent in all V-cycles per simulation, the total number of V-cycles, and the average time per V-cycle. $p$ is fixed at 32 while $N$ is varied.}
  \label{tab:fix_p_vary_N}
\end{table}

\begin{figure}
    \centering
    \begin{tabular}{cc}
        \begin{overpic}[width=\subpanelwid]{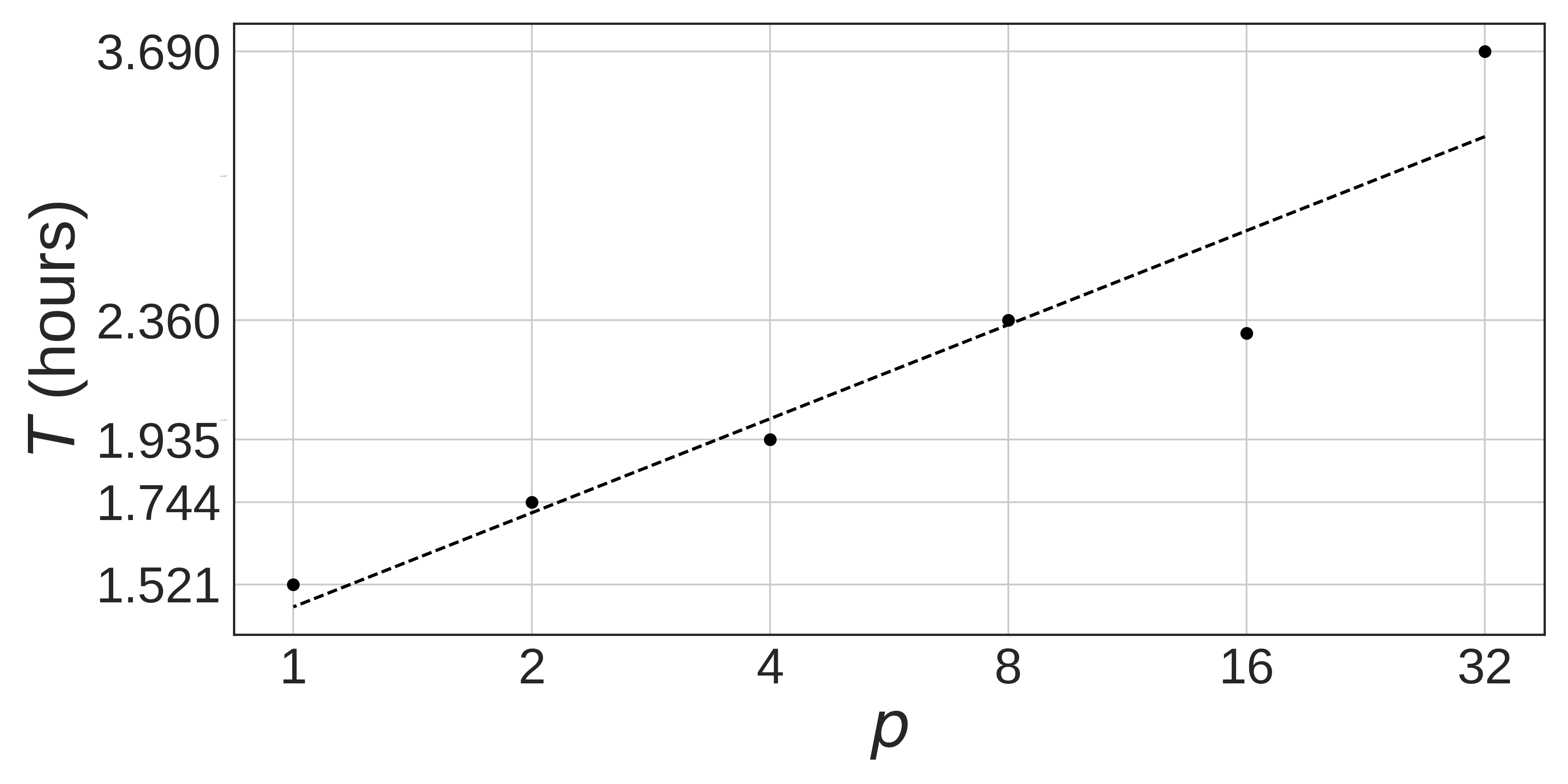}
        \put(0, 52.5){(a)}
        \end{overpic} &
        \begin{overpic}[width=\subpanelwid]{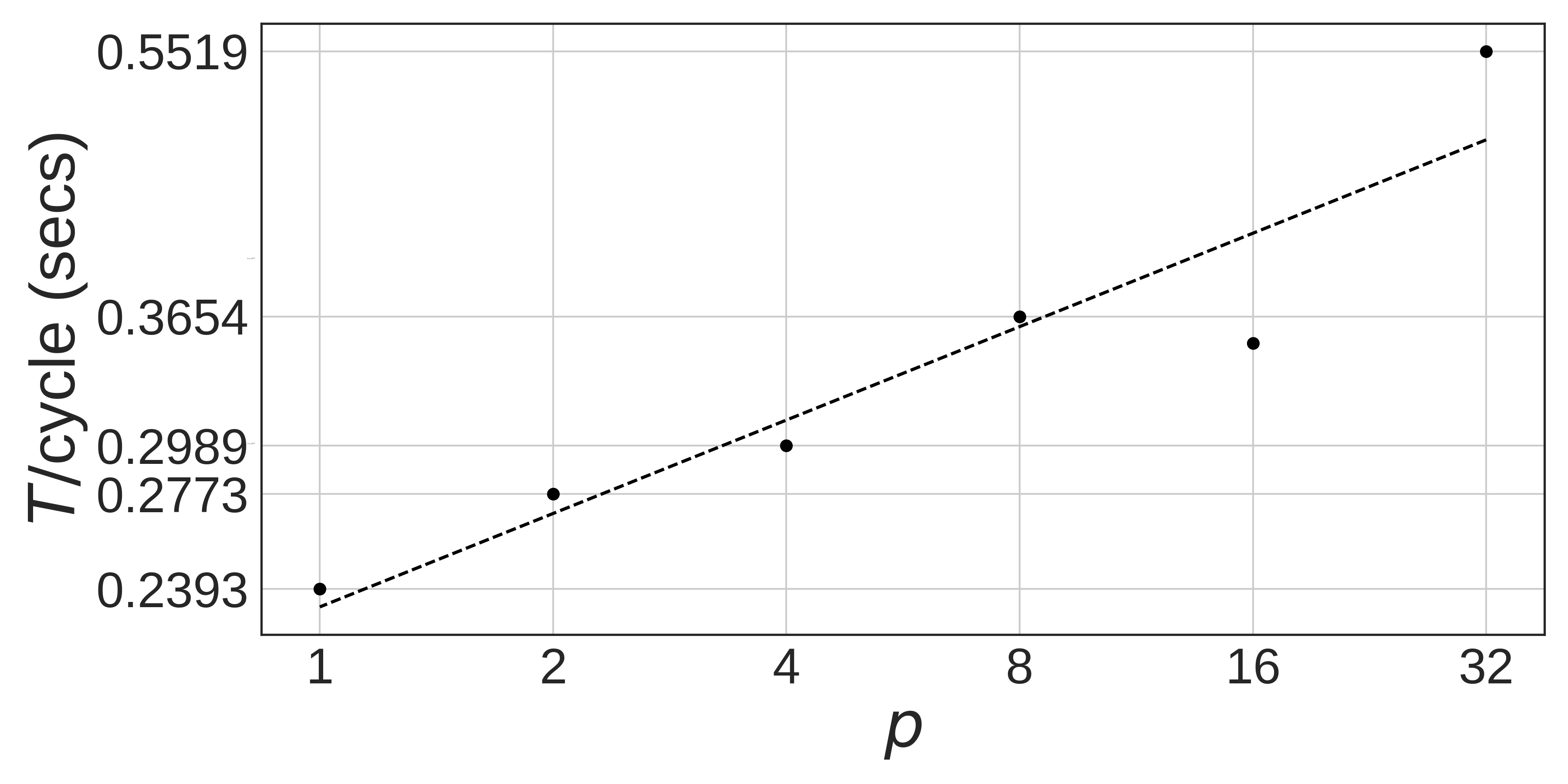}
        \put(0, 52.5){(b)}
        \end{overpic}\\
        \begin{overpic}[width=\subpanelwid]{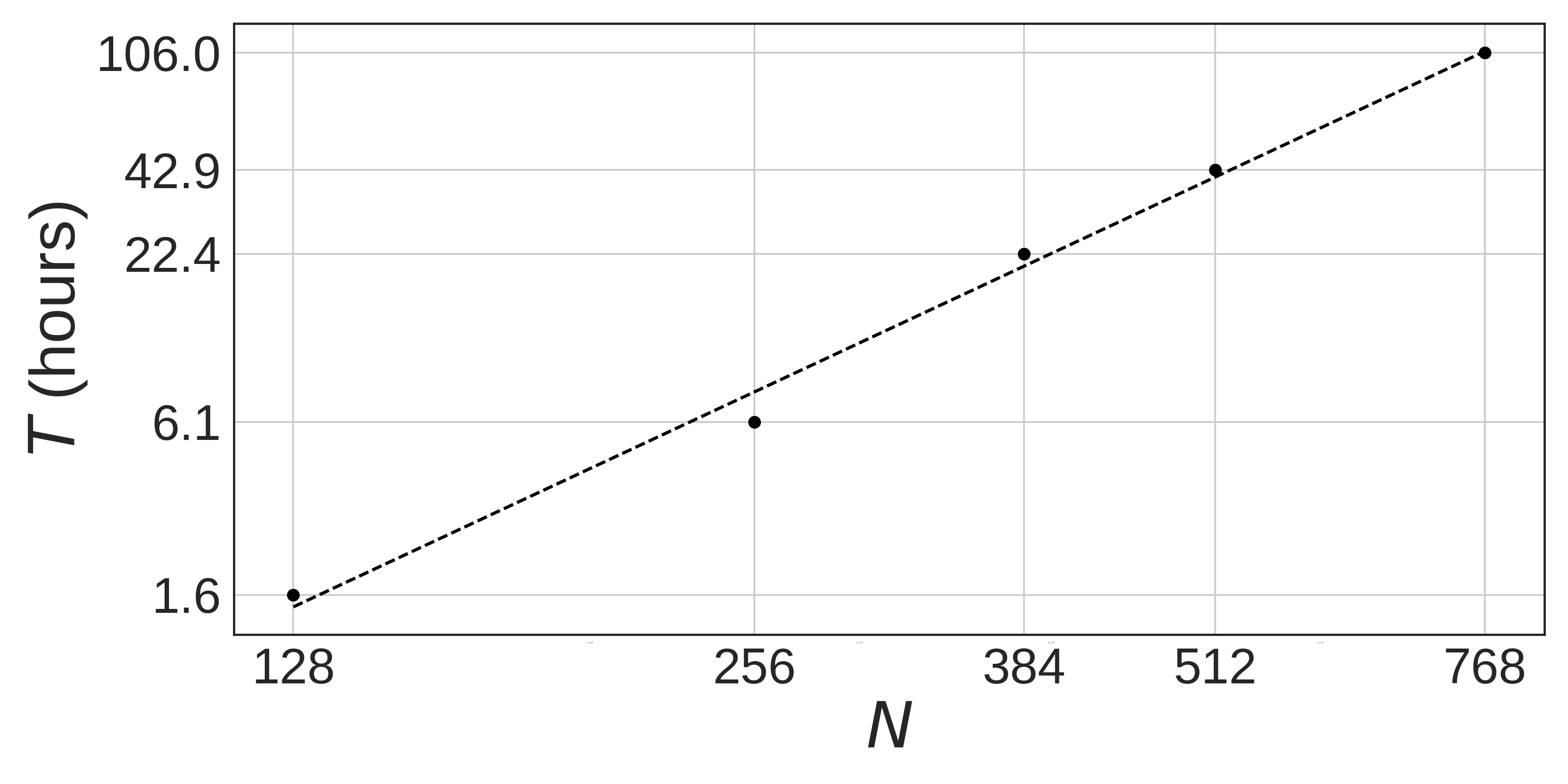}
        \put(0, 52.5){(c)}
        \end{overpic} &
        \begin{overpic}[width=\subpanelwid]{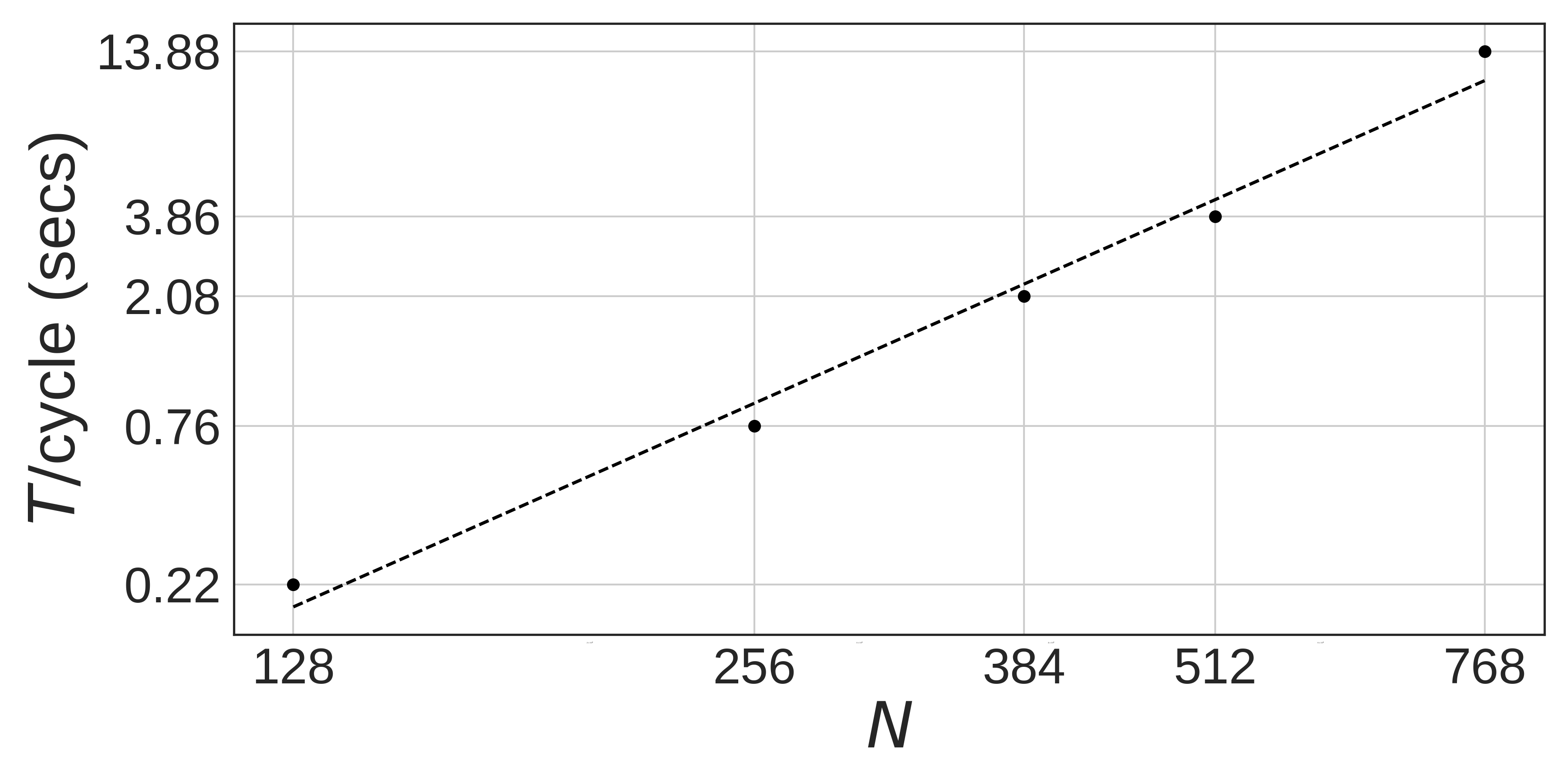}
        \put(0, 52.5){(d)}
        \end{overpic}
    \end{tabular}
    \vspace{5mm}
    \caption{(a,b) Parallel scaling for the total time per simulation and the average time per V-cycle with $\frac{N^3}{2p}$ approximately fixed. (c,d) Parallel scaling for the total time per simulation and average time per V-cycle with $p=32$. (a) Total time. The parallel efficiency exponent (see text) was computed as $\alpha = 0.2258$. $r^2 = 0.8914$ for the linear fit. (b) Average time per V-cycle. $\alpha = 0.2092$. $r^2 = 0.8774$ for the linear fit. (c) Total time. $\beta = 2.395$, $r^2 = 0.9931$ for the linear fit. (d) Average time per V-cycle. $\beta = 2.275$. $r^2 = 0.9860$ for the linear fit.}
    \label{fig:parallel_1}
\end{figure}

To quantitatively determine the parallel scaling efficiency, we assume that the total work required can be written as $W = k\times p$ where $k$ is a constant. We further assume that the total time taken can be computed as $T = W/(p Q E(p))$ where $Q$ is the total power per processor and $E(p)$ describes the parallel efficiency. With $E(p) \propto p^{-\alpha}$ for some exponent $\alpha$, the logarithm of the total time taken is linear in $\log(p)$ with coefficient $\alpha$, $\log T = A + \alpha \log p$ where $A$ is a constant, so that $\alpha$ may be found via linear regression. We have computed the exponent $\alpha$ for the total time taken and the time per V-cycle, as shown in Fig.~\ref{fig:parallel_1}. The $\alpha$ value in each case was found to be $\alpha = 0.2258$ and $\alpha = 0.2092$.

We also considered testing strong scaling performance of our implementation. However, fixing the size of the grid and varying the number of processors is infeasible in this setting. If the simulation is small enough that it can be handled in a reasonable amount of time, going to higher processor counts will not be beneficial due to communication limitations. If the simulation is large enough that higher processor counts such as $p=32$ are useful, the simulation will take too long to run with few processors.

As an alternative measure of parallel efficiency, we consider fixing $p=32$ and varying the number of grid points $N^3/2$. In this experiment, we use the same simulation parameters as for the test of weak scaling, but now with a random (drawn from a normal distribution) initial condition for the effective temperature with mean $550 \text{~K}$ and standard deviation $15 \text{~K}$; more details on the initialization procedure are provided in Sec.~\ref{sssec:rndm}. The results listed in Table \ref{tab:fix_p_vary_N} demonstrate that the number of V-cycles per simulation again increases with $N$. The total time per simulation and time per V-cycle increases greater than linearly. Fitting $\log T = B + \beta \log N$ in Fig.~\ref{fig:parallel_1}, we find that $\beta = 2.395$ for the total simulation time and $\beta = 2.275$ for the average time per V-cycle. Over the range of sizes considered, our implementation has more favorable scaling than the predicted $\mathcal{O}(N^3)$ of multigrid. This is because the communication becomes comparatively more efficient for larger grid sizes, since the ratio of ghost points to total grid points tends to zero as $N$ increases.

\subsection{Gaussian defects}
\label{sssec:defects}
We now turn to simulating realistic physical timescales with the quasi-static method, where we now set the scaling parameter $\zeta = 1$. We first consider the nucleation of shear bands from localized imperfections of higher $\chi$. Physically, these imperfections describe defects within the material structure that may be particularly susceptible to plastic deformation~\cite{li-2002}. To begin, we consider a single defect, corresponding to an initial $\chi$ field of the form
\begin{equation}
    \chi(\bx, t=0) = 550 \text{~K} + (170 \text{~K})\exp\left(-200\frac{\|\bx \|^2}{L^2}\right).
    \label{eqn:case0_chi}
\end{equation}
The simulation is performed on a grid of size $256 \times 256 \times 128$, corresponding to a grid spacing of $h = L/128$. The length scale $l$ appearing in Eq.~\ref{eqn:chi_evo} is fixed at $h$ and sets the width of the shear bands. The boundary velocity is set to a value of $U_b = 10^{-7} L/t_s$, and the simulation is conducted to a final time of $t_f = 10^6 t_s$, using a quasi-static timestep of $\Delta t = 200 t_s$. For three-dimensional visualization, we use the opacity function from Eq.~\ref{eqn:opac}. The simulation takes $7.52$ total hours using 32 processes on an Ubuntu Linux computer with dual 10-core Intel Xeon E5-2630 v4 processors. The total time spent in V-cycle computations is $5.852$ hours. The total number of V-cycles is $17586$.

\begin{figure}
\fcolorbox{black}{black}{
    \begin{tabular}{cc}

        \begin{subfigure}{\subpanelwid}
            \centering
            \includegraphics[width=\textwidth]{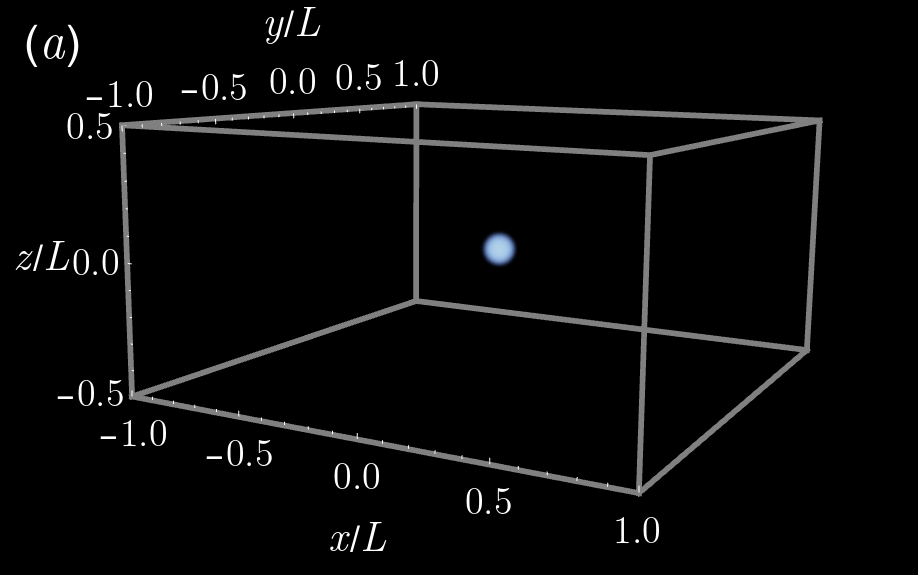}
        \end{subfigure} &

        \begin{subfigure}{\subpanelwid}
            \centering
            \includegraphics[width=\textwidth]{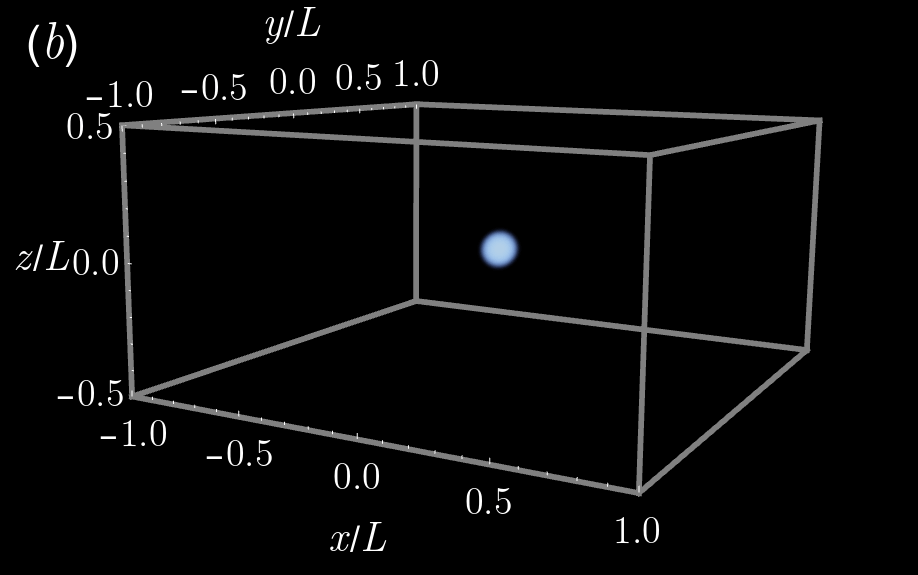}
        \end{subfigure}\\

        \begin{subfigure}{\subpanelwid}
            \centering
            \includegraphics[width=\textwidth]{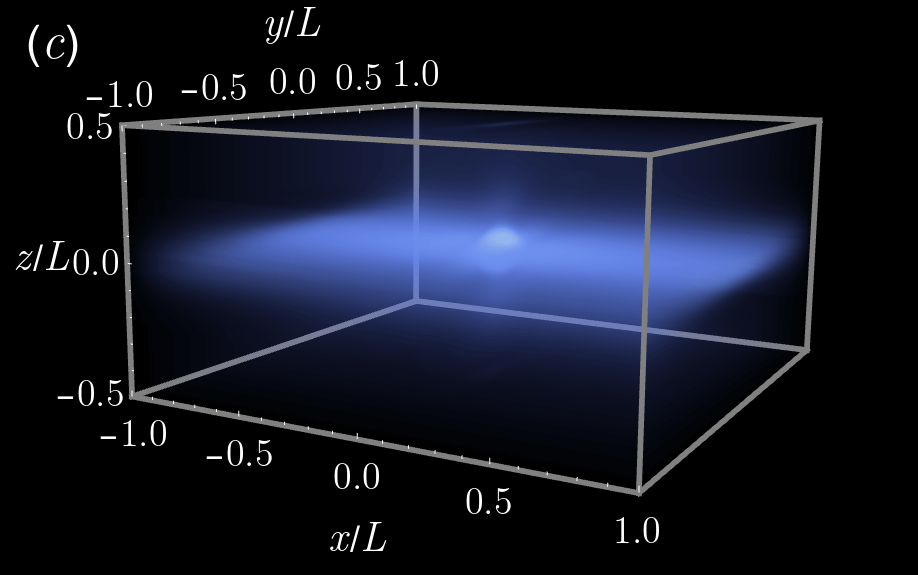}
        \end{subfigure} &

        \begin{subfigure}{\subpanelwid}
            \centering
            \includegraphics[width=\textwidth]{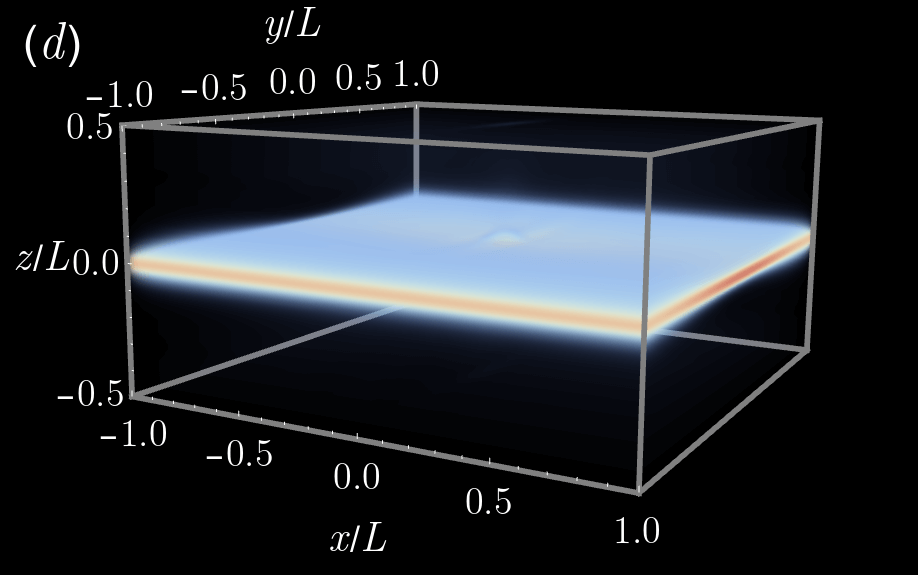}
        \end{subfigure}\\

        \begin{subfigure}{\subpanelwid}
            \centering
            \includegraphics[width=\textwidth]{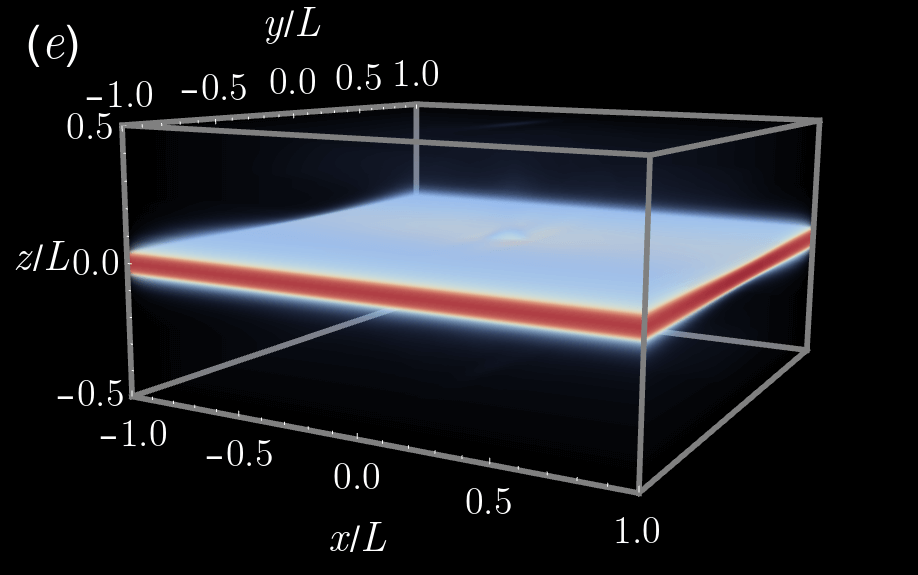}
        \end{subfigure} &

        \begin{subfigure}{\subpanelwid}
            \centering
            \includegraphics[width=\textwidth]{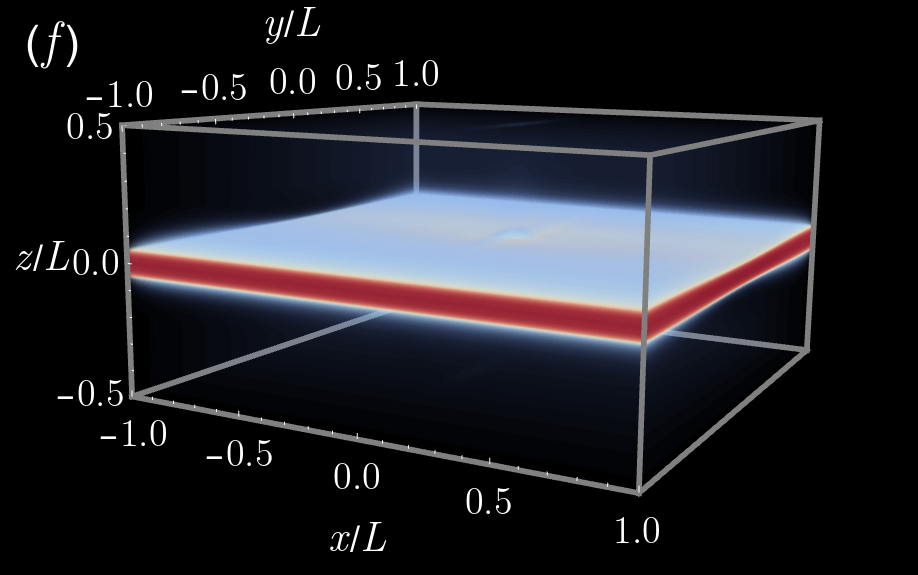}
        \end{subfigure}\\
    \end{tabular}}
    \begin{subfigure}{\textwidth}
        \centering
        \vspace{5mm}
        \includegraphics[width=.75\textwidth]{imgs/rslt_figs/colorbar_2019.png}
        \vspace{5mm}
    \end{subfigure}
    \caption{Snapshots of the effective temperature distribution $\chi(\bx, t)$ for a quasi-static simulation with $\zeta = 1$. The initial condition is given in Eq.~\ref{eqn:case0_chi}, corresponding to a small Gaussian defect at the center of the material. $a = 0.3$ and $\eta = 1.2$ for plots (a)--(c). $a = 0.45$ and $\eta = 1.35$ for plots (d)--(f). (a) $t = 0 t_s$. (b) $t = 2\times10^5 t_s$. (c) $t = 4\times 10^5 t_s$. (d) $t = 6\times 10^5 t_s$. (e) $t = 8\times 10^5 t_s$. (f) $t = 10^6t_s$.}
    \label{fig:case0}
\end{figure}

Snapshots of the effective temperature field at various time points are shown in Fig.~\ref{fig:case0}. The initial condition is shown in Fig.~\ref{fig:case0}(a). At $t=2\times10^5 t_s$ in Fig.~\ref{fig:case0}(b), the defect has started to expand. By $t=4\times 10^5 t_s$ in Fig.~\ref{fig:case0}(c), a shear band begins to nucleate, indicated by a quadrupolar structure emanating from the defect. The background $\chi$ field also begins to increase, as demonstrated by the presence of the transparent light blue background. By $t=6\times 10^5 t_s$ in Fig.~\ref{fig:case0}(d), a distinct system-spanning band has become clear. The band displays no curvature in either of the $x$ or $y$ directions. By $t=8\times 10^5 t_s$ in Fig.~\ref{fig:case0}(e), a prominent band has formed. $t = 10^6 t_s$ in Fig.~\ref{fig:case0}(f) is similar, with a thicker and stronger band.

\begin{figure}
\fcolorbox{black}{black}{
    \begin{tabular}{cc}

        \begin{subfigure}{\subpanelwid}
            \centering
            \includegraphics[width=\textwidth]{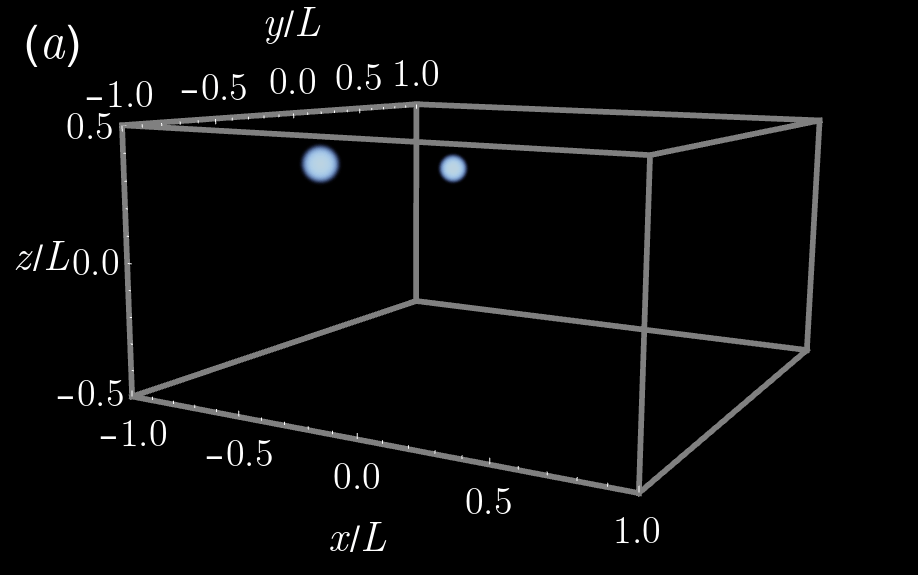}
        \end{subfigure} &

        \begin{subfigure}{\subpanelwid}
            \centering
            \includegraphics[width=\textwidth]{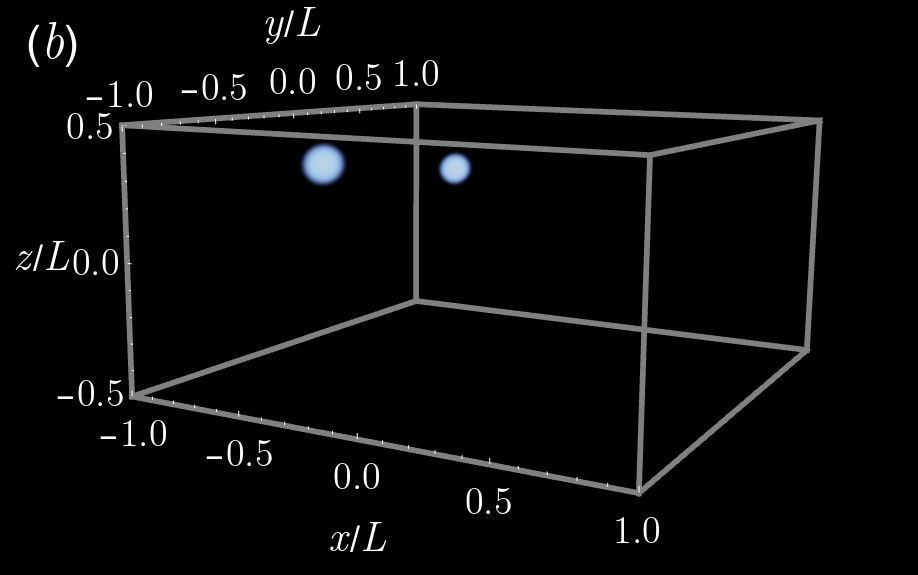}
        \end{subfigure}\\

        \begin{subfigure}{\subpanelwid}
            \centering
            \includegraphics[width=\textwidth]{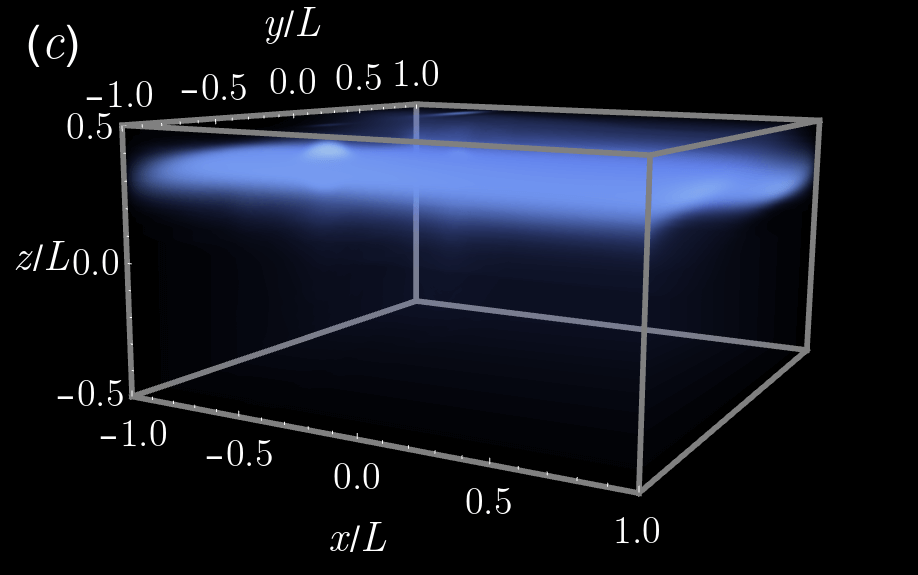}
        \end{subfigure} &

        \begin{subfigure}{\subpanelwid}
            \centering
            \includegraphics[width=\textwidth]{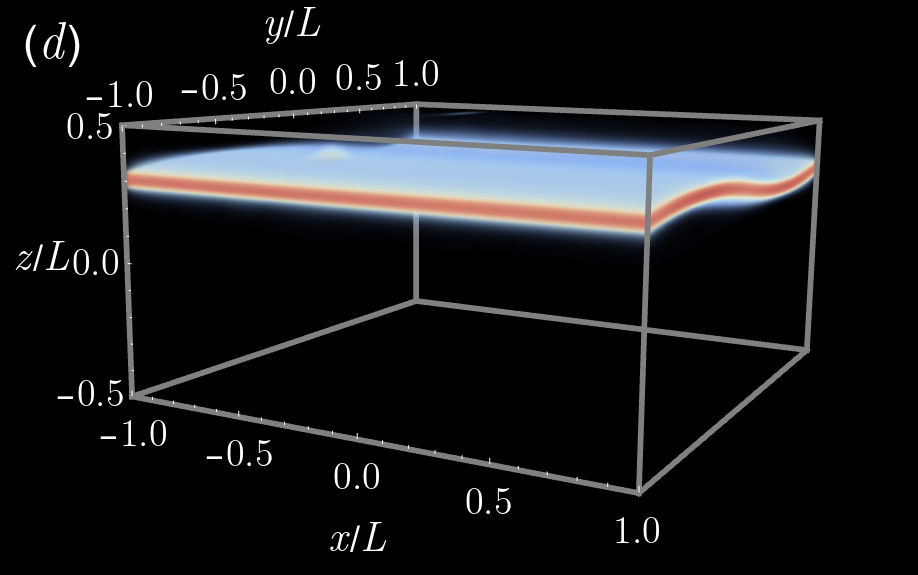}
        \end{subfigure}\\

        \begin{subfigure}{\subpanelwid}
            \centering
            \includegraphics[width=\textwidth]{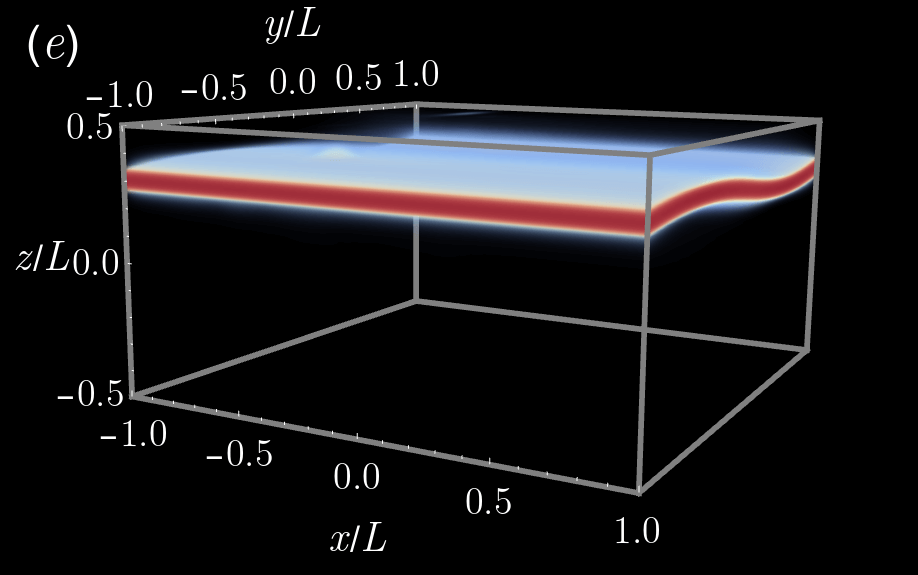}
        \end{subfigure} &

        \begin{subfigure}{\subpanelwid}
            \centering
            \includegraphics[width=\textwidth]{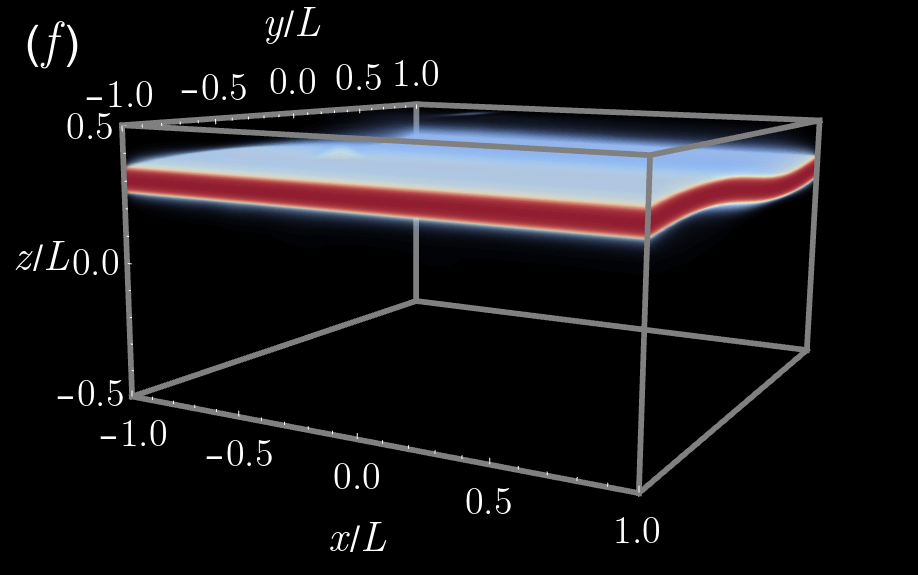}
        \end{subfigure}\\
    \end{tabular}}
    \begin{subfigure}{\textwidth}
        \centering
        \vspace{5mm}
        \includegraphics[width=.75\textwidth]{imgs/rslt_figs/colorbar_2019.png}
        \vspace{5mm}
    \end{subfigure}

    \caption{Snapshots of the effective temperature distribution $\chi(\bx, t)$ for a quasi-static simulation with $\zeta = 1$. The initial condition is given in Eq.~\ref{eqn:case_1_8_chi} with $\bX_1 = (-0.5, -0.5, 0.35)$ and $\bX_2 = (-0.5, 0.5, 0.25)$. $a = 0.3$ and $\eta = 1.2$ for plots (a)--(c). $a = 0.45$ and $\eta = 1.35$ for plots (d)--(f). (a) $t = 0 t_s$. (b) $t = 2\times10^5 t_s$. (c) $t = 4\times 10^5 t_s$. (d) $t = 6\times 10^5 t_s$. (e) $t = 8\times 10^5 t_s$. (f) $t = 10^6t_s$.}
    \label{fig:case1}
\end{figure}

\begin{figure}
\fcolorbox{black}{black}{
    \begin{tabular}{cc}

        \begin{subfigure}{\subpanelwid}
            \centering
            \includegraphics[width=\textwidth]{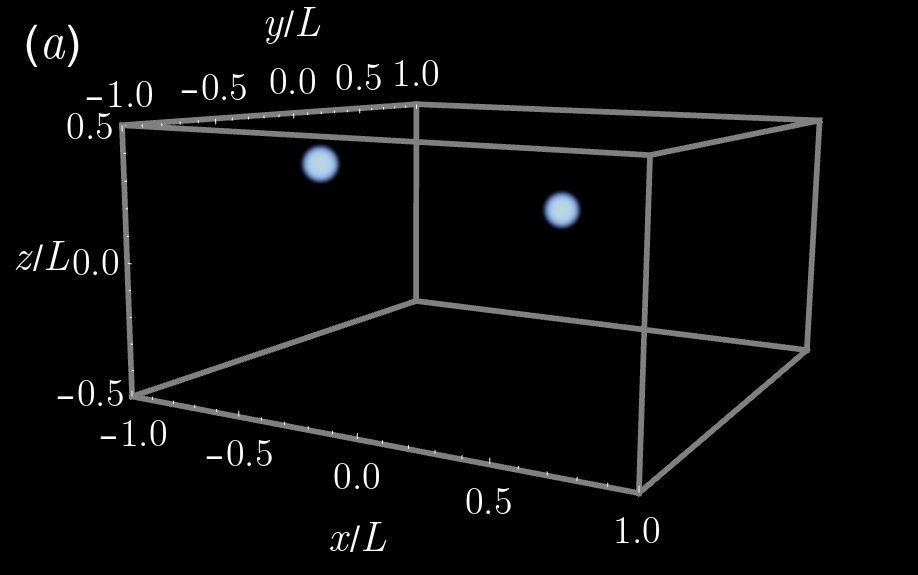}
        \end{subfigure} &

        \begin{subfigure}{\subpanelwid}
            \centering
            \includegraphics[width=\textwidth]{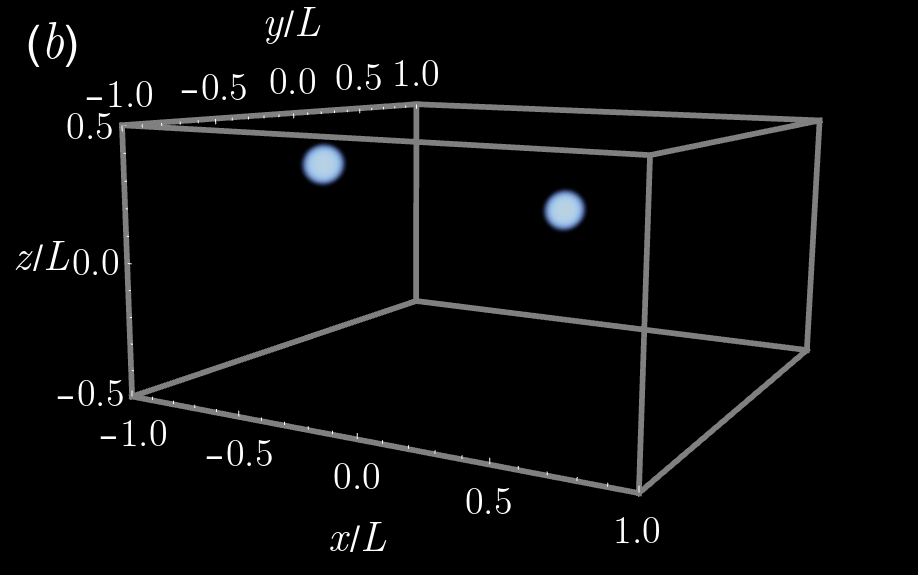}
        \end{subfigure}\\

        \begin{subfigure}{\subpanelwid}
            \centering
            \includegraphics[width=\textwidth]{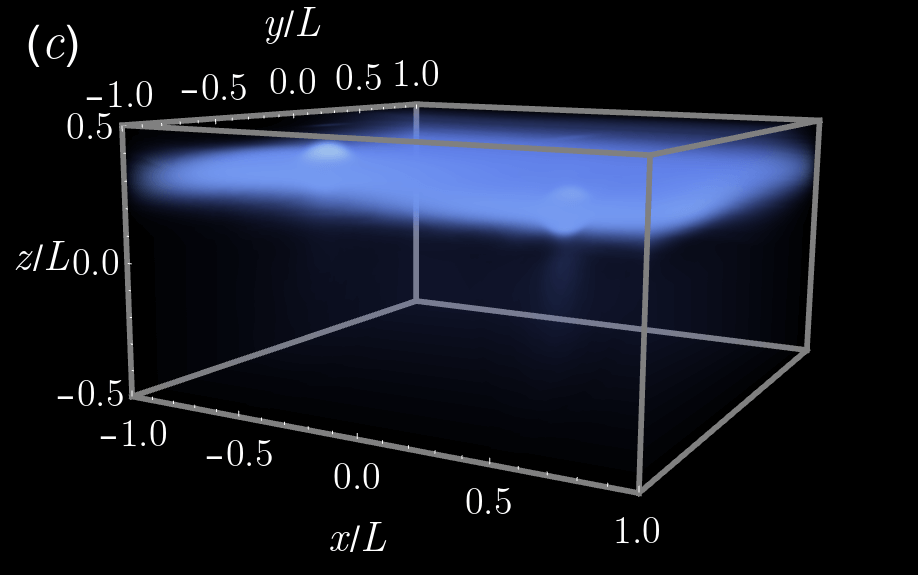}
        \end{subfigure} &

        \begin{subfigure}{\subpanelwid}
            \centering
            \includegraphics[width=\textwidth]{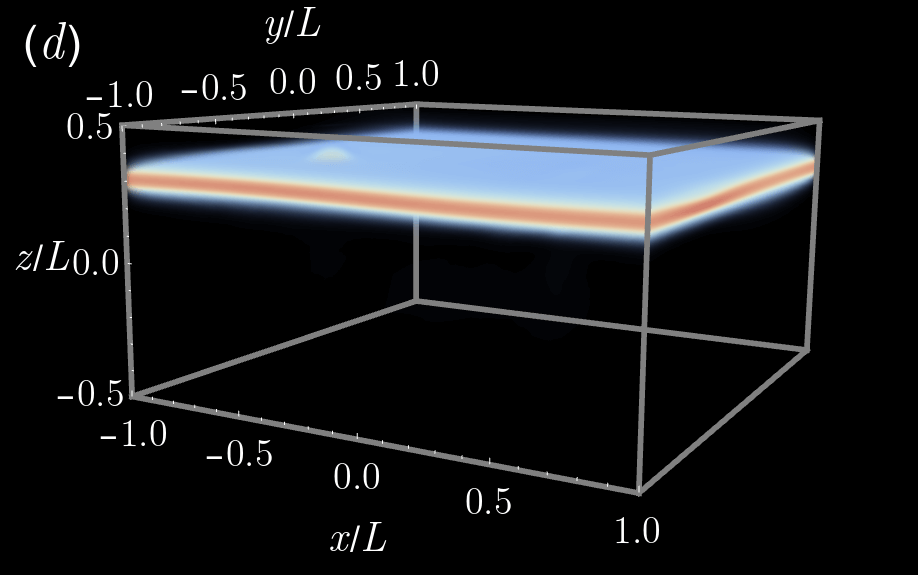}
        \end{subfigure}\\

        \begin{subfigure}{\subpanelwid}
            \centering
            \includegraphics[width=\textwidth]{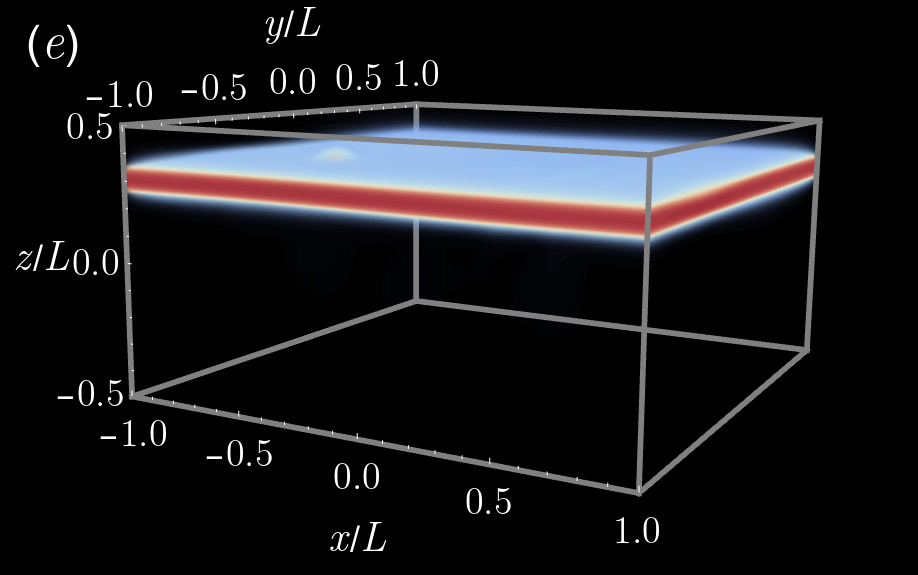}
        \end{subfigure} &

        \begin{subfigure}{\subpanelwid}
            \centering
            \includegraphics[width=\textwidth]{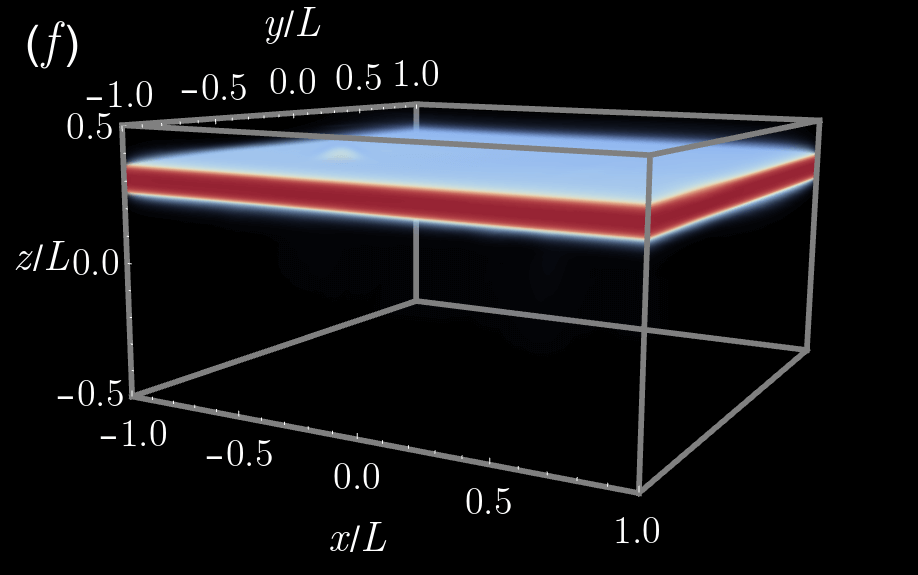}
        \end{subfigure}\\
    \end{tabular}}
    \begin{subfigure}{\textwidth}
        \centering
        \vspace{5mm}
        \includegraphics[width=.75\textwidth]{imgs/rslt_figs/colorbar_2019.png}
        \vspace{5mm}
    \end{subfigure}

    \caption{Snapshots of the effective temperature distribution $\chi(\bx, t)$ for a quasi-static simulation with $\zeta = 1$. The initial condition is given in Eq.~\ref{eqn:case_1_8_chi} with $\bX_1 = (-0.5, -0.5, 0.35)$ and $\bX_2 = (0.5, -0.5, 0.25)$. $a = 0.3$ and $\eta = 1.2$ for plots (a)--(c). $a = 0.45$ and $\eta = 1.35$ for plots (d)--(f). (a) $t = 0 t_s$. (b) $t = 2\times10^5 t_s$. (c) $t = 4\times 10^5 t_s$. (d) $t = 6\times 10^5 t_s$. (e) $t = 8\times 10^5 t_s$. (f) $t = 10^6t_s$.}
    \label{fig:case8}
\end{figure}

We now introduce a second defect to highlight some three-dimensional characteristics of shear banding. We expect that the relative size and displacement between the defects will determine the dynamics, with the possibility of forming a single shear band that connects the two. The initial effective temperature field is
\begin{equation}
  \chi(\bx, t) = 550\text{~K} + \left(200\text{~K}\right) \left(\exp\left(-200\frac{\left\|\left(\bx - \bX_1\right)\right\|^2}{L^2}\right) + \exp\left(-250\frac{\left\|\left(\bx - \bX_2\right)\right\|^2}{L^2}\right)\right).
    \label{eqn:case_1_8_chi}
\end{equation}
Two cases of Eq.~\ref{eqn:case_1_8_chi} are considered. First, we take $\bX_1 = (-0.5, -0.5, 0.35)$ and $\bX_2 = (-0.5, 0.5, 0.25)$, corresponding to two defects symmetric about the $y=0$ plane with the same $x$ coordinate, a slight offset in $z$, and different sizes. The results for this case using the same simulation parameters as Fig.~\ref{fig:case0} are shown in Fig.~\ref{fig:case1}. Second, we take $\bX_1 = (-0.5, -0.5, 0.35)$ and $\bX_2 = (0.5, -0.5, 0.25)$; this is the same as the previous case, but with the roles of $x$ and $y$ interchanged. The results for this case again with the same grid size, quasi-static timestep, and boundary velocity as for the single defect are shown in Fig.~\ref{fig:case8}. The initial configurations are shown in Figs.~\ref{fig:case1}(a) and \ref{fig:case8}(a). The first arrangement takes $5.61$ total hours using 32 processes on a Ubuntu Linux computer with dual 10-core 2.20~GHz Intel Xeon E5-2630 v4 processors. The total time spent in multigrid V-cycles is $4.25$ hours. The total number of V-cycles is $17782$. The second arrangement takes $5.27$ hours using 32 processes on an Ubuntu Linux computer with dual 10-core 2.20~GHz Intel Xeon Silver 4114 processors. The total time spent in multigrid V-cycles is $3.98$ hours. The total number of V-cycles is $17647$.

There is an interesting contrast between the time sequences displayed in Figs.~\ref{fig:case1} and \ref{fig:case8}. Much like in the single defect simulations, $t = 2\times 10^5 t_s$ in Figs.~\ref{fig:case1}(b) and \ref{fig:case8}(b) displays expansion of the defects, and $t = 4\times 10^5 t_s$ in Figs.~\ref{fig:case1}(c) and \ref{fig:case8}(c) shows the initiation of shear band nucleation. In Fig.~\ref{fig:case1}(d), we see the formation of a single curved band connecting the two defects, while in Fig.~\ref{fig:case8}(d), the band is flat. Figs.~\ref{fig:case1}(e), \ref{fig:case1}(f), \ref{fig:case8}(e), and \ref{fig:case8}(f) make this more clear as the band becomes more defined. The curvature seen in Fig.~\ref{fig:case1} is in the direction orthogonal to shear.

The dependence of band curvature on the relative orientation of the two defects can be best understood in terms of the qualitative structure of Fig.~\ref{fig:case0}(c). There is a substantial extension of elevated $\chi$ along the $x$ direction (parallel to shear, in-plane), a small extension along the $y$ direction (orthogonal to shear, in-plane), and a moderate extension along the $z$ direction (orthogonal to shear, out-of-plane). In Fig.~\ref{fig:case1}, the defects are offset in $y$ and $z$. Because the $\chi$ field is stronger in $z$ than in $y$, this relative placement of the defects can accomodate curvature along the $y$ direction. On the other hand, in Fig.~\ref{fig:case8}, the defects are offset in $x$ and $z$. The strength of the $\chi$ field extension in the $x$ direction is great enough that the flat, horizontal regions of the two forming bands reach each other. The two bands thus join into one fatter flat band.

Taken together, Figs.~\ref{fig:case0}, \ref{fig:case1}, \& \ref{fig:case8} provide insight into the structure and nucleation of shear bands from localized material defects. They help understand experimentally observed band curvature and raise the possibility that the placement and orientation of microscopic material properties can influence the qualitative structure of macroscopic shear bands. Finally, they provide intuition for more complex initial conditions, such as the random initializations considered later in this work, as a superposition of many defects.

\subsubsection{Circular defects}
\begin{figure}
\fcolorbox{black}{black}{
    \begin{tabular}{cc}
        \begin{subfigure}{\subpanelwid}
            \centering
            \includegraphics[width=\textwidth]{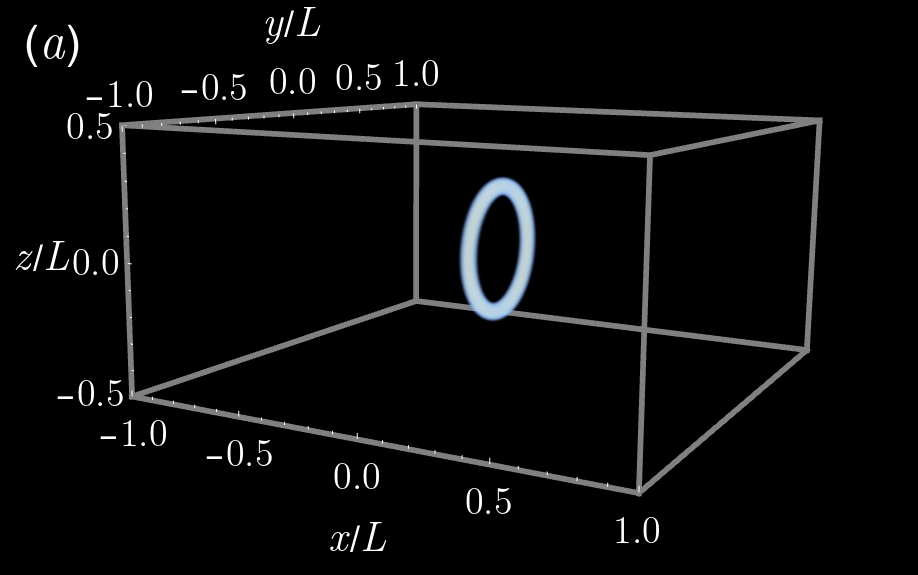}
        \end{subfigure} &

        \begin{subfigure}{\subpanelwid}
            \centering
            \includegraphics[width=\textwidth]{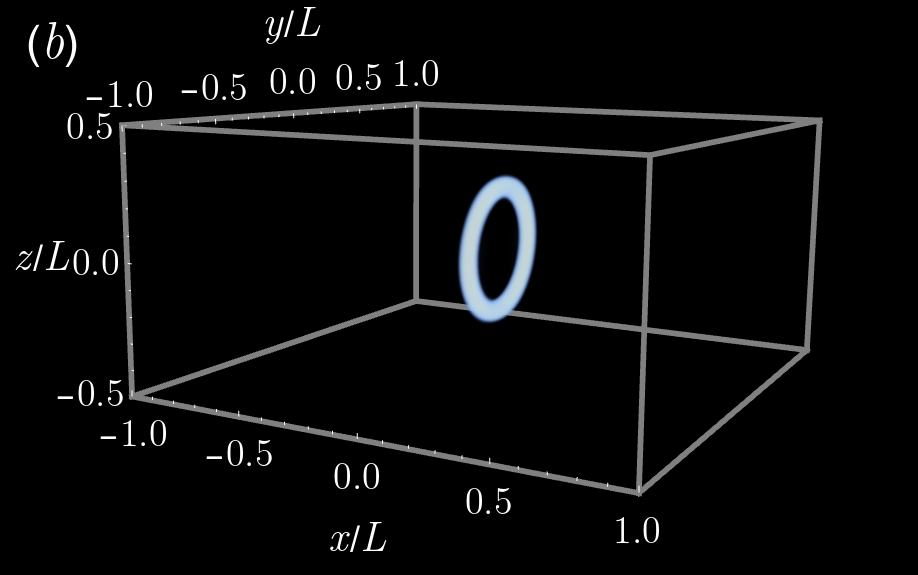}
        \end{subfigure}\\

        \begin{subfigure}{\subpanelwid}
            \centering
            \includegraphics[width=\textwidth]{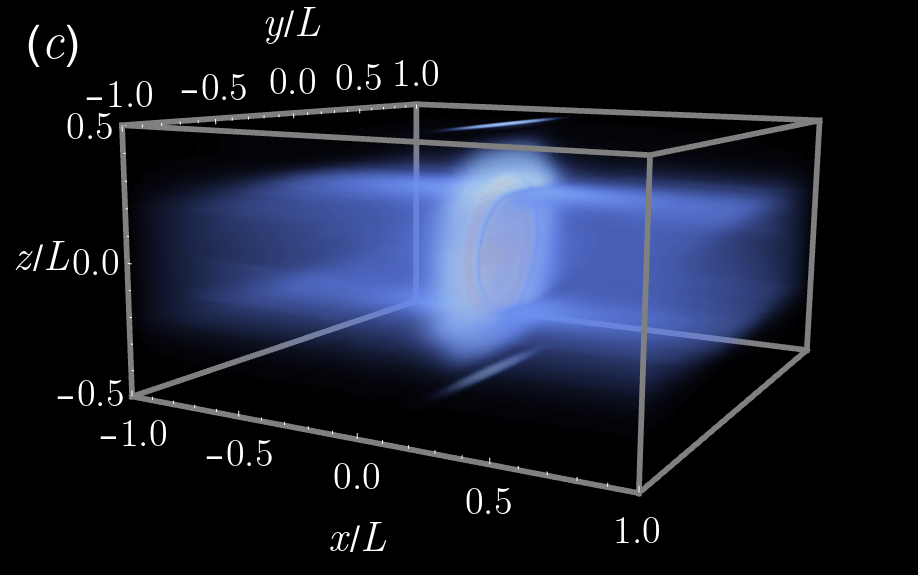}
        \end{subfigure} &

        \begin{subfigure}{\subpanelwid}
            \centering
            \includegraphics[width=\textwidth]{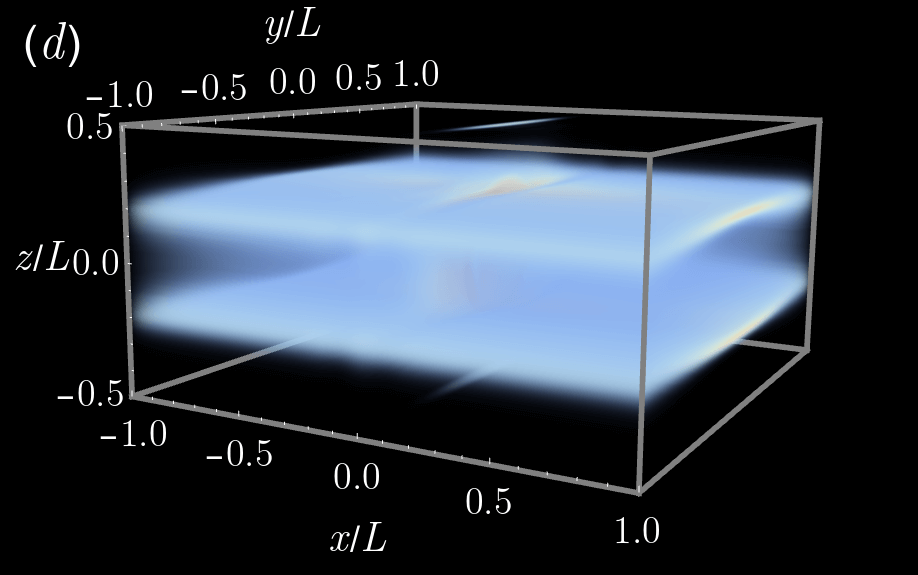}
        \end{subfigure}\\

        \begin{subfigure}{\subpanelwid}
            \centering
            \includegraphics[width=\textwidth]{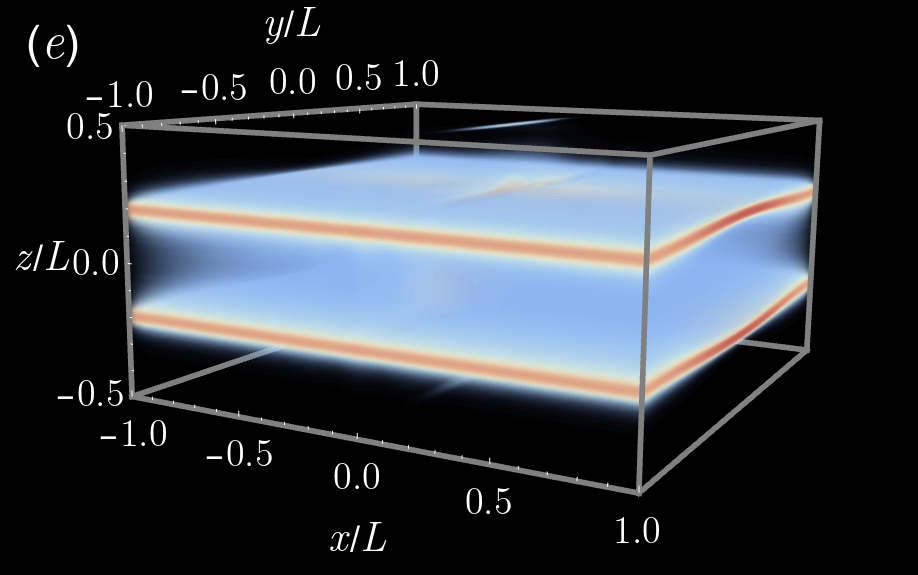}
        \end{subfigure} &

        \begin{subfigure}{\subpanelwid}
            \centering
            \includegraphics[width=\textwidth]{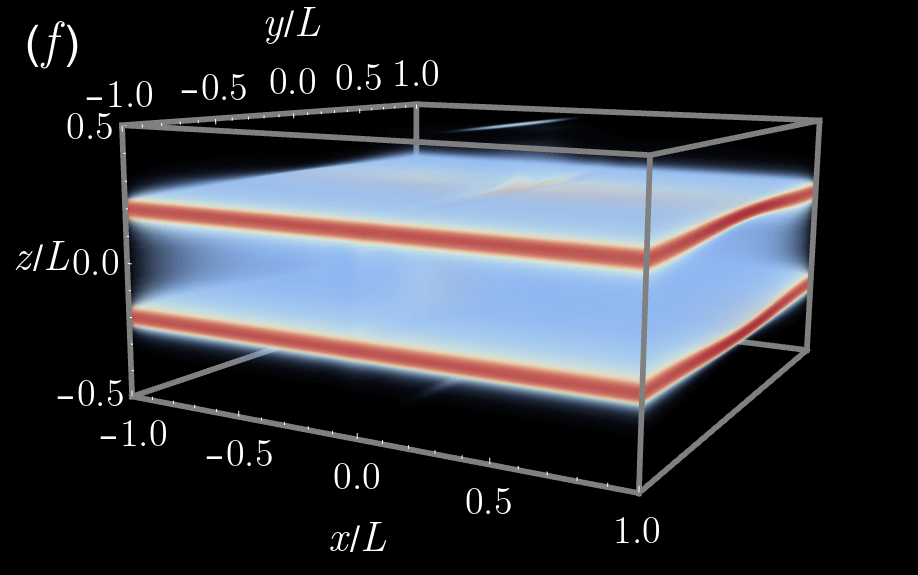}
        \end{subfigure}\\

    \end{tabular}}
    \begin{subfigure}{\textwidth}
        \centering
        \vspace{5mm}
        \includegraphics[width=.75\textwidth]{imgs/rslt_figs/colorbar_2019.png}
        \vspace{5mm}
    \end{subfigure}
    \caption{Snapshots of the effective temperature distribution $\chi(\bx, t)$ for a quasi-static simulation with $\zeta = 1$. The initial condition is given by Eq.~\ref{eqn:case13_chi} with the circle oriented along the $yz$ plane. $a = 0.3$ and $\eta = 1.2$ for plots (a)--(c). $a = 0.45$ and $\eta = 1.35$ for plots (d)--(f). (a) $t = 0 t_s$. (b) $t = 2\times10^5 t_s$. (c) $t = 4\times 10^5 t_s$. (d) $t = 6\times 10^5 t_s$. (e) $t = 8\times 10^5 t_s$. (f) $t = 10^6t_s$.}
    \label{fig:case13}
\end{figure}

\begin{figure}

\fcolorbox{black}{black}{
    \begin{tabular}{cc}
        \begin{subfigure}{\subpanelwid}
            \centering
            \includegraphics[width=\textwidth]{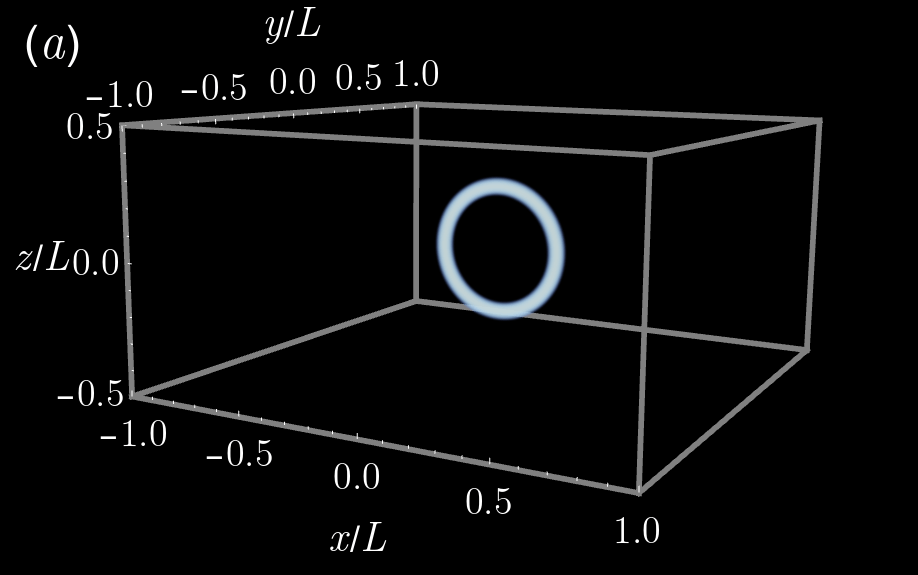}
        \end{subfigure} &

        \begin{subfigure}{\subpanelwid}
            \centering
            \includegraphics[width=\textwidth]{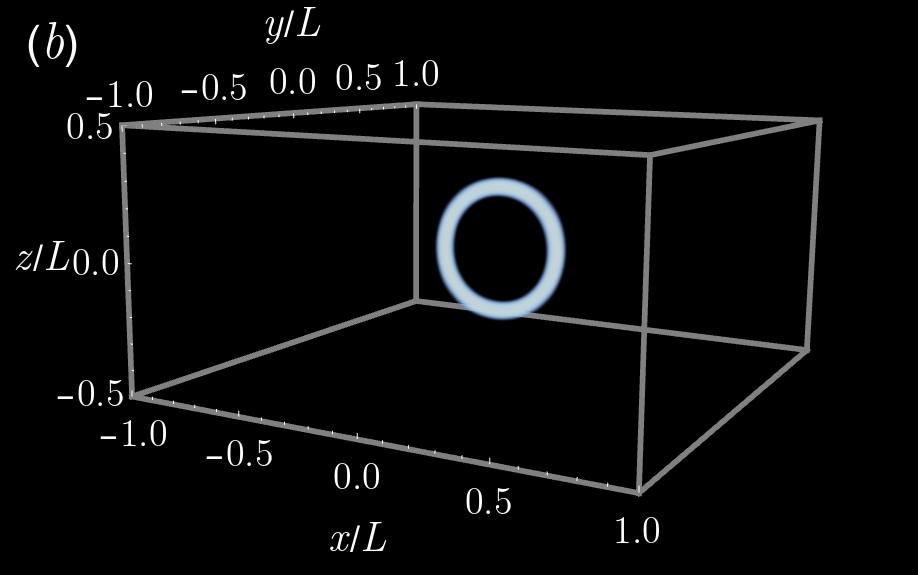}
        \end{subfigure}\\

        \begin{subfigure}{\subpanelwid}
            \centering
            \includegraphics[width=\textwidth]{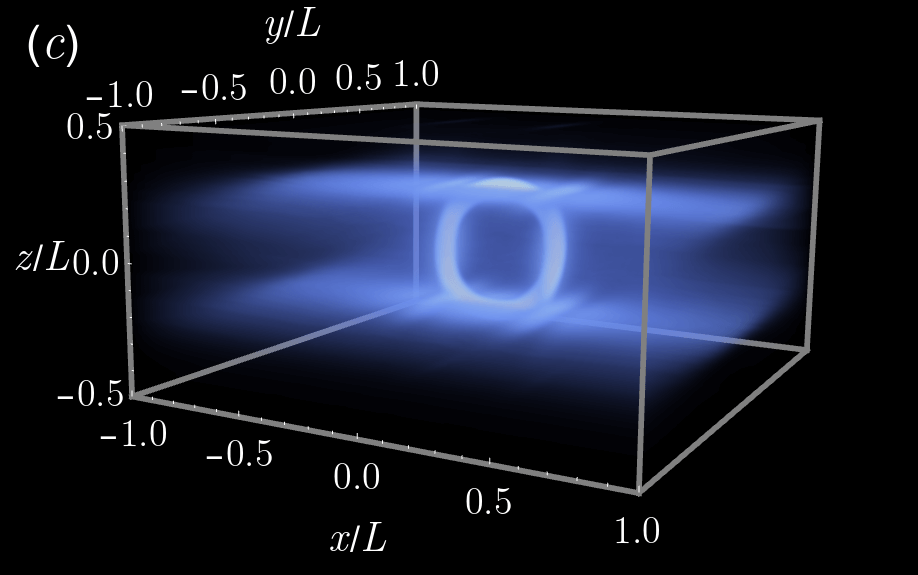}
        \end{subfigure} &

        \begin{subfigure}{\subpanelwid}
            \centering
            \includegraphics[width=\textwidth]{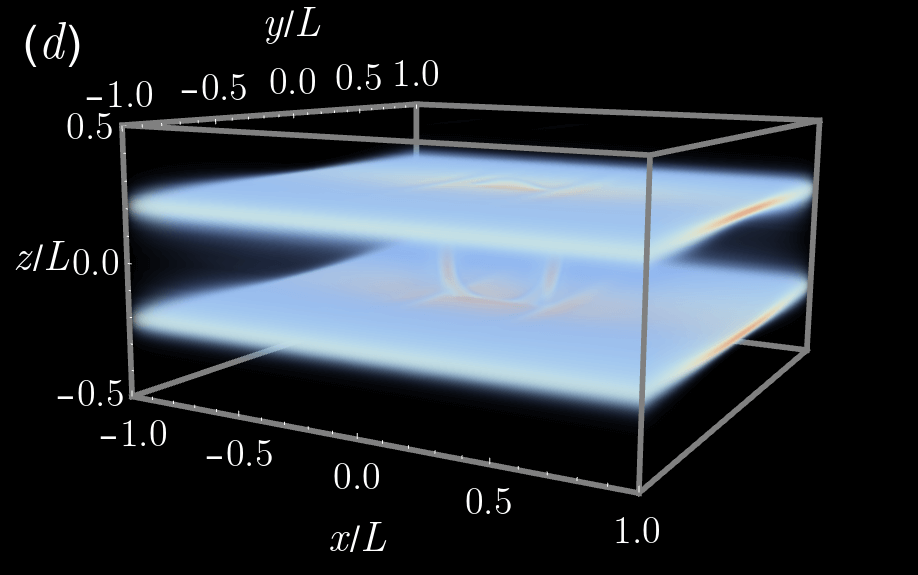}
        \end{subfigure}\\

        \begin{subfigure}{\subpanelwid}
            \centering
            \includegraphics[width=\textwidth]{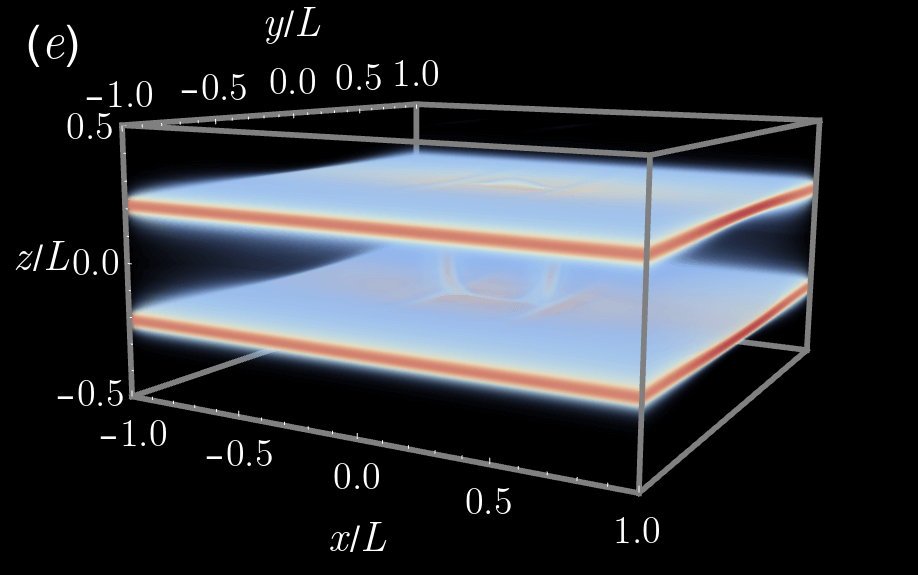}
        \end{subfigure} &

        \begin{subfigure}{\subpanelwid}
            \centering
            \includegraphics[width=\textwidth]{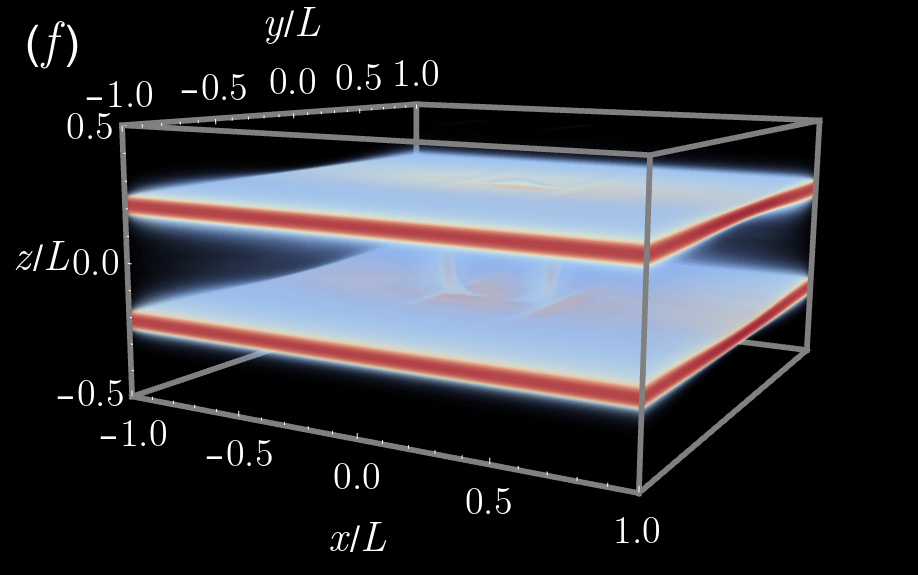}
        \end{subfigure}\\
    \end{tabular}}
    \begin{subfigure}{\textwidth}
        \centering
        \vspace{5mm}
        \includegraphics[width=.75\textwidth]{imgs/rslt_figs/colorbar_2019.png}
        \vspace{5mm}
    \end{subfigure}
    \caption{Snapshots of the effective temperature distribution $\chi(\bx, t)$ for a quasi-static simulation with $\zeta = 1$. The initial condition is given by Eq.~\ref{eqn:case14_chi} with the circle oriented along the $xz$ plane. $a = 0.3$ and $\eta = 1.2$ for plots (a)--(c). $a = 0.45$ and $\eta = 1.35$ for plots (d)--(f). (a) $t = 0 t_s$. (b) $t = 2\times10^5 t_s$. (c) $t = 4\times 10^5 t_s$. (d) $t = 6\times 10^5 t_s$. (e) $t = 8\times 10^5 t_s$. (f) $t = 10^6t_s$.}
    \label{fig:case14}
\end{figure}

We now turn to a set of more complex initial conditions in the effective temperature field. Results for initial conditions corresponding to a circular region of elevated $\chi$ parallel and perpendicular to the direction of shear are shown in Figs.~\ref{fig:case13} and \ref{fig:case14} respectively. The initial conditions are given by
\begin{align}
    d &= \sqrt{\frac{y^2}{L^2} + \frac{z^2}{L^2}} - \frac{1}{4},\nonumber\\
    \chi(\bx, t=0) &= 550\text{~K} + (200 \text{~K}) \exp\left(-750\left(d^2 + x^2\right)\right),
    \label{eqn:case13_chi}
\end{align}
and
\begin{align}
    d &= \sqrt{\frac{x^2}{L^2} + \frac{z^2}{L^2}} - \frac{1}{4},\nonumber\\
    \chi(\bx, t=0) &= 550\text{~K} + (200 \text{~K}) \exp\left(-750\left(d^2 + y^2\right)\right),
    \label{eqn:case14_chi}
\end{align}
representing circles in the $yz$ and $xz$ planes respectively. Simulations are carried out using the same simulation geometry, discretization, quasi-static timestep, and boundary velocity as in the previous section. The initial conditions in Eqs.~\ref{eqn:case13_chi} and \ref{eqn:case14_chi} are displayed in Figs.~\ref{fig:case13}(a), \ref{fig:case14}(a) respectively. The first arrangement takes $5.66$ total hours using 32 processes on a Ubuntu Linux computer with dual 14-core 1.70~GHz Intel Xeon E5-2650L v4 processors. $4.30$ hours are spent in multigrid V-cycles. There are $16980$ total V-cycles. The second arrangement takes $5.46$ total hours using 32 processes on an Ubuntu Linux computer with dual 8-core 2.40~GHz Intel Xeon E5-2630 v3 processors. $4.22$ hours are spent in multigrid V-cycles. There are $17153$ total V-cycles.


By $t = 2\times 10^5 t_s$ in Figs.~\ref{fig:case13}(b) and \ref{fig:case14}(b), little has changed, though the circles have expanded slightly. At $t=4\times 10^5t_s$ in Figs.~\ref{fig:case13}(c) and \ref{fig:case14}(c), differences due to the orientation of the circles become clear. The circle oriented perpendicular to shear closes vertically into a disk. The circle oriented along shear exhibits signatures of shear band nucleation at four equally-spaced points. At $t=6\times 10^5 t_s$ in Fig.~\ref{fig:case13}(d), the disk has expanded and has developed two diffuse shear bands. In Fig.~\ref{fig:case14}(d), there are two thinner, sharper, more well-separated bands forming off the top and bottom of the circle. By $t=8\times 10^5 t_s$ in Figs.~\ref{fig:case13}(e) and \ref{fig:case14}(e), these differences have become even more prominent. The bands are seen to have a curved structure in the $y$ direction in Fig.~\ref{fig:case13}(e) which is not as clear in Fig.~\ref{fig:case14}(e). The interim region is of higher $\chi$ than for the bands seen in Fig.~\ref{fig:case14}(e), where the circular defect is still mostly visible. These features continue to develop into the final pane at $t = 10^6 t_s$. Taken together, Figs.~\ref{fig:case13} and \ref{fig:case14} demonstrate another example of the dependence of shear banding structure and dynamics on the orientation of initial conditions in the $\chi$ field with respect to shear.

\subsubsection{A randomly fluctuating effective temperature field}
\label{sssec:rndm}
In this section, we consider the case of a randomly distributed initial effective temperature field. The initial conditions presented in the previous sections provide insight into the dynamics of shear banding, but it is unlikely that they have exact physical correspondences. The STZ theory postulates that STZs are randomly distributed throughout the material, and a random initial condition in $\chi$ is most faithful to this fundamental assumption~\cite{hinkle15}. Random initial conditions are thus expected to shed the most light on the structure of shear bands observed in experiments. The randomly fluctuating $\chi$ field leads to the formation of multiple shear bands, and may enable the study of shear band interactions in the STZ model~\cite{multi-bands}.

\begin{figure}
\fcolorbox{black}{black}{%
    \begin{tabular}{cc}
        \begin{subfigure}{\subpanelwid}
            \centering
            \includegraphics[width=\textwidth]{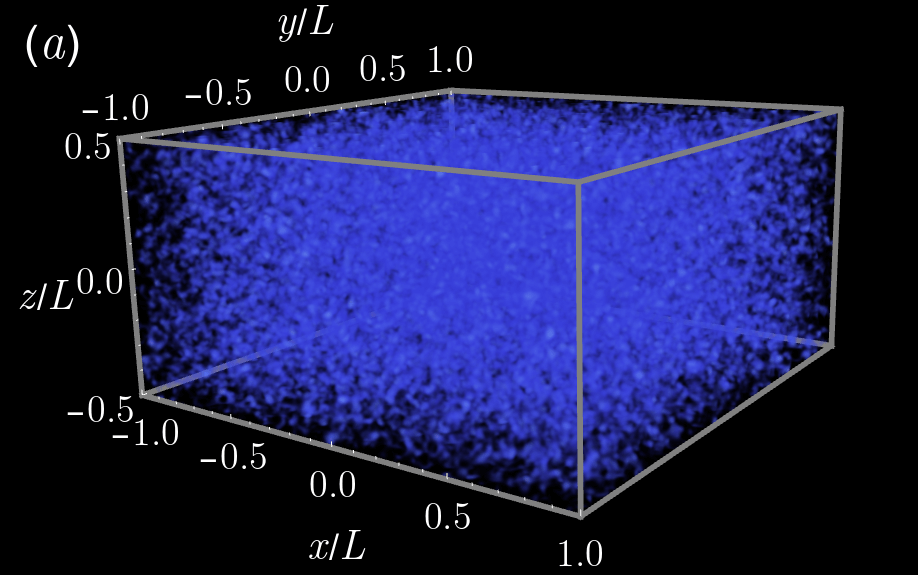}
        \end{subfigure} &
        \begin{subfigure}{\subpanelwid}
            \centering
            \includegraphics[width=\textwidth]{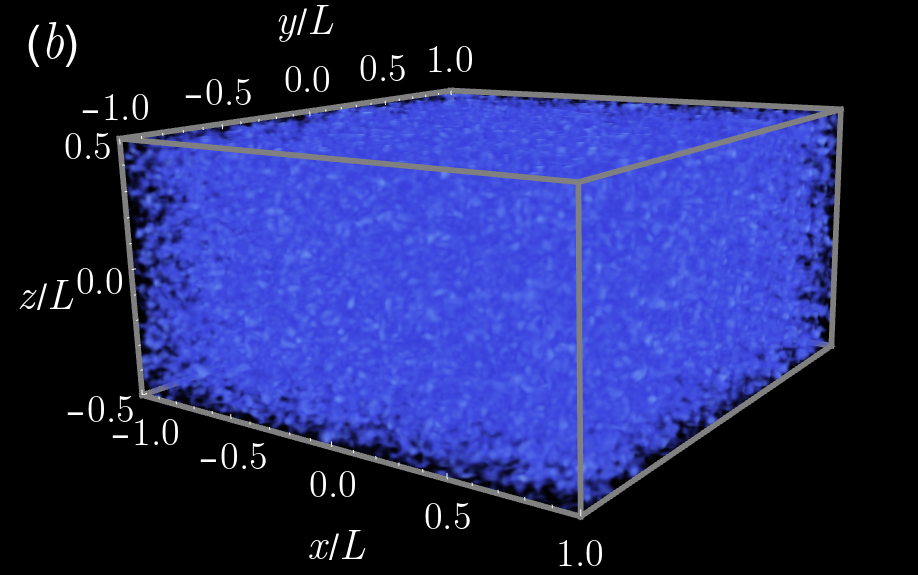}
        \end{subfigure}\\
        \begin{subfigure}{\subpanelwid}
            \centering
            \includegraphics[width=\textwidth]{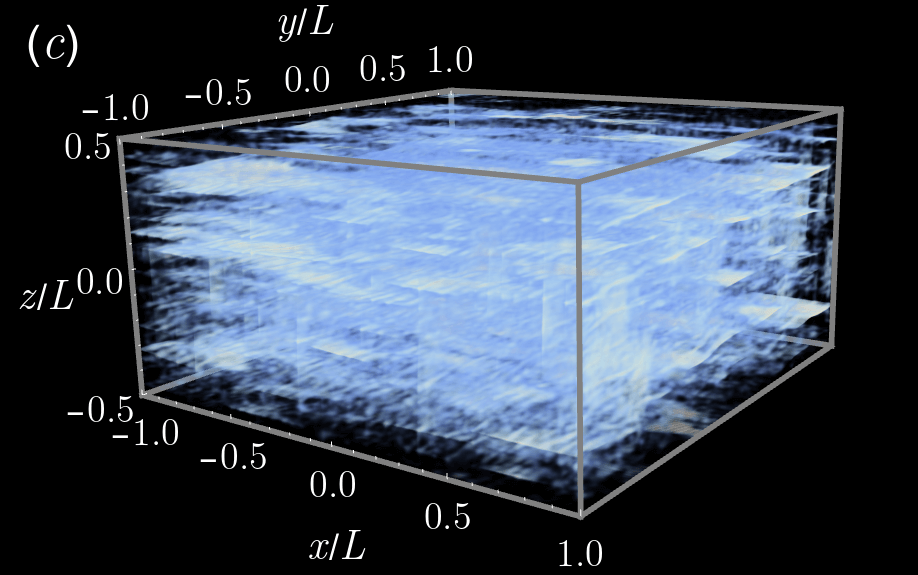}
        \end{subfigure} &
        \begin{subfigure}{\subpanelwid}
            \centering
            \includegraphics[width=\textwidth]{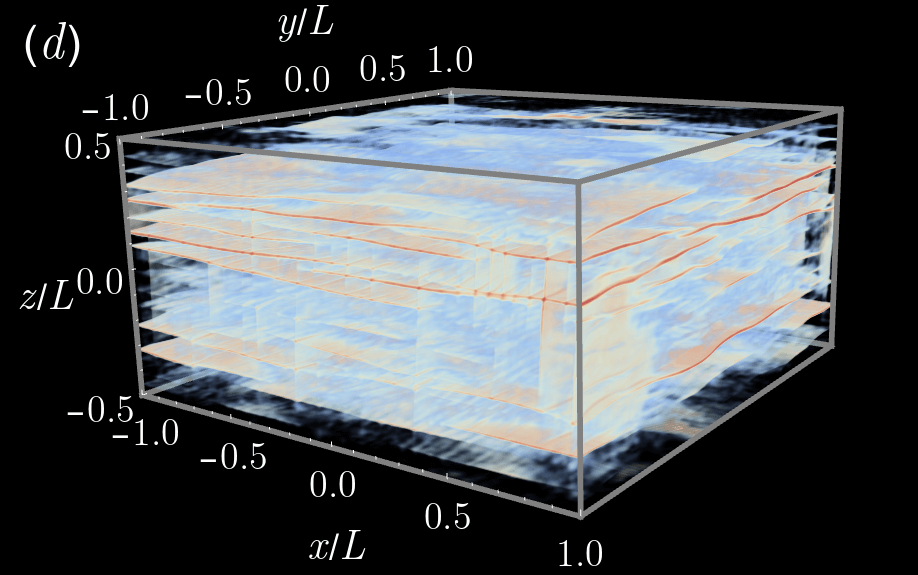}
        \end{subfigure} \\
        \begin{subfigure}{\subpanelwid}
            \centering
            \includegraphics[width=\textwidth]{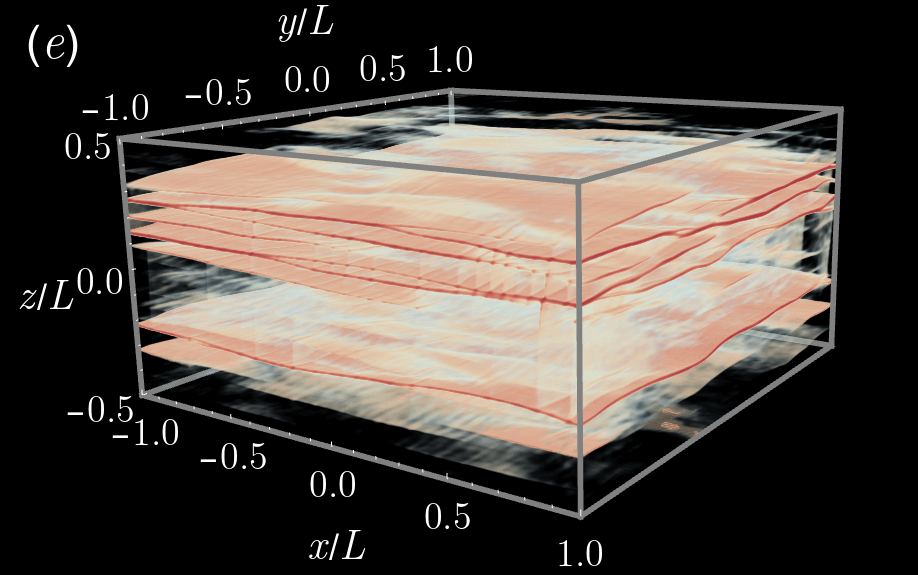}
        \end{subfigure} &
        \begin{subfigure}{\subpanelwid}
            \centering
            \includegraphics[width=\textwidth]{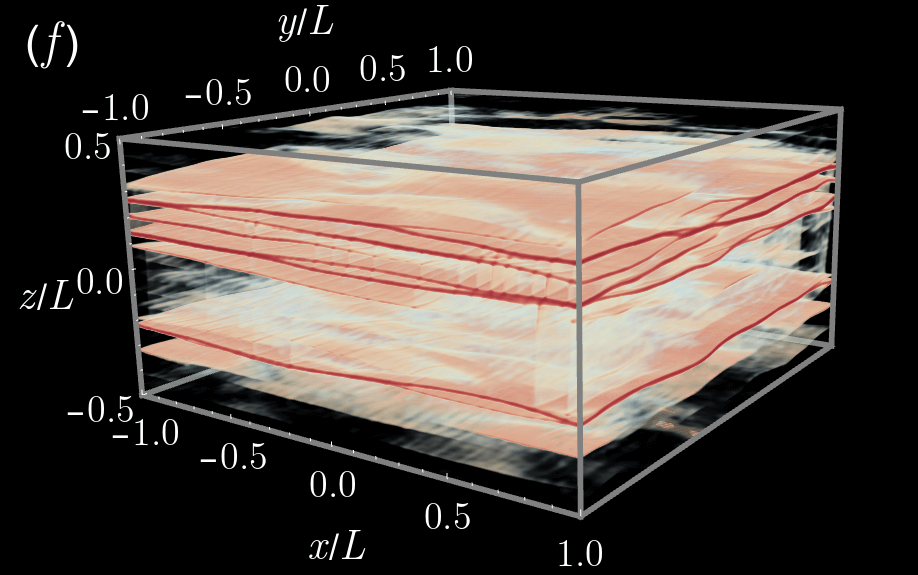}
        \end{subfigure} \\
    \end{tabular}}
    \begin{subfigure}{\textwidth}
        \centering
        \vspace{5mm}
        \includegraphics[width=.75\textwidth]{imgs/rslt_figs/colorbar_2019.png}
        \vspace{5mm}
    \end{subfigure}
    \caption{Snapshots of the effective temperature distribution $\chi(\bx, t)$ for a quasi-static simulation with $\zeta = 1$. The initial conditions are normally distributed in $\chi$ with mean $\mu_\chi = 550 \text{~K}$ and standard deviation $\sigma_\chi = 15 \text{~K}$. (a) $t = 0 t_s$, $a = 0.25$, and $\eta = 1.3$. (b) $t = 2\times 10^5 t_s$, $a = 0.25$, and $\eta = 1.3$. (c) $t = 4\times 10^5 t_s$, $a = 0.45$, and $\eta = 1.75$. (d) $t = 6\times 10^5 t_s$, $a = 0.5$, and $\eta = 1.8$. (e) $t = 8\times 10^5 t_s$, $a = 0.75$, and $\eta = 2.0$. (f) $t = 10^6 t_s$, $a = 0.75$, and $\eta = 2.0$.}
    \label{fig:rndm}
\end{figure}

We first populate the grid and a shell of ghost points with random variables $\chi_\zeta(\mathbf{x})$ using the Box--Muller algorithm. With $\mu_\chi$ and $\sigma_\chi$ respectively denoting the desired mean and standard deviation, we perform the convolution
\begin{equation}
    \chi(\mathbf{x}) = \frac{\sigma_\chi}{N} \sum_{\mathbf{r} \in V'} e^{-\frac{\left\|\mathbf{x} - \mathbf{r}\right\|^2}{l_c^2}}\chi_\zeta(\mathbf{r}) + \mu_\chi, \qquad
    N = \sqrt{\sum_{\mathbf{r} \in V} e^{-2\frac{\|\mathbf{r}\|^2}{l_c^2}}}.
    \label{eqn:convolve}
\end{equation}
where $V$ denotes the set of grid points and $V'$ denotes the set of grid points with the addition of the ghost points. Equation \ref{eqn:convolve} ensures that the effective temperature value at each point is normally distributed with mean $\mu_\chi$ and standard deviation $\sigma_\chi$. In practice, the sums in Eq.~\ref{eqn:convolve} are performed with a cutoff length scale specified as a multiplicative factor of the convolution length scale $l_c$, and the number of ghost points in $V'$ is set by the choice of cutoff length scale. Results for a random initialization with $\mu_\chi = 550\text{~K}$, $\sigma_\chi = 15\text{~K}$, $l_c = 10h$ and a cutoff factor of $5$ (leading to $50$ ghost points in each direction for the convolution) are shown in Fig.~\ref{fig:rndm}. The grid is of size $768\times 768 \times 384$. The simulation geometry, quasi-static timestep, and boundary velocity are the same as in previous sections. The initial conditions are shown in Fig.~\ref{fig:rndm}(a). The simulation takes 105.970 total hours using 32 processes on an Ubuntu Linux computer with dual 16-core Intel Xeon E5-2683 v4 processors. The total time spent in V-cycles is 75.90 hours. The total number of V-cycles is 19689.

By $t = 2\times 10^5 t_s$ in Fig.~\ref{fig:rndm}(b), the effective temperature has increased somewhat uniformly across the grid. At $t = 4\times 10^5 t_s$ in Fig.~\ref{fig:rndm}(c), both horizontal and vertical shear bands begin to nucleate throughout the simulation. Slightly later at $t = 6\times 10^5t_s$ in Fig.~\ref{fig:rndm}(d), a multitude of thin, system-spanning horizontal bands connected by vertical bands have begun to emerge. Curvature is present in the horizontal bands both parallel and orthogonal to the direction of shear. The shear bands become increasingly prominent and grow in number by $t = 8\times 10^5 t_s$ and $t = 10^6 t_s$ in Figs.~\ref{fig:rndm}(e) and (f) respectively, where crossing and branching patterns in the many bands are observed.

\section{Friction welding}
In this section, we investigate the shear banding dynamics for a set of boundary conditions inspired by friction welding, a process of physical interest and engineering relevance \cite{weld1, weld2, weld3}. A typical friction welding example consists of pushing a rapidly spinning cylinder into a block of material. The friction generated at the block--cylinder interface causes the material to heat up and melt, forming a weld. Inspired by this scenario, we examine how applying spinning-disc boundary conditions to a block of material can generate deformation. Let \smash{$r^2 = \left(\frac{x}{L}\right)^2 + \left(\frac{y}{L}\right)^2$}. We impose a boundary condition corresponding to
\begin{equation}
    u = \begin{cases}
      -\frac{\omega x}{L} & \text{if $r^2 < R^2$,}\\
        -e^{-\lambda_\omega\left(r^2 - R^2\right)^2}\frac{\omega x}{L} & \text{otherwise,}
    \end{cases}
    \label{eqn:u_weld}
\end{equation}
\begin{equation}
    v = \begin{cases}
      \frac{\omega y}{L} & \text{if $r^2 < R^2$,}\\
        e^{-\lambda_\omega\left(r^2 - R^2\right)^2}\frac{\omega y}{L} &\text{otherwise,}
    \end{cases}
    \label{eqn:v_weld}
\end{equation}
on the top boundary. $\lambda_w$ is a parameter that sets the smoothness of the boundary condition to avoid stress concentration at the boundary of the rotating region; in the limit $\lambda_w \rightarrow \infty$, Eqs.~\ref{eqn:u_weld}\&\ref{eqn:v_weld} describe a rotation only within the disc $r^2 < R^2$. In all simulations, we set $\lambda_w = 16\log(10)/(1-R^2)^2$ and \smash{$R = \frac{1}{3}$}. This choice of $\lambda_w$ ensures that one quarter of the way between the boundary of the spinning disc $r^2 < R^2$ and the edge of the simulation, the angular velocity is $10\%$ of its value inside the disc.

To promote shear banding with this boundary condition, we set $\gamma = \frac{1}{4}$, corresponding to a more slender domain than considered in previous sections. All simulations are conducted on a $768\times768\times384$ cell grid with a random initial effective temperature distribution as described in the preceding section. We choose values $\mu_\chi = 450$ and $\sigma_\chi = 15$. The convolution lengthscale and cutoff factor are set to $5h$ and $5$ respectively. We simulate until a final time of $t_f = 10^8t_s$ and use a quasi-static timestep of $\Delta t = 2\times 10^4 t_s$. We choose a value of $\omega = \frac{2\pi}{t_f}$, corresponding to a single rotation over the course of the simulation. We found that lowering the angular velocity by increasing the simulation duration promoted interesting shear banding dynamics, and we increased the timestep accordingly.

\begin{figure}
\fcolorbox{black}{black}{%
    \begin{tabular}{cc}
        \begin{subfigure}{\subpanelwid}
            \centering
            \includegraphics[width=\textwidth]{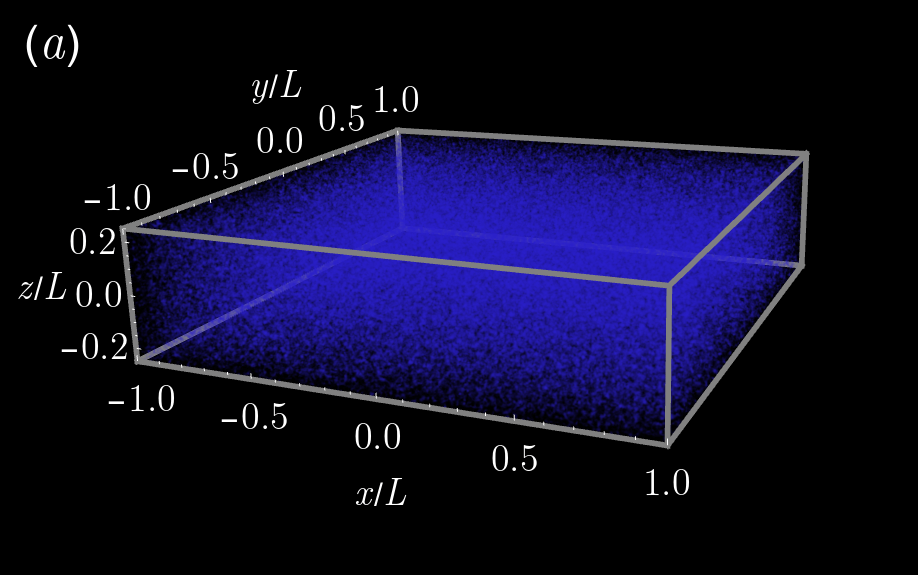}
        \end{subfigure} &
        \begin{subfigure}{\subpanelwid}
            \centering
            \includegraphics[width=\textwidth]{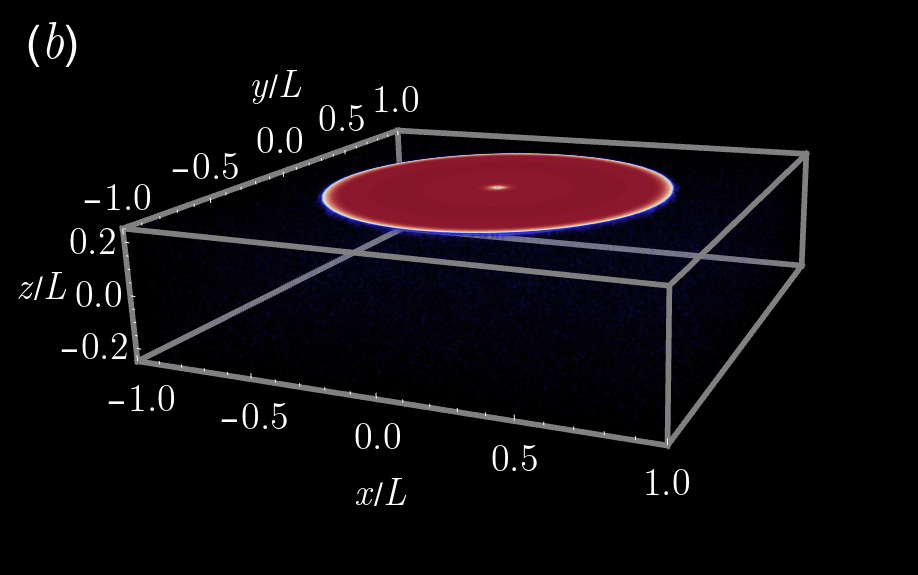}
        \end{subfigure}\\
        \begin{subfigure}{\subpanelwid}
            \centering
            \includegraphics[width=\textwidth]{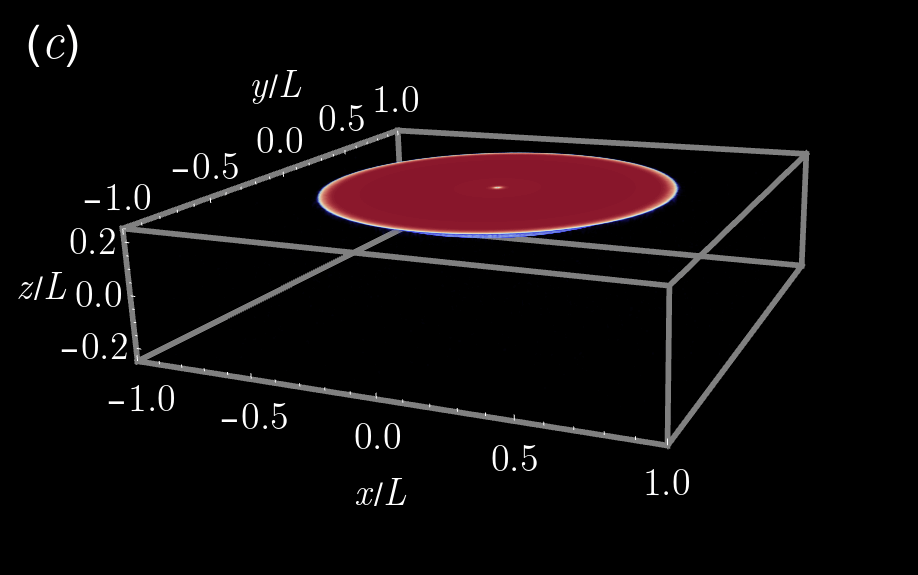}
        \end{subfigure} &
        \begin{subfigure}{\subpanelwid}
            \centering
            \includegraphics[width=\textwidth]{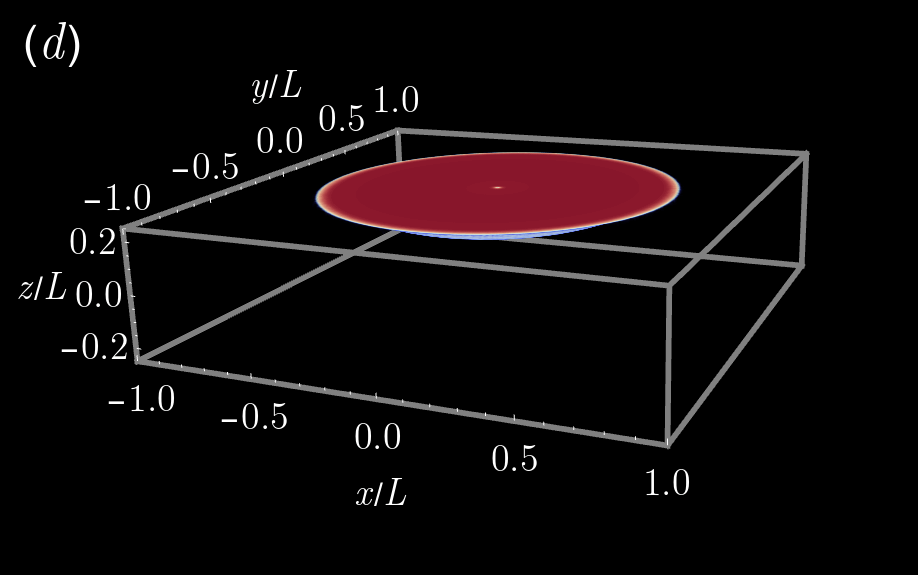}
        \end{subfigure} \\
        \begin{subfigure}{\subpanelwid}
            \centering
            \includegraphics[width=\textwidth]{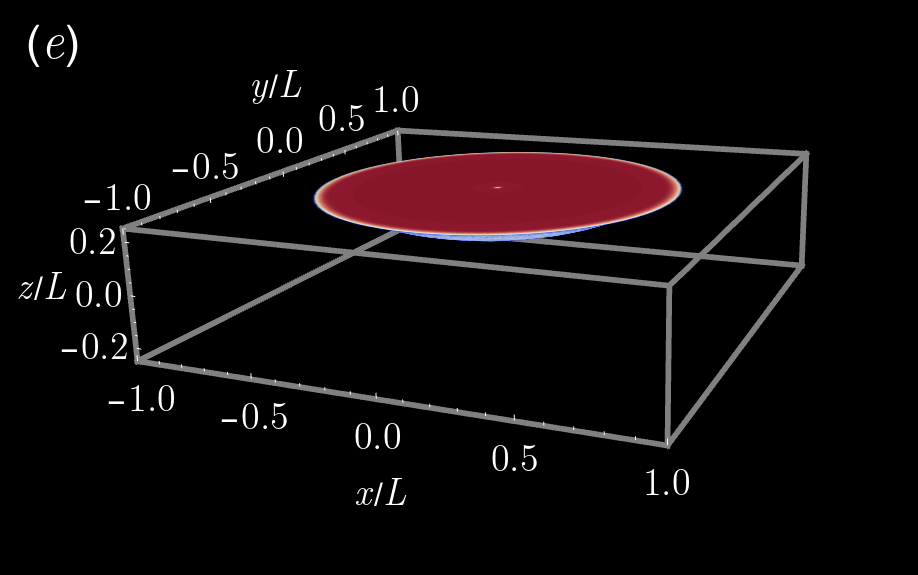}
        \end{subfigure} &
        \begin{subfigure}{\subpanelwid}
            \centering
            \includegraphics[width=\textwidth]{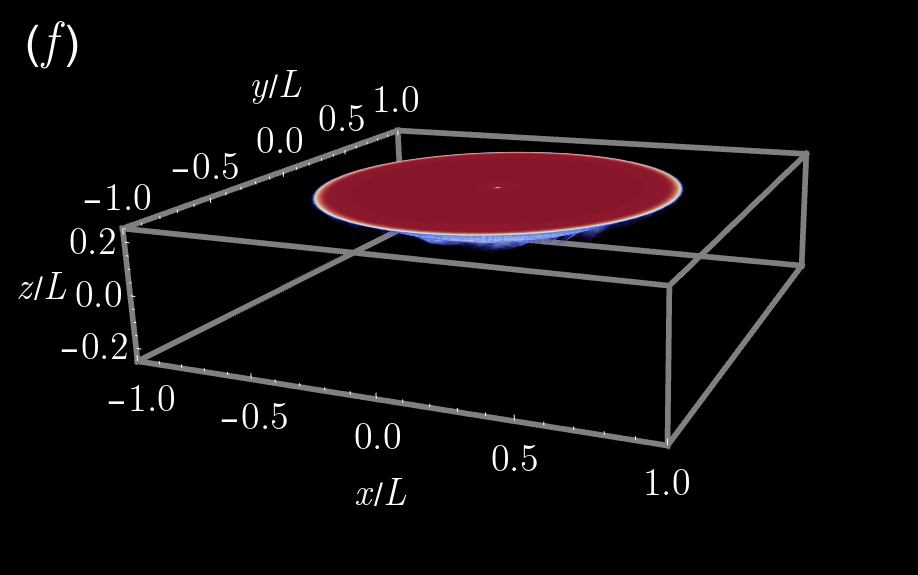}
        \end{subfigure} \\
    \end{tabular}}
    \begin{subfigure}{\textwidth}
        \centering
        \vspace{5mm}
        \includegraphics[width=.75\textwidth]{imgs/rslt_figs/colorbar_2019.png}
        \vspace{5mm}
    \end{subfigure}
    \caption{Snapshots of the effective temperature distribution $\chi(\bx, t)$ for a quasi-static simulation with $\zeta = 1$. The boundary conditions are inspired by friction welding, and Eqs.~\ref{eqn:u_weld} \& \ref{eqn:v_weld} are applied to the top boundary. (a) $t = 0 t_s$, $a = 0.25$, and $\eta = 1.2$. (b) $t = 2\times 10^7 t_s$, $a = 0.35$, and $\eta = 1.3$. (c) $t = 4\times 10^7 t_s$, $a = 0.45$, and $\eta = 1.4$. (d) $t = 6\times 10^7 t_s$, $a = 0.55$, and $\eta = 1.5$. (e) $t = 8\times 10^7 t_s$, $a = 0.55$, and $\eta = 1.5$. (f) $t = 10^8 t_s$, $a = 0.55$, and $\eta = 1.5$.}
    \label{fig:weld_top}
\end{figure}

\begin{figure}
    \centering
    \begin{tabular}{cc}
        \begin{subfigure}{\subpanelwid}
            \centering
            \includegraphics[width=\textwidth]{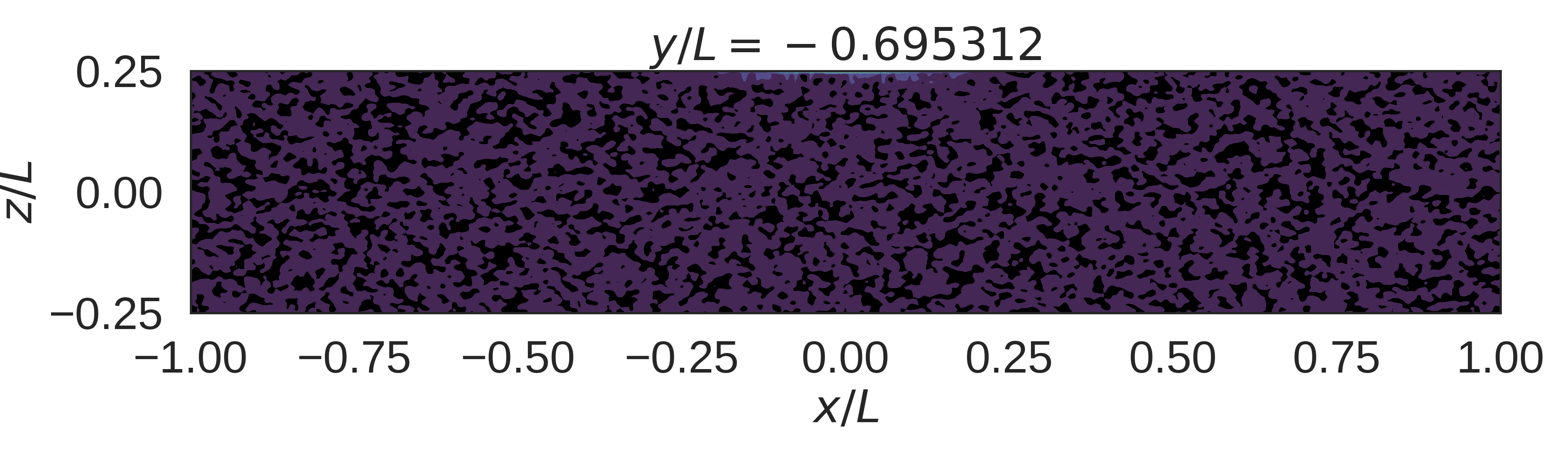}
        \end{subfigure} &
        \begin{subfigure}{\subpanelwid}
            \centering
            \includegraphics[width=\textwidth]{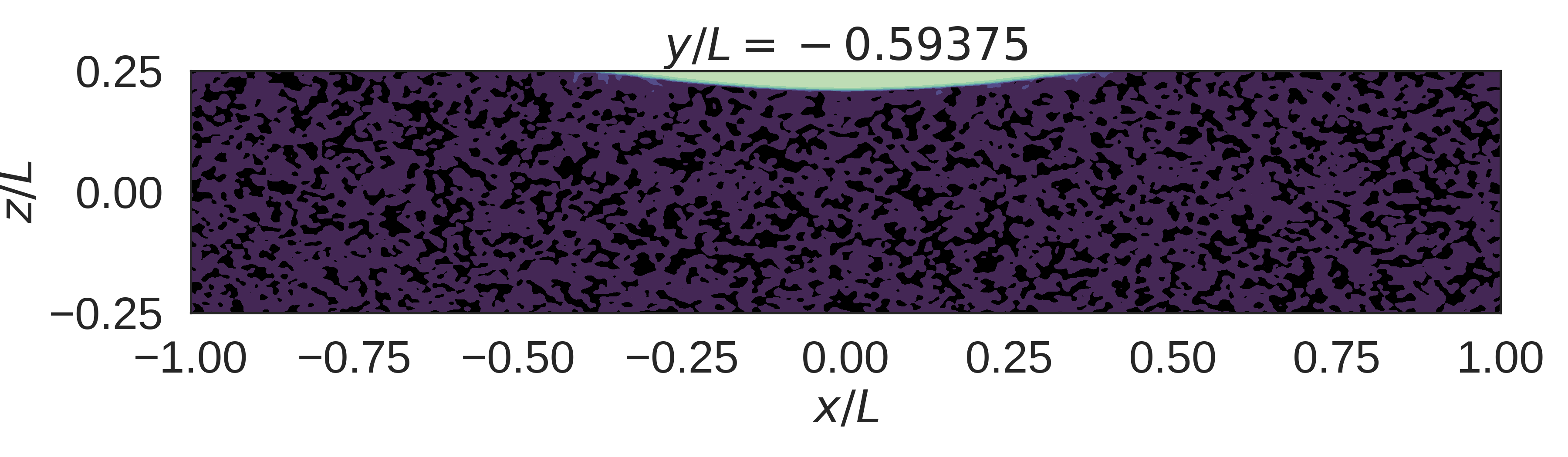}
        \end{subfigure}\\
        \begin{subfigure}{\subpanelwid}
            \centering
            \includegraphics[width=\textwidth]{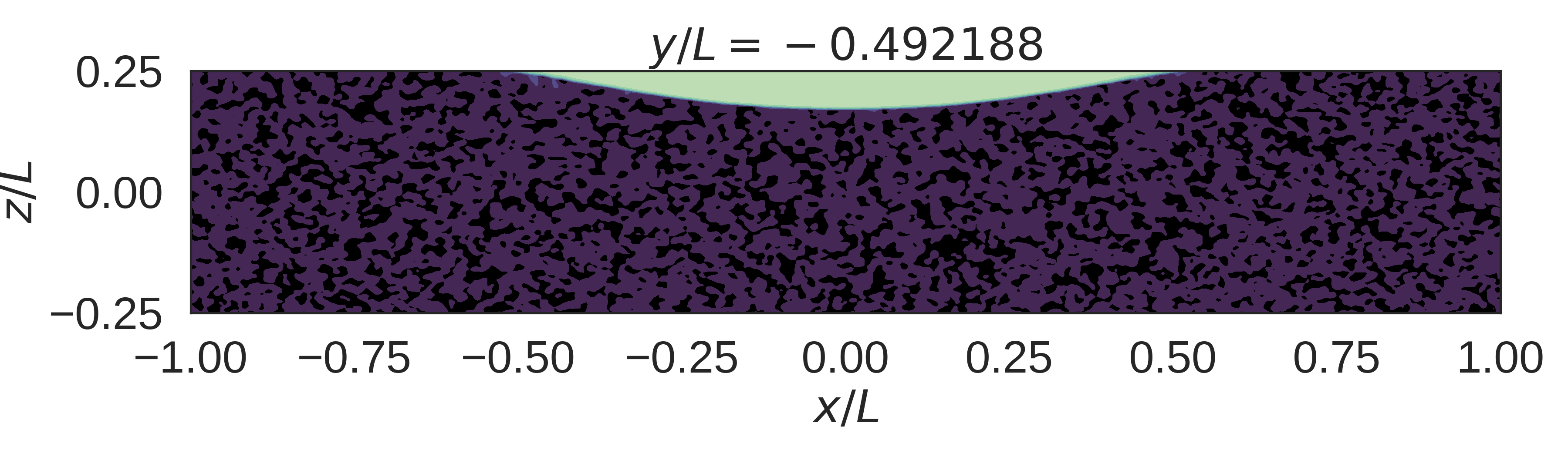}
        \end{subfigure}&
        \begin{subfigure}{\subpanelwid}
            \centering
            \includegraphics[width=\textwidth]{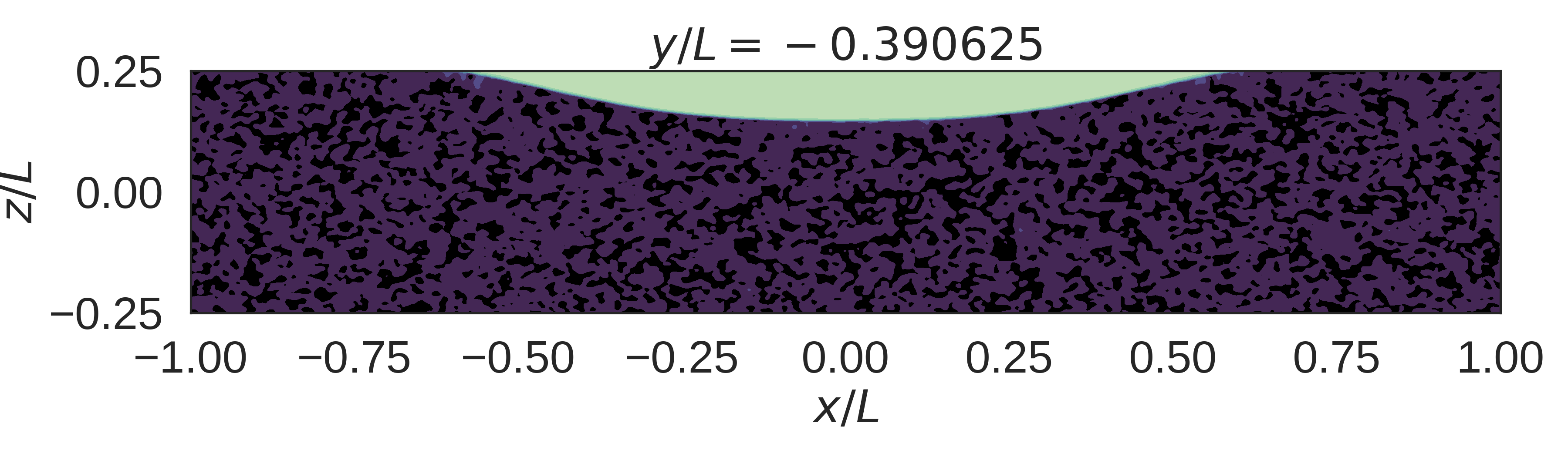}
        \end{subfigure}\\
        \begin{subfigure}{\subpanelwid}
            \centering
            \includegraphics[width=\textwidth]{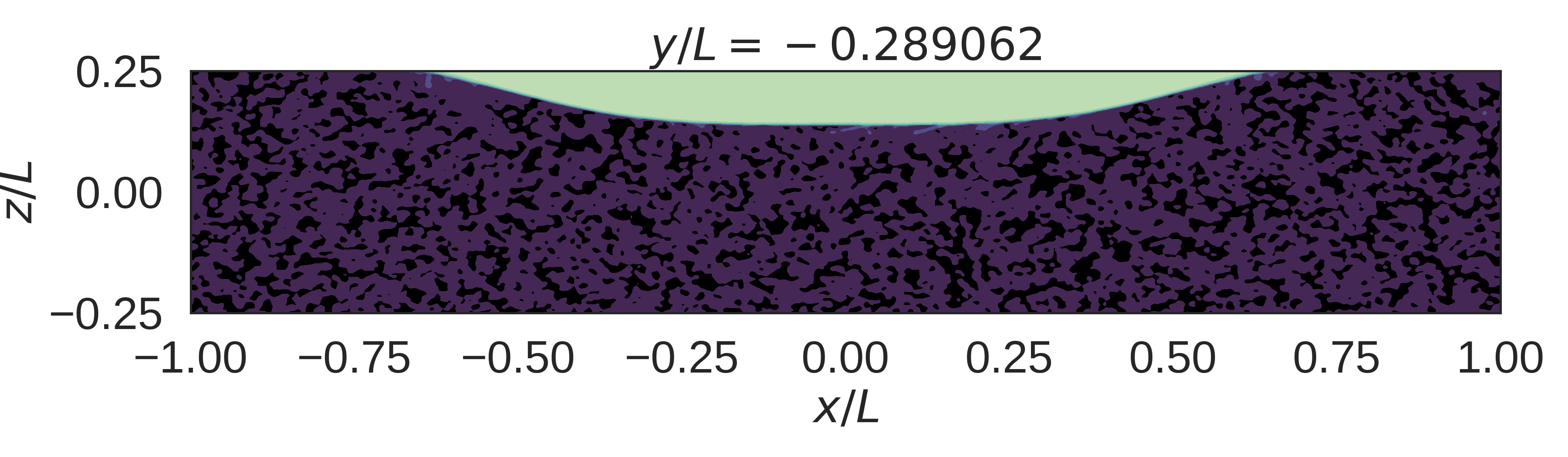}
        \end{subfigure} &
        \begin{subfigure}{\subpanelwid}
            \centering
            \includegraphics[width=\textwidth]{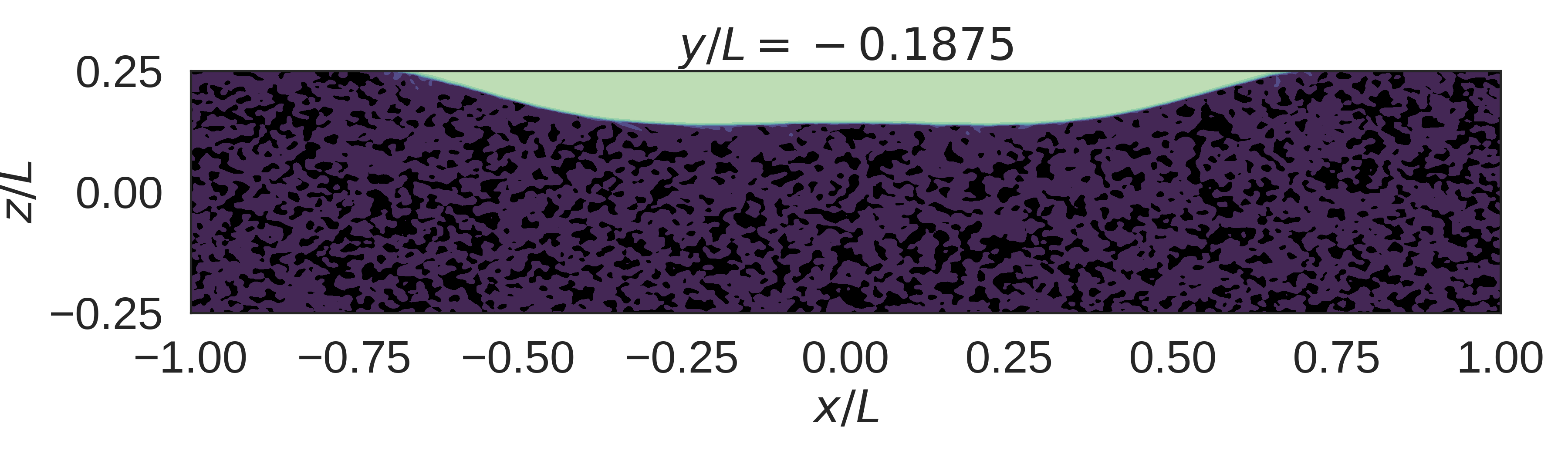}
        \end{subfigure}\\
        \begin{subfigure}{\subpanelwid}
            \centering
            \includegraphics[width=\textwidth]{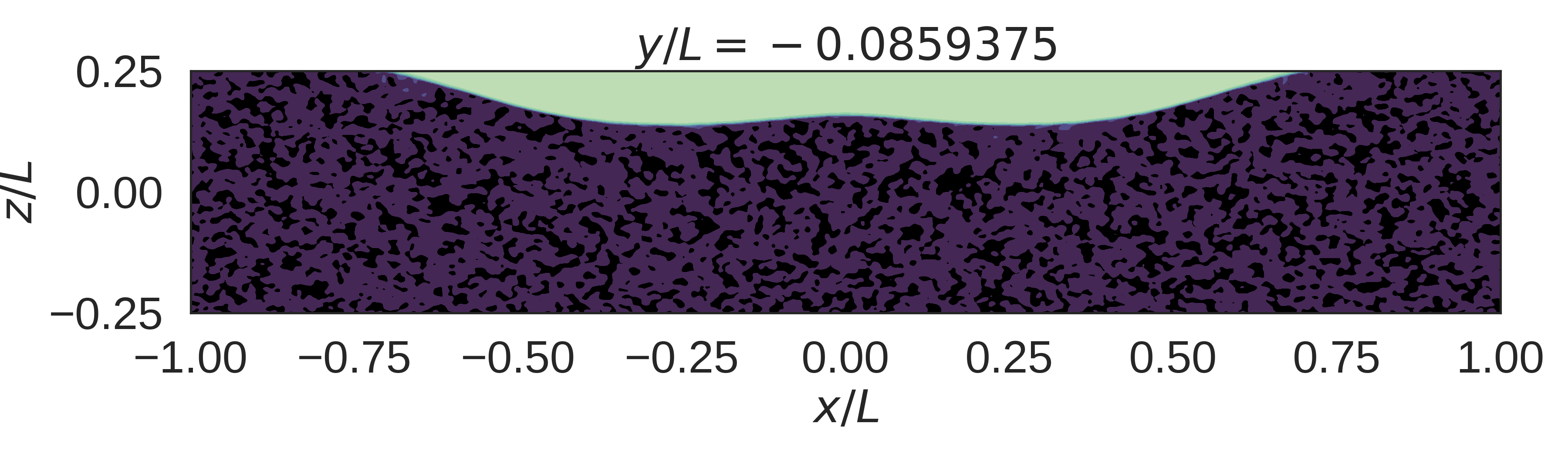}
        \end{subfigure}&
        \begin{subfigure}{\subpanelwid}
            \centering
            \includegraphics[width=\textwidth]{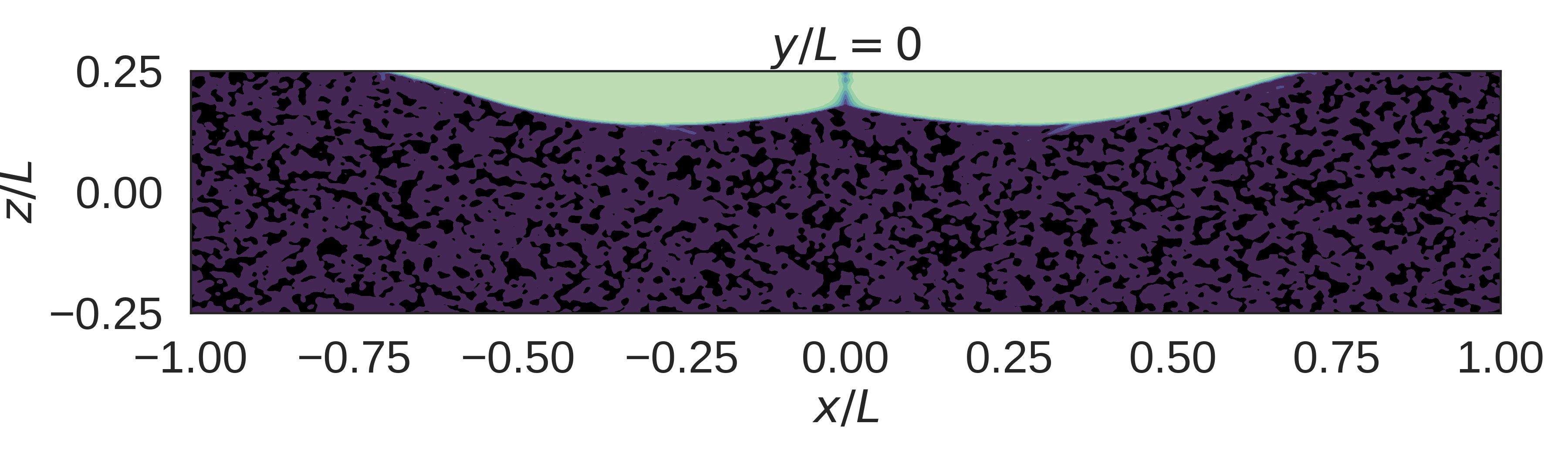}
        \end{subfigure}\\
    \end{tabular}
    \begin{subfigure}{\textwidth}
        \centering
        \includegraphics[width=.6\textwidth]{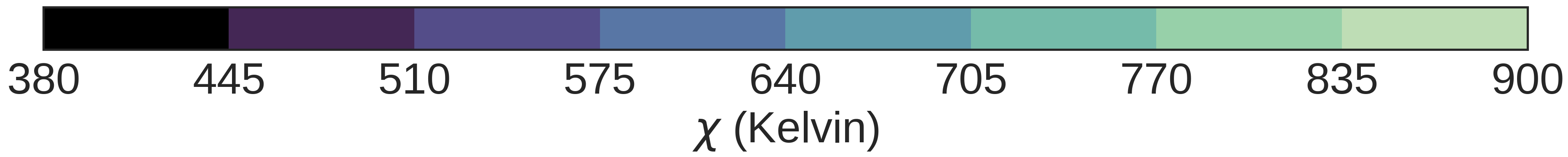}
    \end{subfigure}
    \vspace{5mm}
    \caption{Cross-sections of the $\chi$ field at $t = 5\times 10^7 t_s$ for the friction welding experiment with welding on the top boundary.}
    \label{fig:weld_top_contour}
\end{figure}

Three-dimensional visualizations of the effective temperature distribution are shown in Fig.~\ref{fig:weld_top}. In these figures, it is clear that the plastic deformation is predominantly concentrated on the top boundary, and but propagates a short distance below the disc into the bulk. To better understand the structure of this leakage, we consider $xz$ cross sections of the $\chi$ field for fixed $y$ in Fig.~\ref{fig:weld_top_contour} at $t=5\times 10^7 t_s$. These cross sections are shown at eight roughly equally spaced points on one side of the $y=0$ plane; the dynamics are symmetric up to discrepancies in the random initialization. Here, we see a parabolic region of elevated $\chi$ that begins to pucker as the center of the disc is approached.

\begin{figure}
\fcolorbox{black}{black}{%
    \begin{tabular}{cc}
        \begin{subfigure}{\subpanelwid}
            \centering
            \includegraphics[width=\textwidth]{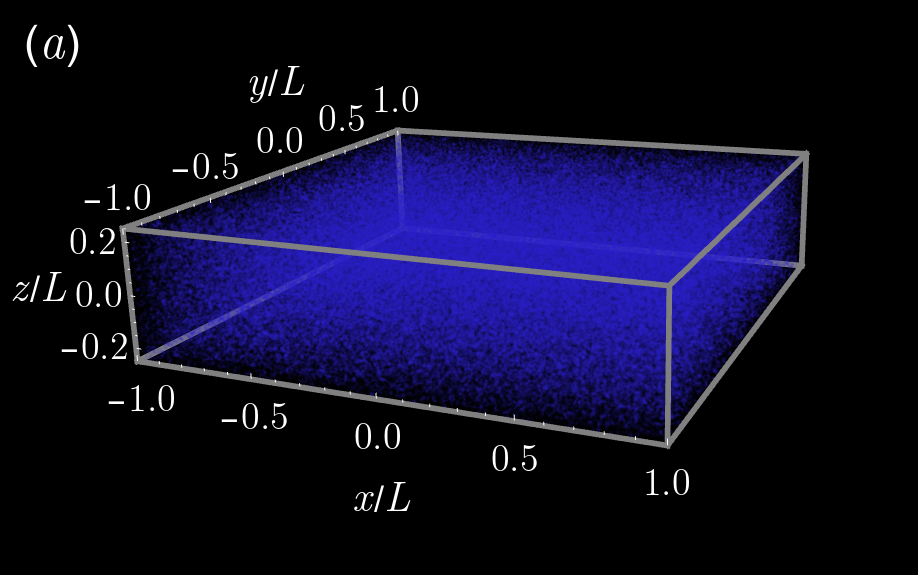}
        \end{subfigure} &
        \begin{subfigure}{\subpanelwid}
            \centering
            \includegraphics[width=\textwidth]{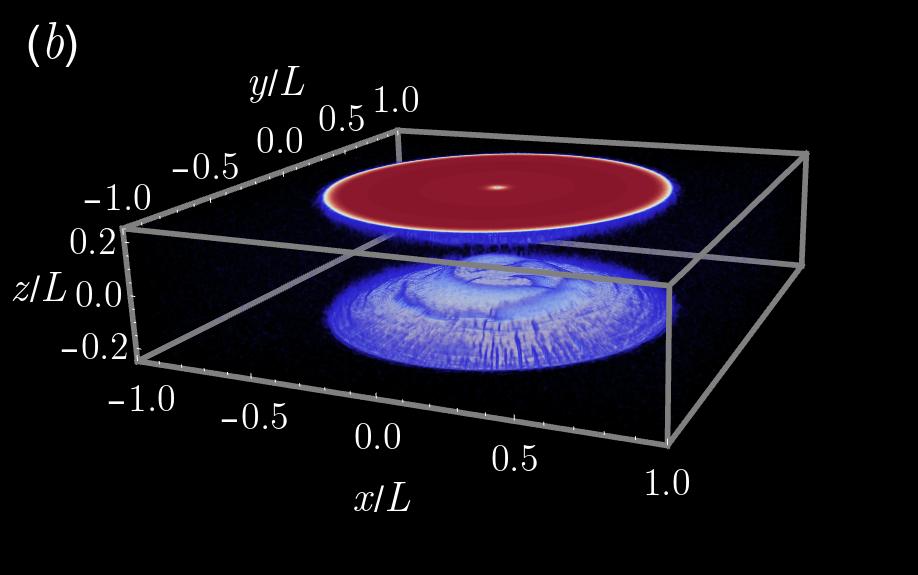}
        \end{subfigure}\\
        \begin{subfigure}{\subpanelwid}
            \centering
            \includegraphics[width=\textwidth]{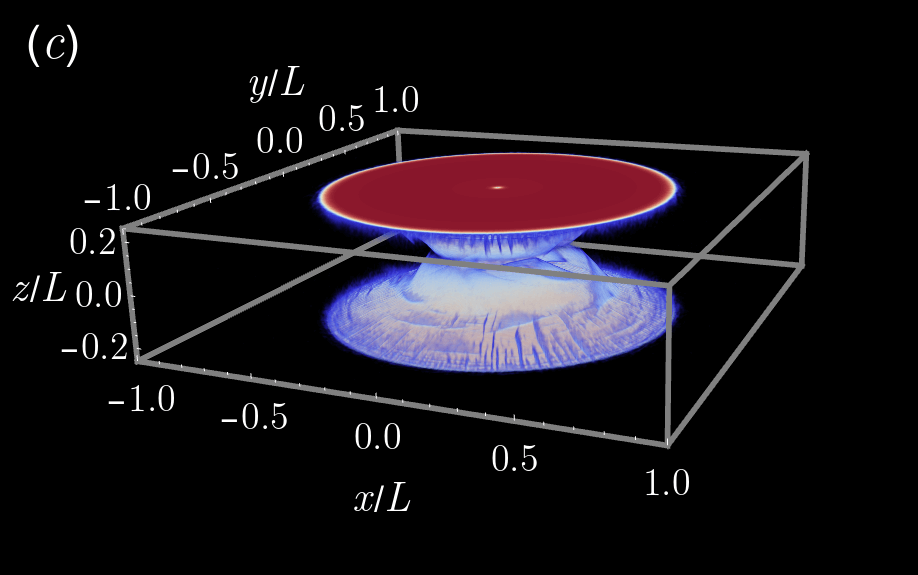}
        \end{subfigure} &
        \begin{subfigure}{\subpanelwid}
            \centering
            \includegraphics[width=\textwidth]{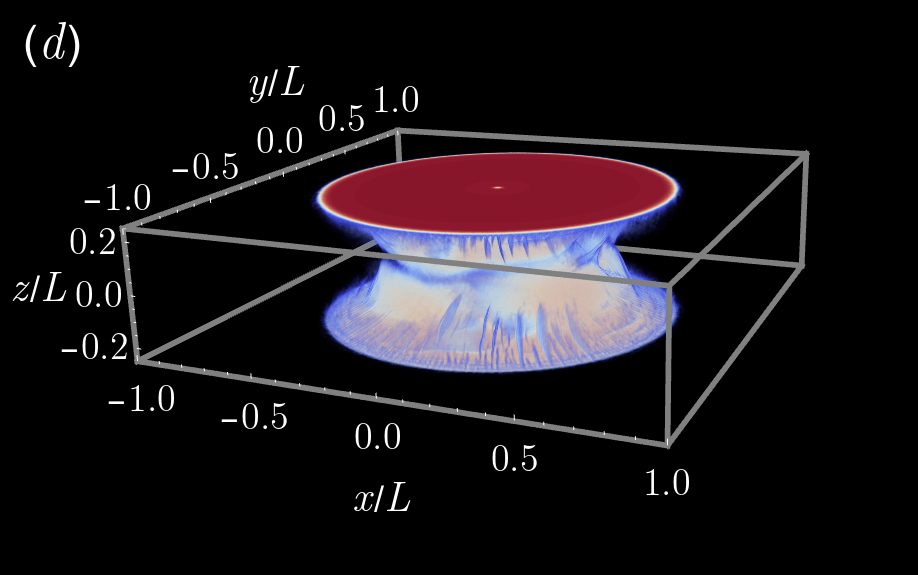}
        \end{subfigure} \\
        \begin{subfigure}{\subpanelwid}
            \centering
            \includegraphics[width=\textwidth]{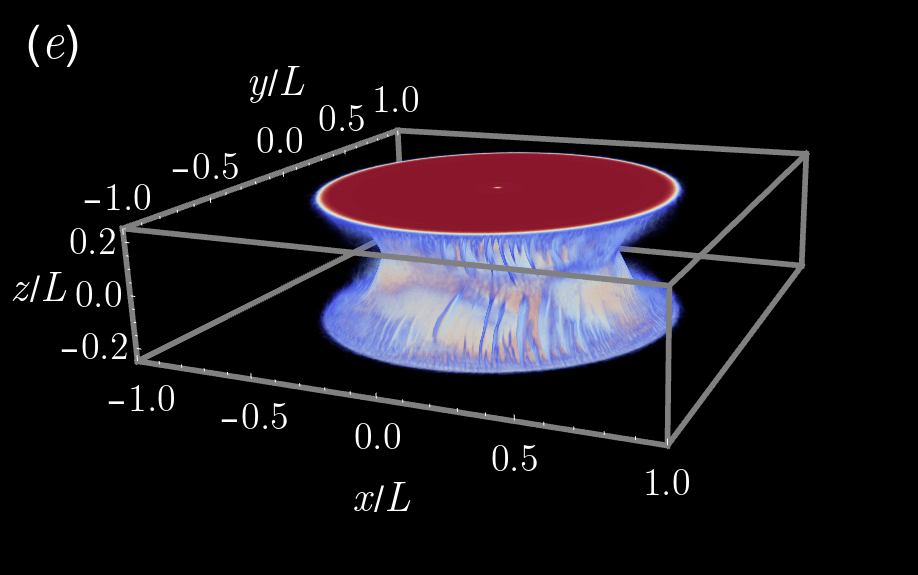}
        \end{subfigure} &
        \begin{subfigure}{\subpanelwid}
            \centering
            \includegraphics[width=\textwidth]{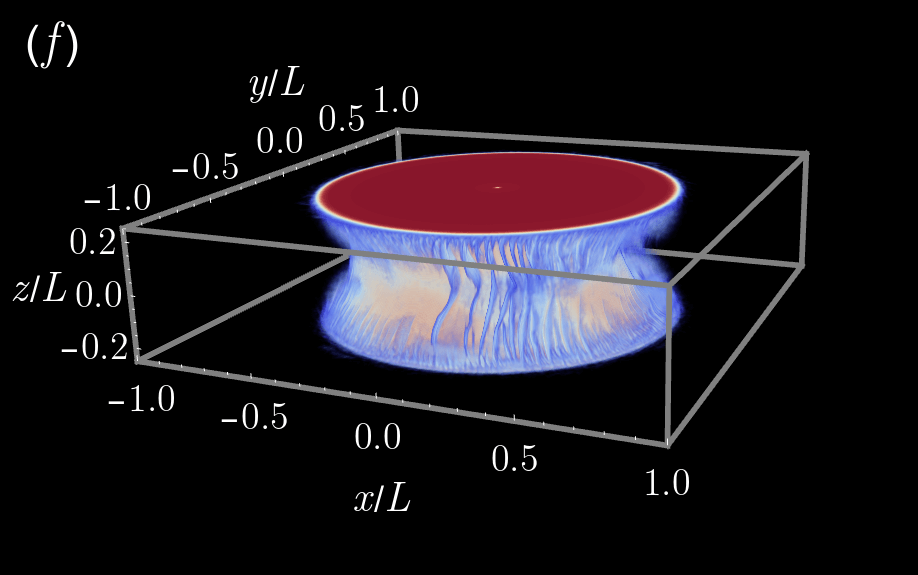}
        \end{subfigure} \\
    \end{tabular}}
    \begin{subfigure}{\textwidth}
        \centering
        \vspace{5mm}
        \includegraphics[width=.75\textwidth]{imgs/rslt_figs/colorbar_2019.png}
        \vspace{5mm}
    \end{subfigure}
    \caption{Snapshots of the effective temperature distribution $\chi(\bx, t)$ for a quasi-static simulation with $\zeta = 1$. The boundary conditions are inspired by friction welding, and Eqs.~\ref{eqn:u_weld} \& \ref{eqn:v_weld} are applied to the top and bottom boundaries. (a) $t = 0 t_s$, $a = 0.25$, and $\eta = 1.2$. (b) $t = 2\times 10^7 t_s$, $a = 0.35$, and $\eta = 1.3$. (c) $t = 4\times 10^7 t_s$, $a = 0.45$, and $\eta = 1.4$. (d) $t = 6\times 10^7 t_s$, $a = 0.55$, and $\eta = 1.5$. (e) $t = 8\times 10^7 t_s$, $a = 0.55$, and $\eta = 1.5$. (f) $t = 10^8 t_s$, $a = 0.55$, and $\eta = 1.5$.}
    \label{fig:weld_both}
\end{figure}

Given the structure of the $\chi$ field in the cross sections in Fig.~\ref{fig:weld_top_contour}, we may expect more interesting dynamics if both the top and bottom surfaces are welded. In Fig.~\ref{fig:weld_both}, we show three-dimensional visualizations of the shear banding dynamics when Eqs.~\ref{eqn:u_weld} and \ref{eqn:v_weld} are applied to both the top and bottom boundaries. Here, the deformation dynamics are much richer, and complex structure unlike that produced by simple shear can be seen. In Fig.~\ref{fig:weld_both}(b), the discs of elevated $\chi$ on the top and bottom boundaries have both begun to propagate into the bulk. Even at this stage, the leakage is more complicated than in Fig.~\ref{fig:weld_top}, and includes thin extensions of elevated $\chi$. In Fig.~\ref{fig:weld_both}(c)--(f), the structures on the top and bottom begin to connect with each other, forming a hyperboloidal region of elevated $\chi$ spanning the system in $z$.

\begin{figure}
    \centering
    \begin{tabular}{cc}
        \begin{subfigure}{\subpanelwid}
            \centering
            \includegraphics[width=\textwidth]{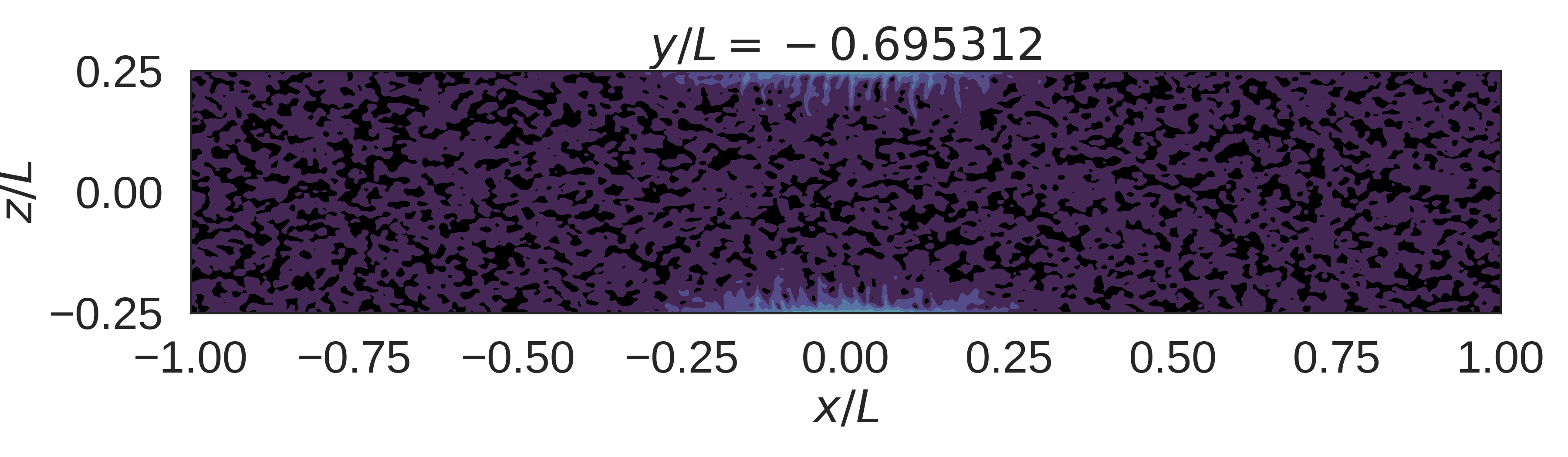}
        \end{subfigure} &
        \begin{subfigure}{\subpanelwid}
            \centering
            \includegraphics[width=\textwidth]{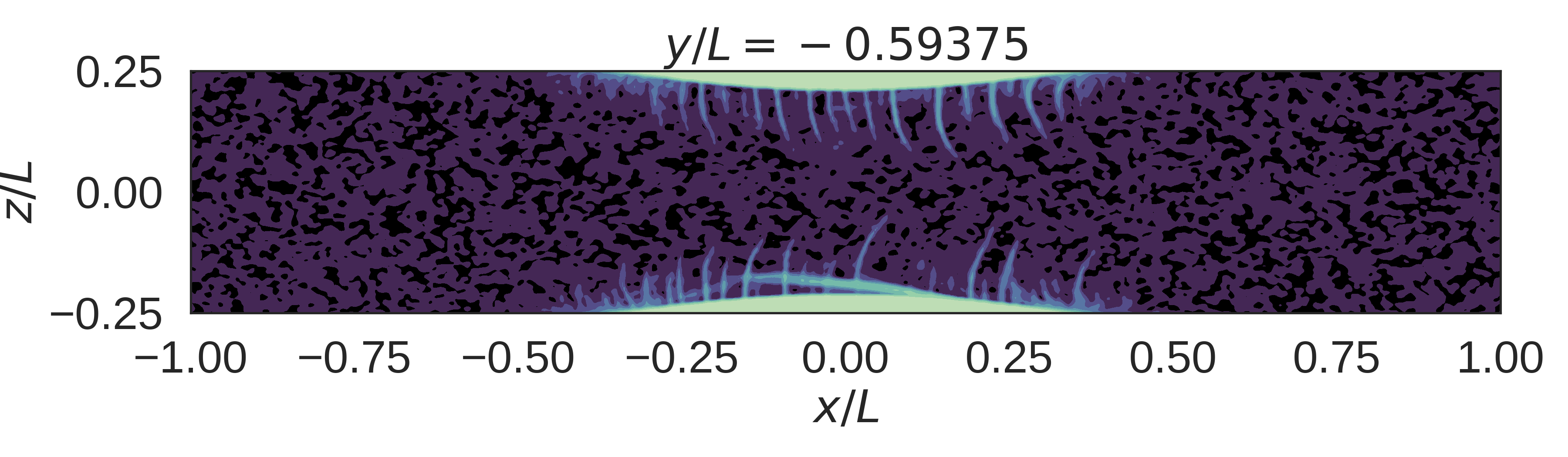}
        \end{subfigure}\\
        \begin{subfigure}{\subpanelwid}
            \centering
            \includegraphics[width=\textwidth]{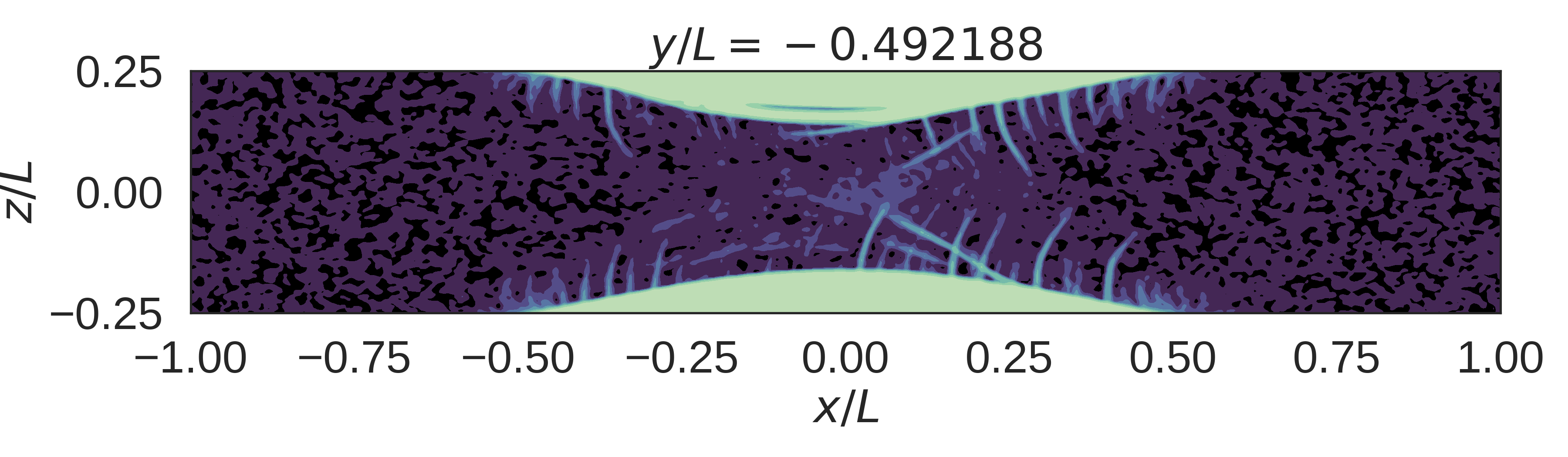}
        \end{subfigure}&
        \begin{subfigure}{\subpanelwid}
            \centering
            \includegraphics[width=\textwidth]{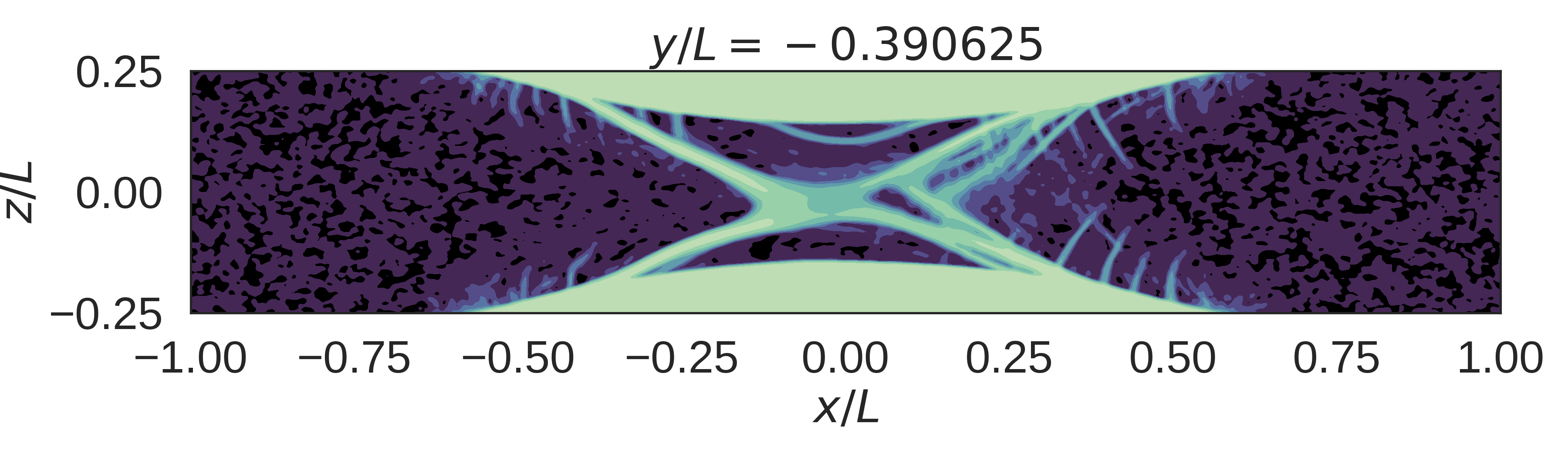}
        \end{subfigure}\\
        \begin{subfigure}{\subpanelwid}
            \centering
            \includegraphics[width=\textwidth]{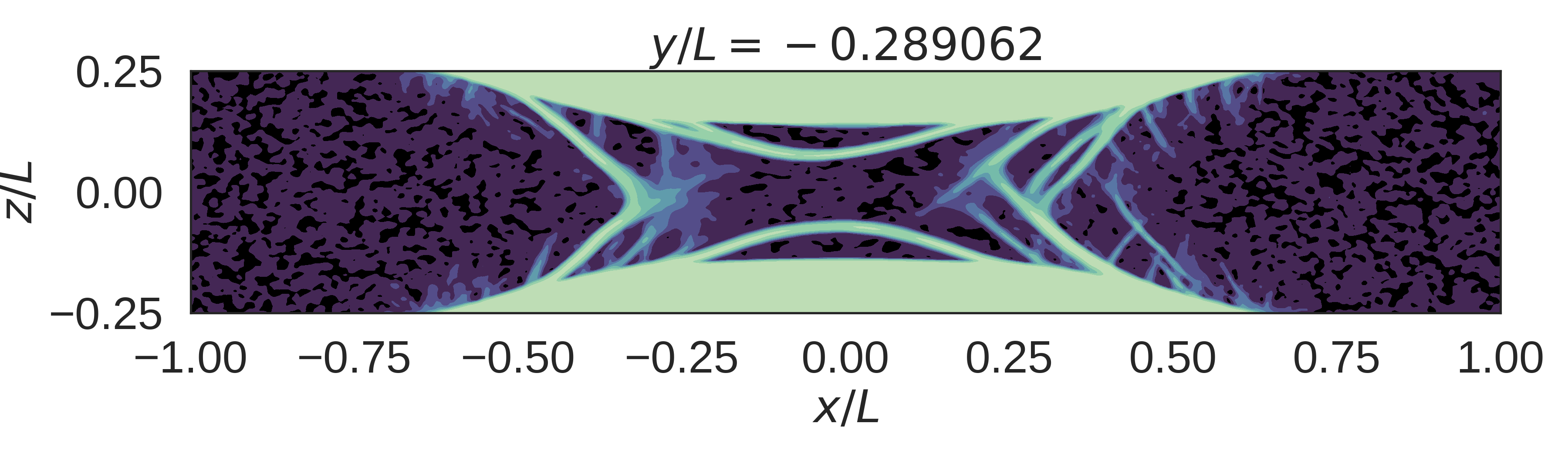}
        \end{subfigure} &
        \begin{subfigure}{\subpanelwid}
            \centering
            \includegraphics[width=\textwidth]{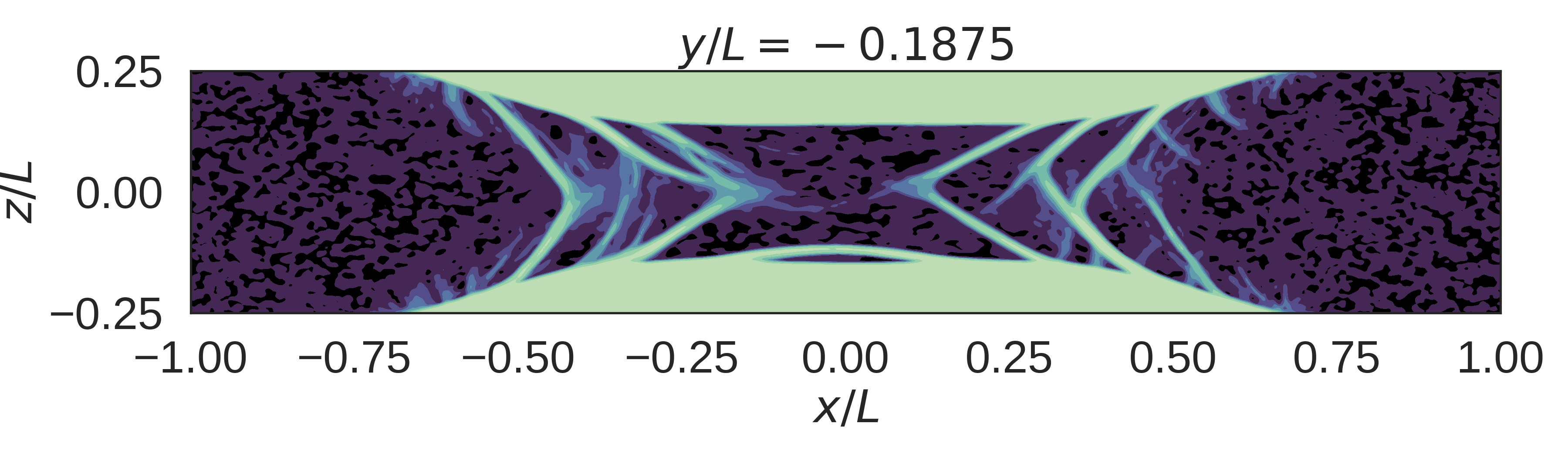}
        \end{subfigure}\\
        \begin{subfigure}{\subpanelwid}
            \centering
            \includegraphics[width=\textwidth]{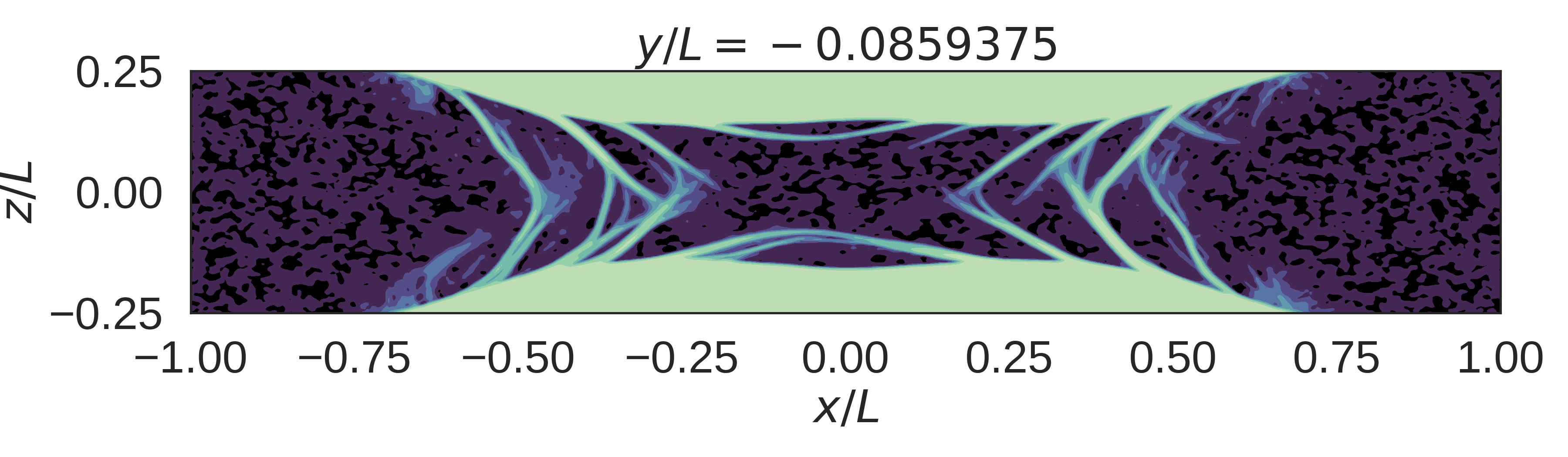}
        \end{subfigure}&
        \begin{subfigure}{\subpanelwid}
            \centering
            \includegraphics[width=\textwidth]{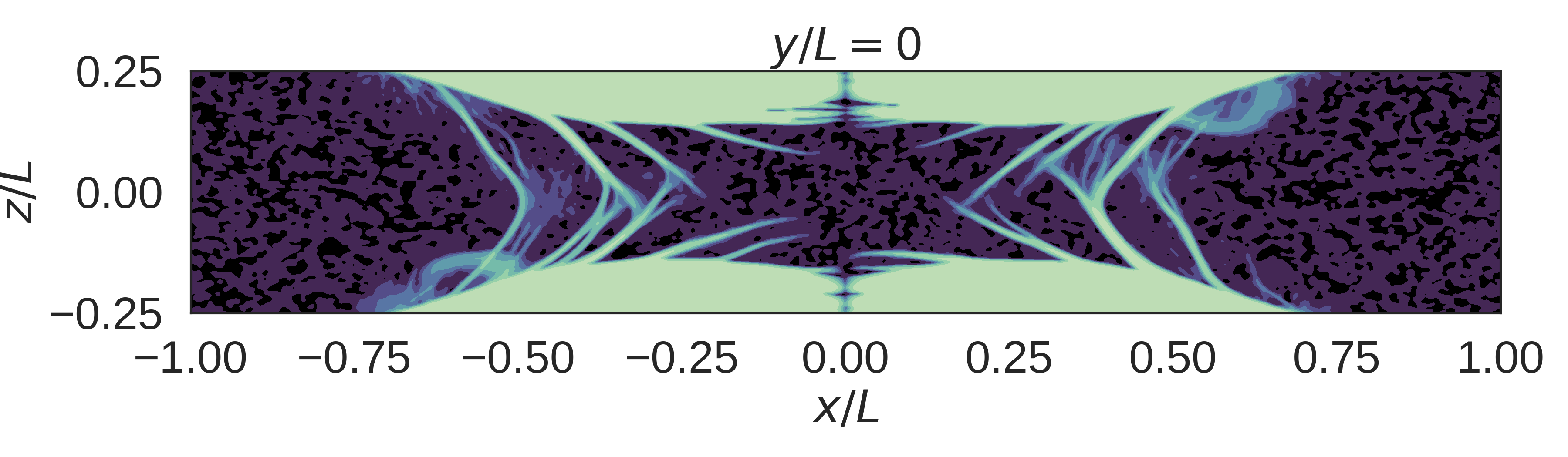}
        \end{subfigure}\\
    \end{tabular}
    \begin{subfigure}{\textwidth}
        \centering
        \includegraphics[width=.6\textwidth]{imgs/rslt_figs/part_one/fric_768_both/contour_weld_cb.png}
    \end{subfigure}
    \vspace{5mm}
    \caption{Cross-sections of the $\chi$ field at $t=5\times 10^7 t_s$ for the friction welding experiment with welding on both the top and bottom. The angular velocity is the same on both the top and bottom.}
    \label{fig:weld_both_contour}
\end{figure}

To probe the structure of this hyperboloid, we again consider two-dimensional cross sections in Fig.~\ref{fig:weld_both_contour} at $t = 5\times 10^7 t_s$. The hyperboloid does not extend entirely through the bulk, but instead has a gap between the two regions of elevated $\chi$ created by the spinning discs. In the gap are thin, tendril-like shear bands that work to connect the two regions and close the gap.

\begin{figure}
\fcolorbox{black}{black}{%
    \begin{tabular}{cc}
        \begin{subfigure}{\subpanelwid}
            \centering
            \includegraphics[width=\textwidth]{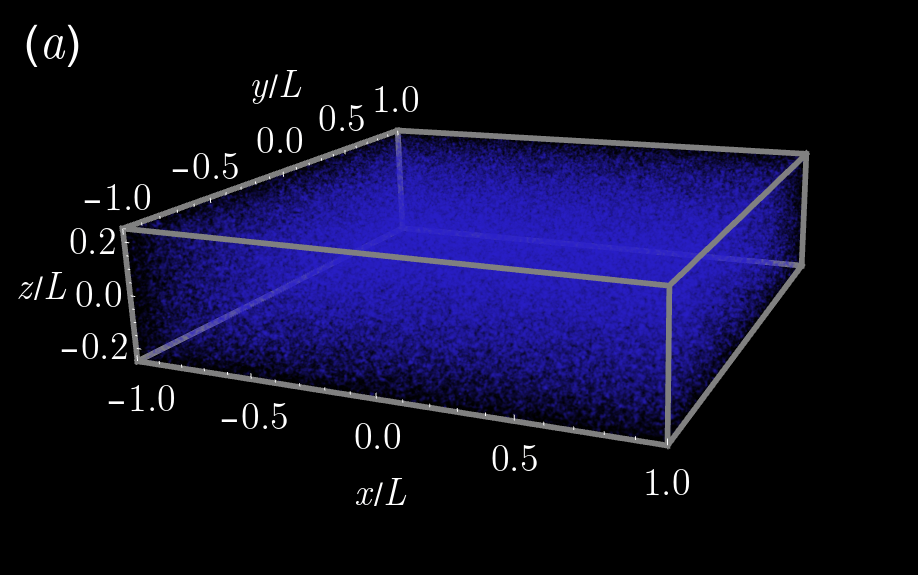}
        \end{subfigure} &
        \begin{subfigure}{\subpanelwid}
            \centering
            \includegraphics[width=\textwidth]{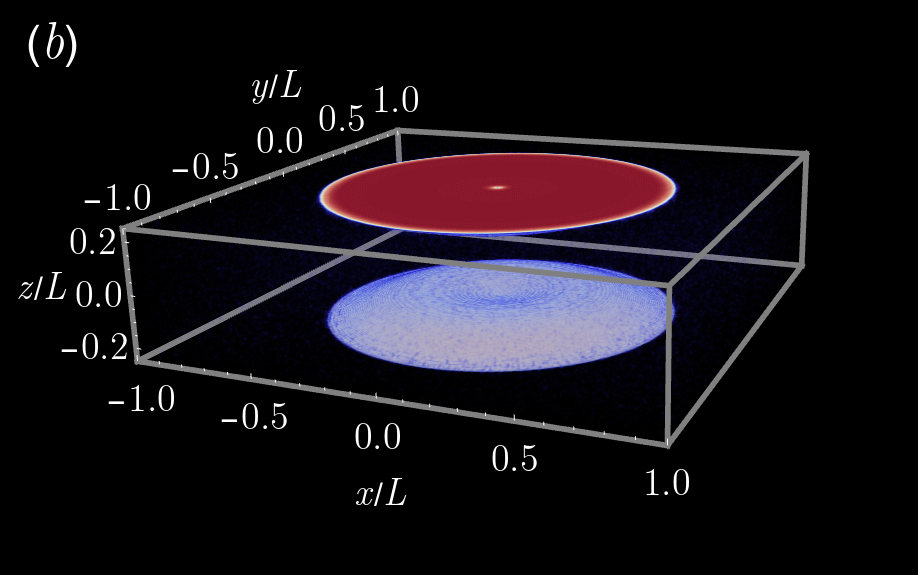}
        \end{subfigure}\\
        \begin{subfigure}{\subpanelwid}
            \centering
            \includegraphics[width=\textwidth]{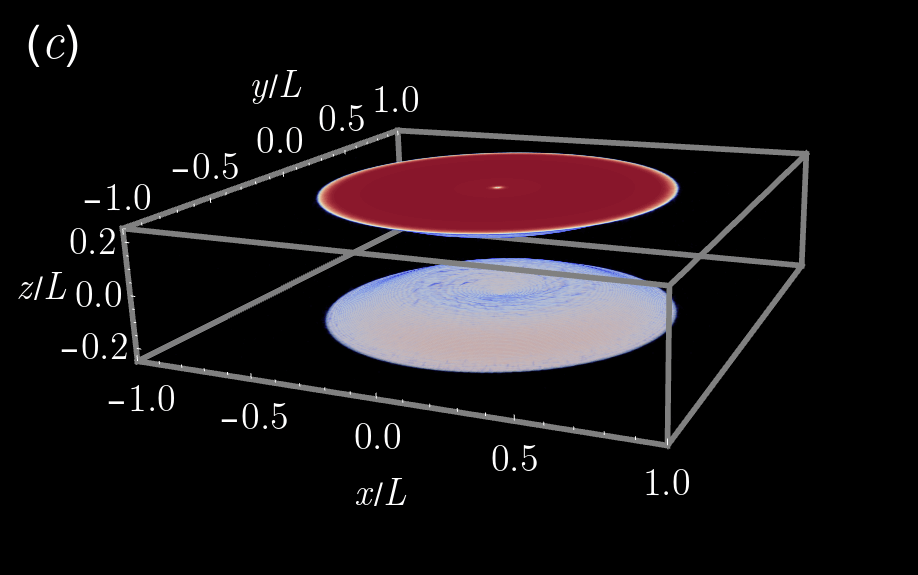}
        \end{subfigure} &
        \begin{subfigure}{\subpanelwid}
            \centering
            \includegraphics[width=\textwidth]{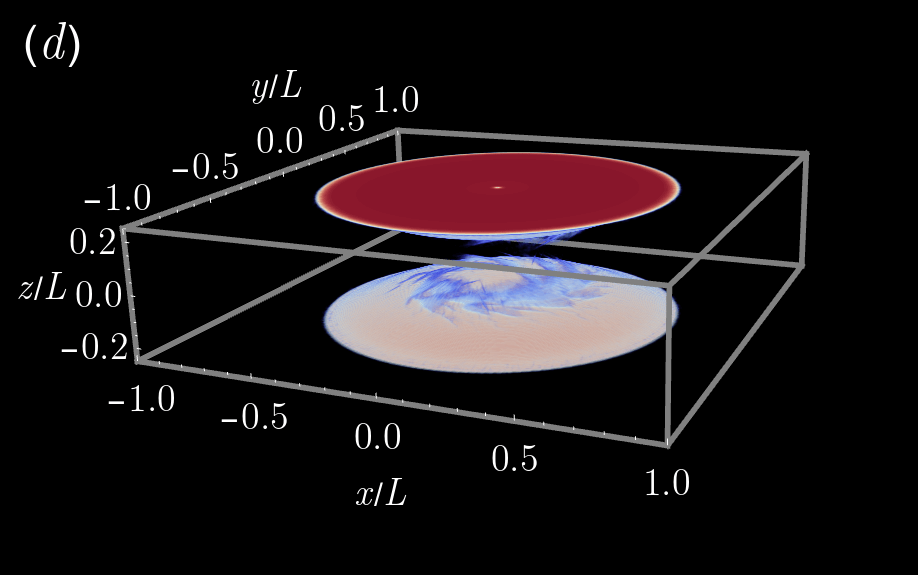}
        \end{subfigure} \\
        \begin{subfigure}{\subpanelwid}
            \centering
            \includegraphics[width=\textwidth]{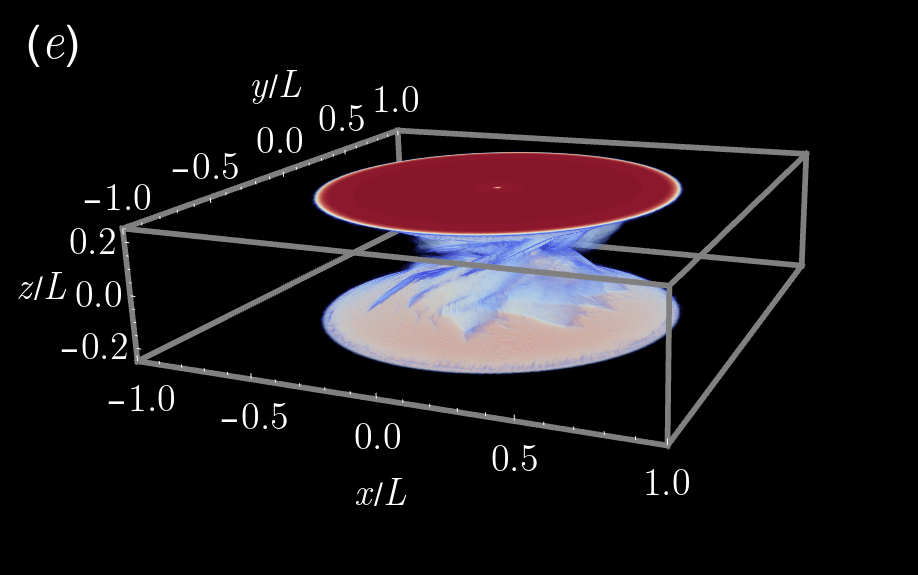}
        \end{subfigure} &
        \begin{subfigure}{\subpanelwid}
            \centering
            \includegraphics[width=\textwidth]{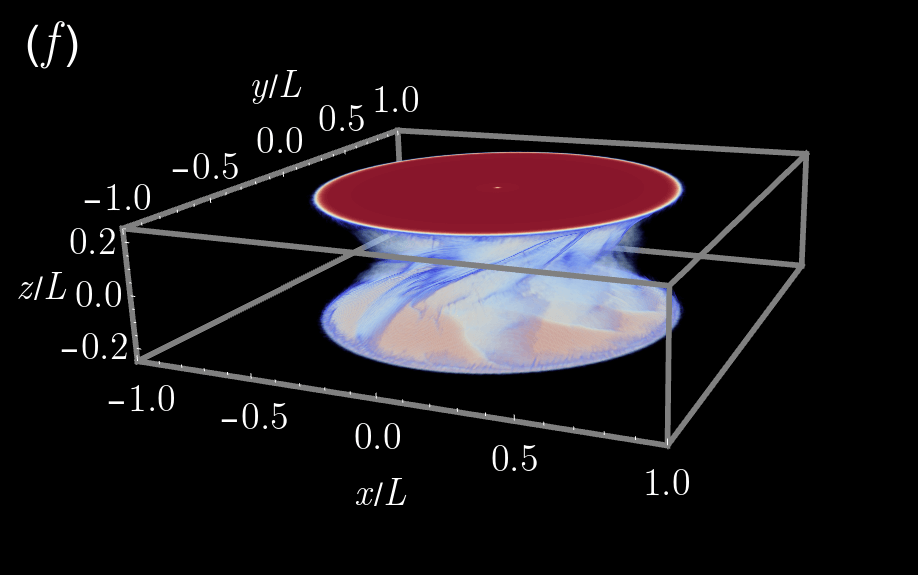}
        \end{subfigure} \\
    \end{tabular}}
    \begin{subfigure}{\textwidth}
        \centering
        \vspace{5mm}
        \includegraphics[width=.75\textwidth]{imgs/rslt_figs/colorbar_2019.png}
        \vspace{5mm}
    \end{subfigure}
    \caption{Snapshots of the effective temperature distribution $\chi(\bx, t)$ for a quasi-static simulation with $\zeta = 1$. The boundary conditions are inspired by friction welding, and Eqs.~\ref{eqn:u_weld} \& \ref{eqn:v_weld} are applied to the top boundary, with equivalent conditions on the bottom boundary with a negative value of $\omega$. (a) $t = 0 t_s$, $a = 0.25$, and $\eta = 1.2$. (b) $t = 2\times 10^7 t_s$, $a = 0.35$, and $\eta = 1.3$. (c) $t = 4\times 10^7 t_s$, $a = 0.45$, and $\eta = 1.4$. (d) $t = 6\times 10^7 t_s$, $a = 0.55$, and $\eta = 1.5$. (e) $t = 8\times 10^7 t_s$, $a = 0.55$, and $\eta = 1.5$. (f) $t = 10^8 t_s$, $a = 0.55$, and $\eta = 1.5$.}
    \label{fig:weld_opp}
\end{figure}

\begin{figure}
    \centering
    \begin{tabular}{cc}
        \begin{subfigure}{\subpanelwid}
            \centering
            \includegraphics[width=\textwidth]{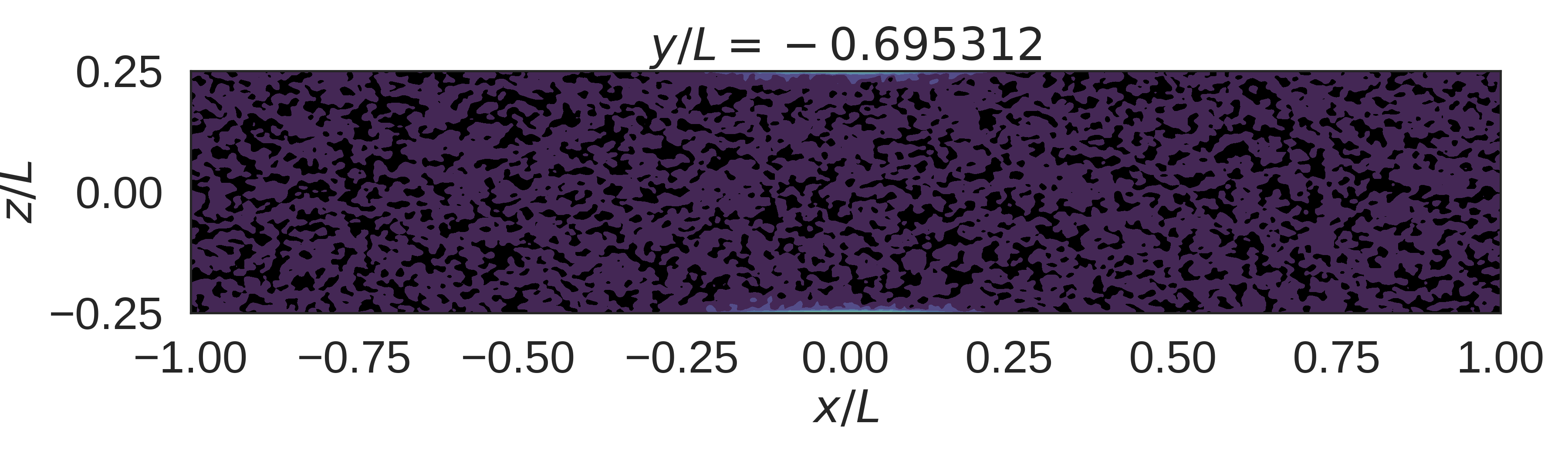}
        \end{subfigure} &
        \begin{subfigure}{\subpanelwid}
            \centering
            \includegraphics[width=\textwidth]{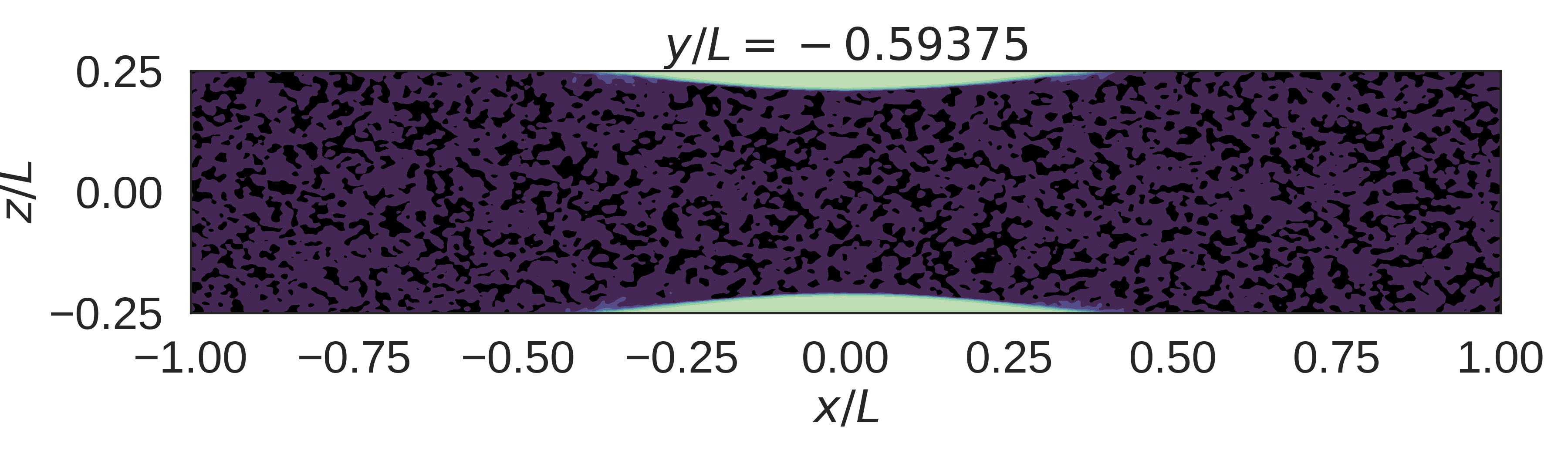}
        \end{subfigure}\\
        \begin{subfigure}{\subpanelwid}
            \centering
            \includegraphics[width=\textwidth]{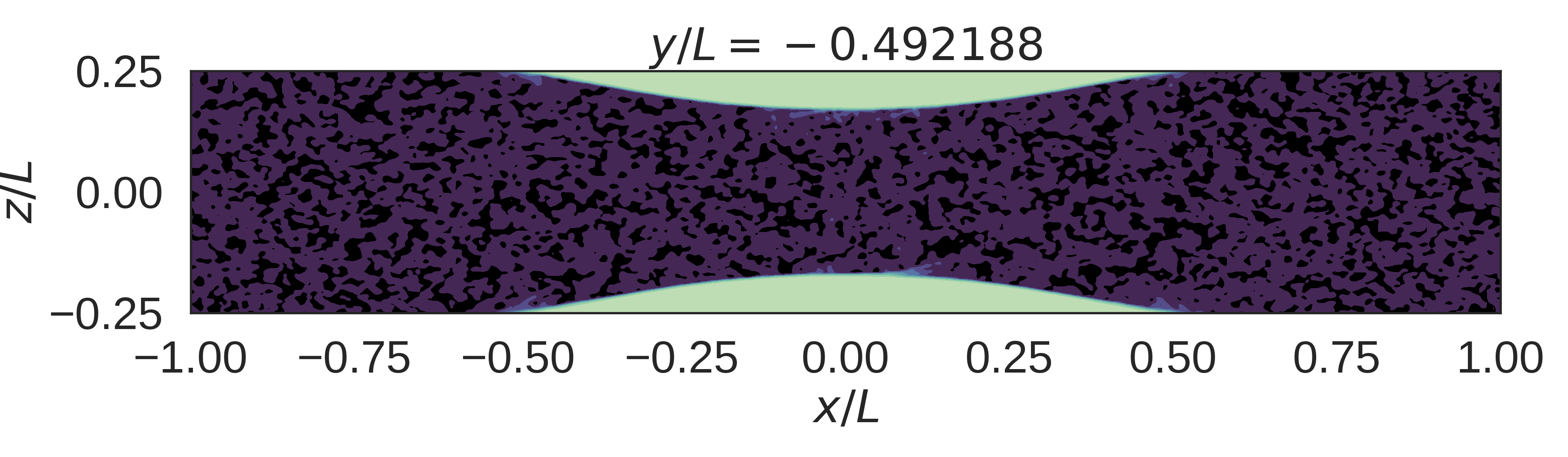}
        \end{subfigure}&
        \begin{subfigure}{\subpanelwid}
            \centering
            \includegraphics[width=\textwidth]{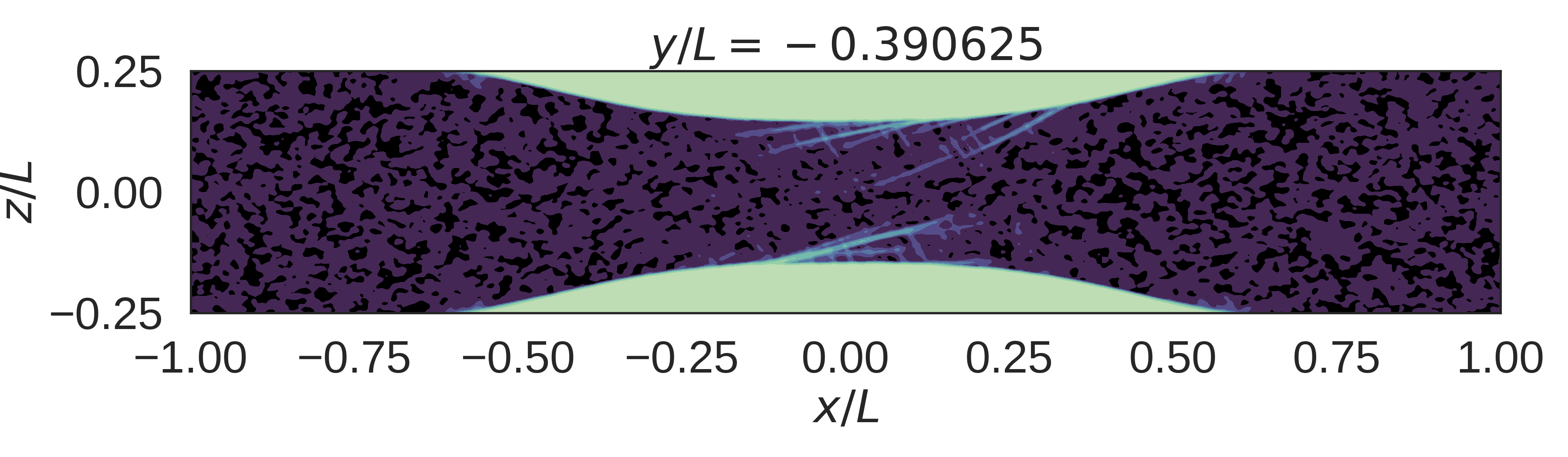}
        \end{subfigure}\\
        \begin{subfigure}{\subpanelwid}
            \centering
            \includegraphics[width=\textwidth]{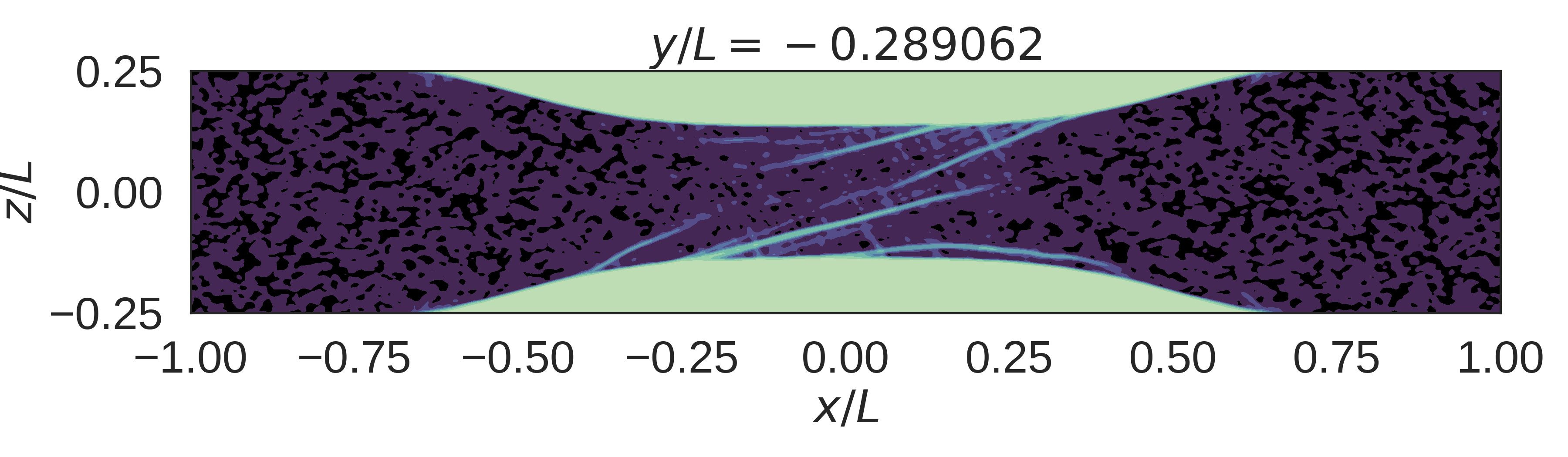}
        \end{subfigure} &
        \begin{subfigure}{\subpanelwid}
            \centering
            \includegraphics[width=\textwidth]{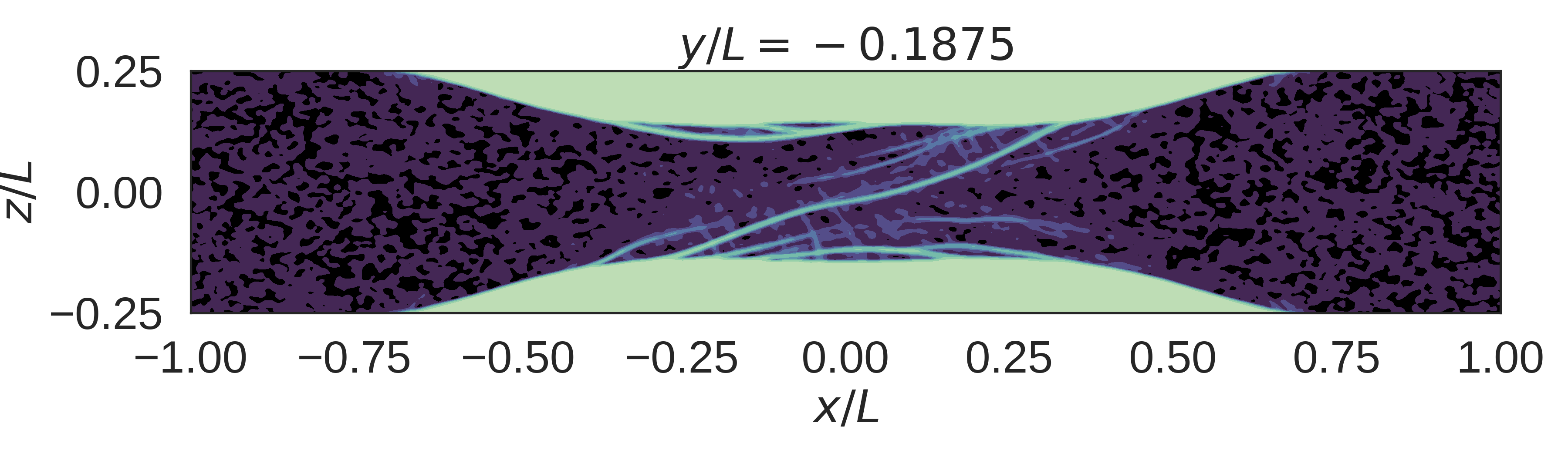}
        \end{subfigure}\\
        \begin{subfigure}{\subpanelwid}
            \centering
            \includegraphics[width=\textwidth]{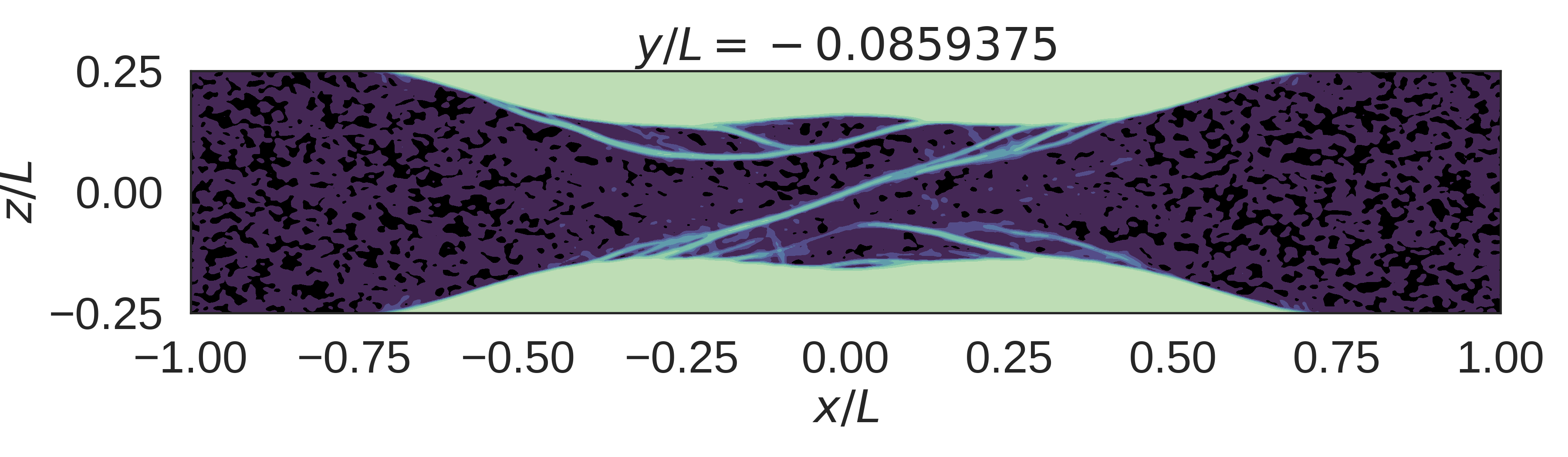}
        \end{subfigure}&
        \begin{subfigure}{\subpanelwid}
            \centering
            \includegraphics[width=\textwidth]{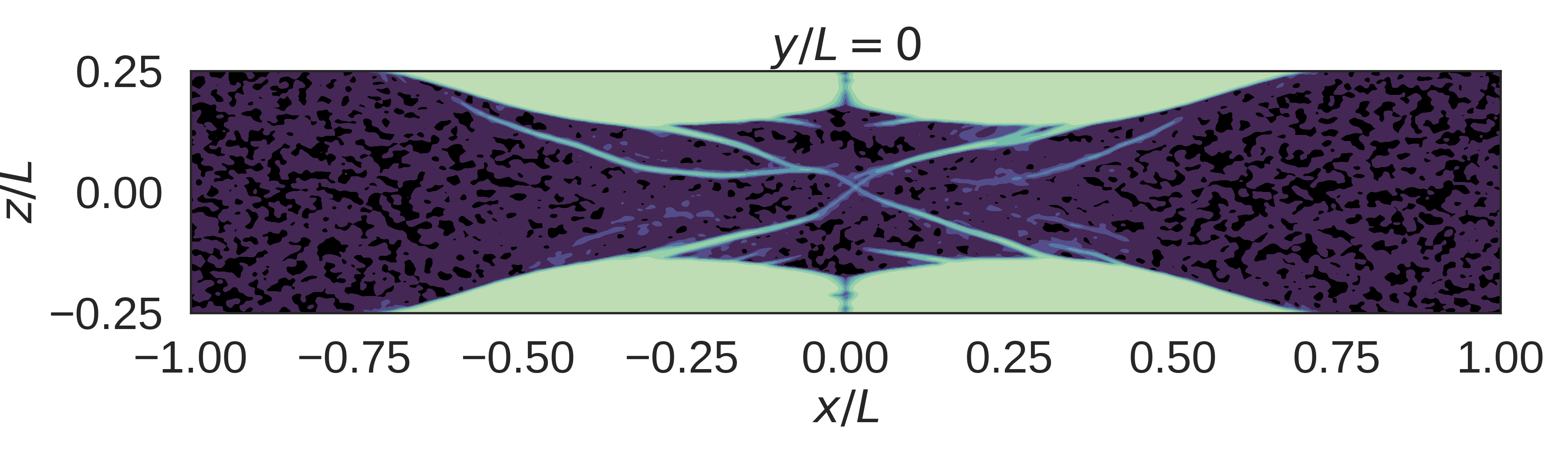}
        \end{subfigure}\\
    \end{tabular}
    \begin{subfigure}{\textwidth}
        \centering
        \includegraphics[width=.6\textwidth]{imgs/rslt_figs/part_one/fric_768_both/contour_weld_cb.png}
    \end{subfigure}
    \vspace{5mm}
    \caption{Cross-sections of the $\chi$ field at $t=5\times 10^7 t_s$ for the friction welding experiment with welding on both the top and bottom. The angular velocity is equal in magnitude but opposite in sign on the top and bottom.}
    \label{fig:weld_opp_contour}
\end{figure}

Finally, we may ask how the structure will change if the top and bottom are anti-rotated with respect to each other. Three-dimensional visualizations for this setup are shown in Fig.~\ref{fig:weld_opp}. A similar hyperboloid region to Fig.~\ref{fig:weld_both} is seen, though it develops later in the simulation, and does not occupy as much of the disc $r^2 < R^2$. This structure is made more clear in the contour plots in Fig.~\ref{fig:weld_opp_contour}, where the connecting tendrils only form near the center of the disc and are thinner.

\section{Conclusion}
\label{sec:conc}
Expanding on prior two-dimensional work~\cite{rycroft15}, we have developed a three-dimensional numerical method for simulating quasi-static elastoplastic materials by analogy with the projection method of Chorin for incompressible fluid dynamics~\cite{chorin67, chorin68}. The method is particularly suitable for stiff materials with small elastic deformation and high elastic wave speeds, where plastic deformation is often quasi-static, and the hypo-elastoplastic assumption is valid. In these materials, the timestep of an explicit method is limited by the CFL condition---making macroscopic timescales and realistic loading rates prohibitive to simulate---while the quasi-static method has no such restriction.

The method was used to examine the properties of shear banding in a metallic glass with the STZ plasticity model, focusing on uniquely three-dimensional features such as curvature development, nucleation off of complex geometries, dependence of shear band structure on the orientation of features in the initial $\chi$ distribution, and high-resolution banding with a random initialization in $\chi$. Essential to these studies was the development of a high performance, parallelized, three-dimensional geometric multigrid solver, along with an MPI and C++-based implementation of the projection algorithm. We provided a weak parallel scaling analysis for our implementation and showed that it exhibits favorable weak scaling over a range of processor counts. The method, though applied to the study of BMGs in this work, is independent of the plasticity model.

The method is based on a correspondence between the variables $(\vv, p)$ in incompressible fluid dynamics and the variables $(\bsig, \vv)$ in quasi-static hypo-elastoplasticity. As part of the development of the method, we introduced an auxiliary vector field $\bPhi$, which plays a role analogous to the auxiliary scalar field $\phi$ used in projection algorithms for fluid dynamics. The choice of $\phi$, along with careful consideration of its boundary conditions, leads to higher-order projection and gauge methods in fluid dynamics.  The analogy between the two auxiliary fields is reminiscent of the analogy between the physical fields that led to the development of the quasi-static method, and it suggests that similar approaches may generalize to quasi-static hypo-elastoplasticity~\cite{brown01, saye_dg1, saye_dg2, saye_gauge}. Gauge methods couple naturally with discontinuous Galerkin discretization (dG) methods, and the theoretical developments presented here thus pave the way for the possibility of similar dG methods in quasi-static hypo-elastoplasticity. Such methods may help resolve the fine-scale features of instabilities like shear bands, and the development of these higher-order methods is an interesting avenue for future work.

\appendix
\section{Connection to the continuous-time framework}
\label{app:alg_conn}
We can make a connection to the general continuous-time framework presented in Sec.~\ref{sec:qs_alg_gen} as follows. By comparison of Eqs.~\ref{eqn:Phi} and \ref{eqn:proj}, we can identify $\bPhi = \dt\ \vv$. Equation~\ref{eqn:D_recov} then says that
\begin{equation}
    \tC : \tD^{n+1} = \tC : \left(\nabla\mathbf{q} + \dt \frac{\p \tD^{n+1}}{\p t}\right).
    \label{eqn:connect}
\end{equation}
Recall that $\mathbf{q}$ is chosen to be the best available approximation to $\vv^{n+1}$, and note that by symmetry of $\tC$,
\begin{equation}
    \tC : \nabla q = \tC : \frac{1}{2}\left(\nabla \mathbf{q} + \nabla \left(\mathbf{q}\right)^\Trans\right).
\end{equation}
Equation \ref{eqn:connect} thus says the following: $\tC : \tD^{n+1}$ is given by the best available guess before the solve for $\vv^{n+1}$ - $\tC : \nabla \mathbf{q}$ - plus an $\bigO(\dt)$ correction constructed via a first-order Taylor expansion in time.

\section*{Acknowledgments}
The authors thank Eran Bouchbinder and Avraham Moriel (Weizmann Institute of
Science) for useful discussions about this work. This work was supported by the
National Science Foundation under Grant Nos.~DMR-1409560 and DMS-1753203.
N.~M.~Boffi was supported by a Department of Energy Computational Science
Graduate Fellowship under Grant No.~DE-FG02-97ER25308. C.~H.~Rycroft was partially supported by the Applied
Mathematics Program of the U.S. DOE Office of Advanced Scientific Computing
Research under contract number DE-AC02-05CH11231.

\section*{References}

\bibliography{elas}

\begin{thebibliography}{100}
\expandafter\ifx\csname url\endcsname\relax
  \def\url#1{\texttt{#1}}\fi
\expandafter\ifx\csname urlprefix\endcsname\relax\def\urlprefix{URL }\fi
\expandafter\ifx\csname href\endcsname\relax
  \def\href#1#2{#2} \def\path#1{#1}\fi

\bibitem{jirasek02a}
M.~Jir{\'a}sek, Objective modeling of strain localization, Revue Fran{\c c}aise
  de G{\'e}nie Civil 6~(6) (2002) 1119--1132.
\newblock \href {https://doi.org/10.1080/12795119.2002.9692735}
  {\path{doi:10.1080/12795119.2002.9692735}}.

\bibitem{hutchinson74}
J.~Hutchinson, J.~Miles, Bifurcation analysis of the onset of necking in an
  elastic/plastic cylinder under uniaxial tension, J. Mech. Phys. Solids 22~(1)
  (1974) 61--71.
\newblock \href {https://doi.org/https://doi.org/10.1016/0022-5096(74)90014-3}
  {\path{doi:https://doi.org/10.1016/0022-5096(74)90014-3}}.

\bibitem{ghosh77}
A.~K. Ghosh, Tensile instability and necking in materials with strain hardening
  and strain-rate hardening, Acta Metallurgica 25~(12) (1977) 1413--1424.
\newblock \href {https://doi.org/10.1016/0001-6160(77)90072-4}
  {\path{doi:10.1016/0001-6160(77)90072-4}}.

\bibitem{tvergaard93}
V.~Tvergaard, Necking in tensile bars with rectangular cross-section, Computer
  Methods in Applied Mechanics and Engineering 103~(1) (1993) 273--290.
\newblock \href {https://doi.org/10.1016/0045-7825(93)90049-4}
  {\path{doi:10.1016/0045-7825(93)90049-4}}.

\bibitem{guo07}
H.~Guo, P.~F. Yan, Y.~B. Wang, J.~Tan, Z.~F. Zhang, M.~L. Sui, E.~Ma, Tensile
  ductility and necking of metallic glass, Nature Materials 6 (2007) 735--739.
\newblock \href {https://doi.org/10.1038/nmat1984}
  {\path{doi:10.1038/nmat1984}}.

\bibitem{rudnicki75}
J.~W. Rudnicki, J.~R. Rice, Conditions for the localization of deformation in
  pressure-sensitive dilatant materials, Journal of the Mechanics and Physics
  of Solids 23~(6) (1975) 371--394.
\newblock \href {https://doi.org/10.1016/0022-5096(75)90001-0}
  {\path{doi:10.1016/0022-5096(75)90001-0}}.

\bibitem{hutchinson81}
J.~Hutchinson, V.~Tvergaard, Shear band formation in plane strain,
  International Journal of Solids and Structures 17~(5) (1981) 451--470.
\newblock \href {https://doi.org/10.1016/0020-7683(81)90053-6}
  {\path{doi:10.1016/0020-7683(81)90053-6}}.

\bibitem{harren88}
S.~Harren, H.~D\`eve, R.~Asaro, Shear band formation in plane strain
  compression, Acta Metallurgica 36~(9) (1988) 2435--2480.
\newblock \href {https://doi.org/10.1016/0001-6160(88)90193-9}
  {\path{doi:10.1016/0001-6160(88)90193-9}}.

\bibitem{conner04}
R.~Conner, Y.~Li, W.~Nix, W.~Johnson, Shear band spacing under bending of
  zr-based metallic glass plates, Acta Materialia 52~(8) (2004) 2429--2434.
\newblock \href {https://doi.org/10.1016/j.actamat.2004.01.034}
  {\path{doi:10.1016/j.actamat.2004.01.034}}.

\bibitem{bei06}
H.~Bei, S.~Xie, E.~P. George, Softening caused by profuse shear banding in a
  bulk metallic glass, Phys. Rev. Lett. 96 (2006) 105503.
\newblock \href {https://doi.org/10.1103/PhysRevLett.96.105503}
  {\path{doi:10.1103/PhysRevLett.96.105503}}.

\bibitem{asaro77}
R.~Asaro, J.~Rice, Strain localization in ductile single crystals, J. Mech.
  Phys. Solids 25~(5) (1977) 309--338.
\newblock \href {https://doi.org/10.1016/0022-5096(77)90001-1}
  {\path{doi:10.1016/0022-5096(77)90001-1}}.

\bibitem{steif82}
P.~Steif, F.~Spaepen, J.~Hutchinson, Strain localization in amorphous metals,
  Acta Metallurgica 30~(2) (1982) 447--455.
\newblock \href {https://doi.org/10.1016/0001-6160(82)90225-5}
  {\path{doi:10.1016/0001-6160(82)90225-5}}.

\bibitem{rycroft15}
C.~H. Rycroft, Y.~Sui, E.~Bouchbinder, An eulerian projection method for
  quasi-static elastoplasticity, Journal of Computational Physics 300 (2015)
  136--166.
\newblock \href {https://doi.org/10.1016/j.jcp.2015.06.046}
  {\path{doi:10.1016/j.jcp.2015.06.046}}.

\bibitem{gurtin10}
M.~E. Gurtin, E.~Fried, L.~Anand, The Mechanics and Thermodynamics of Continua,
  Cambridge University Press, Cambridge, 2010.

\bibitem{xiao06}
H.~Xiao, O.~Bruhns, A.~Meyers, Elastoplasticity beyond small deformations, Acta
  Mechanica 182~(1-2) (2006) 31--111.
\newblock \href {https://doi.org/10.1007/s00707-005-0282-7}
  {\path{doi:10.1007/s00707-005-0282-7}}.

\bibitem{prager60}
W.~Prager, An elementary discussion of definitions of stress rate, Quart. Appl.
  Math. 18 (1960) 403--407.

\bibitem{reina14}
C.~Reina, S.~Conti, Kinematic description of crystal plasticity in the finite
  kinematic framework: A micromechanical understanding of
  $\ten{F}=\ten{F}_e\ten{F}_p$, Journal of the Mechanics and Physics of Solids
  67 (2014) 40--61.
\newblock \href {https://doi.org/10.1016/j.jmps.2014.01.014}
  {\path{doi:10.1016/j.jmps.2014.01.014}}.

\bibitem{truesdell55}
C.~Truesdell, Hypo-elasticity, Indiana Univ. Math. J. 4 (1955) 83--133.

\bibitem{hill58}
R.~Hill, A general theory of uniqueness and stability in elastic--plastic
  solids, Journal of the Mechanics and Physics of Solids 6~(3) (1958) 236--249.
\newblock \href {https://doi.org/10.1016/0022-5096(58)90029-2}
  {\path{doi:10.1016/0022-5096(58)90029-2}}.

\bibitem{dienes79}
J.~Dienes, On the analysis of rotation and stress rate in deforming bodies,
  Acta Mechanica 32~(4) (1979) 217--232.
\newblock \href {https://doi.org/10.1007/BF01379008}
  {\path{doi:10.1007/BF01379008}}.

\bibitem{nagtegaal81}
J.~C. Nagtegaal, J.~E. De~Jong, Some computational aspects of elastic--plastic
  large strain analysis, International Journal for Numerical Methods in
  Engineering 17~(1) (1981) 15--41.
\newblock \href {https://doi.org/10.1002/nme.1620170103}
  {\path{doi:10.1002/nme.1620170103}}.

\bibitem{hughes80}
T.~J.~R. Hughes, J.~Winget, Finite rotation effects in numerical integration of
  rate constitutive equations arising in large-deformation analysis,
  International Journal for Numerical Methods in Engineering 15~(12) (1980)
  1862--1867.
\newblock \href {https://doi.org/10.1002/nme.1620151210}
  {\path{doi:10.1002/nme.1620151210}}.

\bibitem{eterovic90}
A.~L. Eterovic, K.-J. Bathe, A hyperelastic-based large strain elasto-plastic
  constitutive formulation with combined isotropic-kinematic hardening using
  the logarithmic stress and strain measures, International Journal for
  Numerical Methods in Engineering 30~(6) (1990) 1099--1114.
\newblock \href {https://doi.org/10.1002/nme.1620300602}
  {\path{doi:10.1002/nme.1620300602}}.

\bibitem{bell89}
J.~B. Bell, P.~Colella, H.~M. Glaz, A second-order projection method for the
  incompressible {N}avier--{S}tokes equations, Journal of Computational Physics
  85~(2) (1989) 257--283.
\newblock \href {https://doi.org/10.1016/0021-9991(89)90151-4}
  {\path{doi:10.1016/0021-9991(89)90151-4}}.

\bibitem{almgren96}
A.~Almgren, J.~Bell, W.~Szymczak, A numerical method for the incompressible
  {N}avier--{S}tokes equations based on an approximate projection, SIAM Journal
  on Scientific Computing 17~(2) (1996) 358--369.
\newblock \href {https://doi.org/10.1137/S1064827593244213}
  {\path{doi:10.1137/S1064827593244213}}.

\bibitem{sussman99}
M.~Sussman, A.~S. Almgren, J.~B. Bell, P.~Colella, L.~H. Howell, M.~L. Welcome,
  An adaptive level set approach for incompressible two-phase flows, J. Comput.
  Phys. 148~(1) (1999) 81--124.
\newblock \href {https://doi.org/10.1006/jcph.1998.6106}
  {\path{doi:10.1006/jcph.1998.6106}}.

\bibitem{Briggs2000}
W.~L. Briggs, V.~E. Henson, S.~F. McCormick, A Multigrid Tutorial: Second
  Edition, Society for Industrial and Applied Mathematics, Philadelphia, PA,
  USA, 2000.

\bibitem{demmel}
J.~W. Demmel, Applied Numerical Linear Algebra, SIAM, 1997.

\bibitem{courant67}
R.~Courant, K.~Friedrichs, H.~Lewy, On the partial difference equations of
  mathematical physics, IBM Journal of Research and Development 11~(2) (1967)
  215--234.
\newblock \href {https://doi.org/10.1147/rd.112.0215}
  {\path{doi:10.1147/rd.112.0215}}.

\bibitem{bing-2005}
B.~Yang, M.~L. Morrison, P.~K. Liaw, R.~A. Buchanan, G.~Wang, C.~T. Liu,
  M.~Denda, Dynamic evolution of nanoscale shear bands in a bulk-metallic
  glass, Applied Physics Letters 86~(14) (2005) 141904.
\newblock \href {https://doi.org/10.1063/1.1891302}
  {\path{doi:10.1063/1.1891302}}.

\bibitem{chorin67}
A.~J. Chorin, A numerical method for solving incompressible viscous flow
  problems, Journal of Computational Physics 2~(1) (1967) 12--26.
\newblock \href {https://doi.org/10.1016/0021-9991(67)90037-X}
  {\path{doi:10.1016/0021-9991(67)90037-X}}.

\bibitem{chorin68}
A.~J. Chorin, Numerical solution of the {N}avier--{S}tokes equations,
  Mathematics of Computation 22~(104) (1968) 745--762.
\newblock \href {https://doi.org/10.1090/S0025-5718-1968-0242392-2}
  {\path{doi:10.1090/S0025-5718-1968-0242392-2}}.

\bibitem{vanka}
H.~Wobker, S.~Turek, Numerical studies of vanka-type smoothers in computational
  solid mechanics, Advances in Applied Mathematics and Mechanics February Adv.
  Appl. Math. Mech 1 (2009) 29--55.

\bibitem{vanka2}
F.~Suttmeier, An adaptive displacement/pressure finite element scheme for
  treating incompressibility effects in elasto‐plastic materials, Numerical
  Methods for Partial Differential Equations 17 (2001) 369 -- 382.
\newblock \href {https://doi.org/10.1002/num.1017}
  {\path{doi:10.1002/num.1017}}.

\bibitem{almgren_super}
A.~S. Almgren, J.~B. Bell, C.~A. Rendleman, M.~Zingale, Low mach number
  modeling of type ia supernovae. i. hydrodynamics, The Astrophysical Journal
  637~(2) (2006) 922--936.
\newblock \href {https://doi.org/https://doi.org/10.10862F498426}
  {\path{doi:https://doi.org/10.10862F498426}}.

\bibitem{bell_nuclear}
J.~Bell, M.~Day, C.~Rendleman, S.~Woosley, M.~Zingale, Adaptive low mach number
  simulations of nuclear flame microphysics, Journal of Computational Physics
  195~(2) (2004) 677 -- 694.
\newblock \href {https://doi.org/https://doi.org/10.1016/j.jcp.2003.10.035}
  {\path{doi:https://doi.org/10.1016/j.jcp.2003.10.035}}.

\bibitem{majda_seth}
A.~Majda, J.~Sethian, The derivation and numerical solution of the equations
  for zero mach number combustion, Combustion Science and Technology 42~(3-4)
  (1985) 185--205.
\newblock \href {https://doi.org/https://doi.org/10.1080/00102208508960376}
  {\path{doi:https://doi.org/10.1080/00102208508960376}}.

\bibitem{klein_maj1}
S.~Klainerman, A.~Majda, Compressible and incompressible fluids, Communications
  on Pure and Applied Mathematics 35~(5) (1982) 629--651.
\newblock \href {https://doi.org/10.1002/cpa.3160350503}
  {\path{doi:10.1002/cpa.3160350503}}.

\bibitem{klein_maj2}
S.~Klainerman, A.~Majda, Singular limits of quasilinear hyperbolic systems with
  large parameters and the incompressible limit of compressible fluids,
  Communications on Pure and Applied Mathematics 34~(4) (1981) 481--524.
\newblock \href {https://doi.org/10.1002/cpa.3160340405}
  {\path{doi:10.1002/cpa.3160340405}}.

\bibitem{falk98}
M.~L. Falk, J.~S. Langer, Dynamics of viscoplastic deformation in amorphous
  solids, Phys. Rev. E 57~(6) (1998) 7192--7205.
\newblock \href {https://doi.org/10.1103/PhysRevE.57.7192}
  {\path{doi:10.1103/PhysRevE.57.7192}}.

\bibitem{bouchbinder07}
E.~Bouchbinder, J.~S. Langer, I.~Procaccia, Athermal shear-transformation-zone
  theory of amorphous plastic deformation. {I}. {B}asic principles, Phys. Rev.
  E 75~(3) (2007) 036107.
\newblock \href {https://doi.org/10.1103/PhysRevE.75.036107}
  {\path{doi:10.1103/PhysRevE.75.036107}}.

\bibitem{langer08}
J.~S. Langer, Shear-transformation-zone theory of plastic deformation near the
  glass transition, Phys. Rev. E 77~(2) (2008) 021502.
\newblock \href {https://doi.org/10.1103/PhysRevE.77.021502}
  {\path{doi:10.1103/PhysRevE.77.021502}}.

\bibitem{bouchbinder09}
E.~Bouchbinder, J.~S. Langer, Nonequilibrium thermodynamics of driven amorphous
  materials. {I}. {I}nternal degrees of freedom and volume deformation, Phys.
  Rev. E 80~(3) (2009) 031131.
\newblock \href {https://doi.org/10.1103/PhysRevE.80.031131}
  {\path{doi:10.1103/PhysRevE.80.031131}}.

\bibitem{bmg-struc}
J.~J. Kruzic, Bulk metallic glasses as structural materials: A review, Advanced
  Engineering Materials 18~(8) (2016) 1308--1331.
\newblock \href {https://doi.org/10.1002/adem.201600066}
  {\path{doi:10.1002/adem.201600066}}.

\bibitem{huf-def}
T.~C. Hufnagel, T.~Jiao, Y.~Li, L.-Q. Xing, K.~T. Ramesh, Deformation and
  failure of $\text{Zr}_{57}\text{Ti}_5\text{Cu}_{20}\text{Ni}_8\text{Al}_{10}$
  bulk metallic glass under quasi-static and dynamic compression, Journal of
  Materials Research 17~(6) (2002) 1441--1445.
\newblock \href {https://doi.org/10.1557/JMR.2002.0214}
  {\path{doi:10.1557/JMR.2002.0214}}.

\bibitem{manning07}
M.~L. Manning, J.~S. Langer, J.~M. Carlson, Strain localization in a shear
  transformation zone model for amorphous solids, Phys. Rev. E 76~(5) (2007)
  056106.
\newblock \href {https://doi.org/10.1103/PhysRevE.76.056106}
  {\path{doi:10.1103/PhysRevE.76.056106}}.

\bibitem{manning09}
M.~L. Manning, E.~G. Daub, J.~S. Langer, J.~M. Carlson, Rate-dependent shear
  bands in a shear-transformation-zone model of amorphous solids, Phys. Rev. E
  79~(1) (2009) 016110.
\newblock \href {https://doi.org/10.1103/PhysRevE.79.016110}
  {\path{doi:10.1103/PhysRevE.79.016110}}.

\bibitem{sun-2013}
B.~A. Sun, S.~Pauly, J.~Hu, W.~H. Wang, U.~K\"uhn, J.~Eckert, Origin of
  intermittent plastic flow and instability of shear band sliding in bulk
  metallic glasses, Phys. Rev. Lett. 110 (2013) 225501.
\newblock \href {https://doi.org/10.1103/PhysRevLett.110.225501}
  {\path{doi:10.1103/PhysRevLett.110.225501}}.

\bibitem{antonaglia-2014}
J.~Antonaglia, W.~J. Wright, X.~Gu, R.~R. Byer, T.~C. Hufnagel, M.~LeBlanc,
  J.~T. Uhl, K.~A. Dahmen, Bulk metallic glasses deform via slip avalanches,
  Phys. Rev. Lett. 112 (2014) 155501.
\newblock \href {https://doi.org/10.1103/PhysRevLett.112.155501}
  {\path{doi:10.1103/PhysRevLett.112.155501}}.

\bibitem{falk_langer_rev}
M.~L. Falk, J.~Langer, Deformation and failure of amorphous, solidlike
  materials, Annual Review of Condensed Matter Physics 2~(1) (2011) 353--373.
\newblock \href {https://doi.org/{10.1146/annurev-conmatphys-062910-140452}}
  {\path{doi:{10.1146/annurev-conmatphys-062910-140452}}}.

\bibitem{eastgate03}
L.~O. Eastgate, J.~S. Langer, L.~Pechenik, Dynamics of large-scale plastic
  deformation and the necking instability in amorphous solids, Phys. Rev. Lett.
  90~(4) (2003) 045506.
\newblock \href {https://doi.org/10.1103/PhysRevLett.90.045506}
  {\path{doi:10.1103/PhysRevLett.90.045506}}.

\bibitem{rycroft12}
C.~H. Rycroft, F.~Gibou, Simulations of a stretching bar using a plasticity
  model from the shear transformation zone theory, Journal of Computational
  Physics 231~(5) (2012) 2155--2179.
\newblock \href {https://doi.org/10.1016/j.jcp.2011.10.009}
  {\path{doi:10.1016/j.jcp.2011.10.009}}.

\bibitem{moriel18}
A.~Moriel, E.~Bouchbinder, Necking instabilities in elastoviscoplastic
  materials, Phys. Rev. Materials 2 (2018) 073602.
\newblock \href {https://doi.org/10.1103/PhysRevMaterials.2.073602}
  {\path{doi:10.1103/PhysRevMaterials.2.073602}}.

\bibitem{rycroft12b}
C.~H. Rycroft, E.~Bouchbinder, Fracture toughness of metallic glasses:
  Annealing-induced embrittlement, Phys. Rev. Lett. 109 (2012) 194301.
\newblock \href {https://doi.org/10.1103/PhysRevLett.109.194301}
  {\path{doi:10.1103/PhysRevLett.109.194301}}.

\bibitem{vasoya16}
M.~Vasoya, C.~H. Rycroft, E.~Bouchbinder, Notch fracture toughness of glasses:
  Dependence on rate, age, and geometry, Phys. Rev. Applied 6 (2016) 024008.
\newblock \href {https://doi.org/10.1103/PhysRevApplied.6.024008}
  {\path{doi:10.1103/PhysRevApplied.6.024008}}.

\bibitem{schroers2018}
J.~Ketkaew, W.~Chen, H.~Wang, A.~Datye, M.~Fan, G.~Pereira, U.~D. Schwarz,
  Z.~Liu, R.~Yamada, W.~Dmowski, M.~D. Shattuck, C.~S. O'Hern, T.~Egami,
  E.~Bouchbinder, J.~Schroers,
  \href{https://doi.org/10.1038/s41467-018-05682-8}{Mechanical glass transition
  revealed by the fracture toughness of metallic glasses}, Nature
  Communications 9~(1) (2018) 3271.
\newblock \href {https://doi.org/10.1038/s41467-018-05682-8}
  {\path{doi:10.1038/s41467-018-05682-8}}.
\newline\urlprefix\url{https://doi.org/10.1038/s41467-018-05682-8}

\bibitem{argon79}
A.~S. Argon, Plastic deformation in metallic glasses, Acta Metallurgica 27
  (1979) 47--58.

\bibitem{Wisitsorasak1287}
A.~Wisitsorasak, P.~G. Wolynes, Dynamical theory of shear bands in structural
  glasses, Proc. Natl. Acad. Sci. 114~(6) (2017) 1287--1292.
\newblock \href {https://doi.org/10.1073/pnas.1620399114}
  {\path{doi:10.1073/pnas.1620399114}}.

\bibitem{hypo-elas1}
H.~Xiao, O.~T. Bruhns, A.~Meyers, Hypo-elasticity model based upon the
  logarithmic stress rate, Journal of Elasticity 47~(1) (1997) 51--68.
\newblock \href {https://doi.org/10.1023/A:1007356925912}
  {\path{doi:10.1023/A:1007356925912}}.

\bibitem{hypo-elas2}
C.~Truesdell, Hypo‐elastic shear, Journal of Applied Physics 27~(5) (1956)
  441--447.
\newblock \href {https://doi.org/10.1063/1.1722399}
  {\path{doi:10.1063/1.1722399}}.

\bibitem{hypo-elas3}
J.~Bardet, Finite element analysis of surface instability in hypo-elastic
  solids, Computer Methods in Applied Mechanics and Engineering 78~(3) (1990)
  273--296.
\newblock \href {https://doi.org/10.1016/0045-7825(90)90002-4}
  {\path{doi:10.1016/0045-7825(90)90002-4}}.

\bibitem{hajarolasvadi17}
S.~Hajarolasvadi, A.~E. Elbanna, {A new hybrid numerical scheme for modelling
  elastodynamics in unbounded media with near-source heterogeneities},
  Geophysical Journal International 211~(2) (2017) 851--864.
\newblock \href {https://doi.org/10.1093/gji/ggx337}
  {\path{doi:10.1093/gji/ggx337}}.

\bibitem{ma18}
X.~Ma, A.~Elbanna, Strain localization in dry sheared granular materials: A
  compactivity-based approach, Phys. Rev. E 98 (2018) 022906.
\newblock \href {https://doi.org/10.1103/PhysRevE.98.022906}
  {\path{doi:10.1103/PhysRevE.98.022906}}.

\bibitem{rate-ind1}
T.~J. Hughes, Numerical Implementation of Constitutive Models: Rate-Independent
  Deviatoric Plasticity, Springer, 1984.

\bibitem{rate-ind2}
L.~Anand, M.~Kothari, A computational procedure for rate-independent crystal
  plasticity, Journal of the Mechanics and Physics of Solids 44~(4) (1996)
  525--558.
\newblock \href {https://doi.org/10.1016/0022-5096(96)00001-4}
  {\path{doi:10.1016/0022-5096(96)00001-4}}.

\bibitem{rate-ind3}
G.~Puglisi, L.~Truskinovsky, Thermodynamics of rate-independent plasticity,
  Journal of the Mechanics and Physics of Solids 53~(3) (2005) 655--679.
\newblock \href {https://doi.org/10.1016/j.jmps.2004.08.004}
  {\path{doi:10.1016/j.jmps.2004.08.004}}.

\bibitem{rate-ind4}
J.~Simo, R.~Taylor, Consistent tangent operators for rate-independent
  elastoplasticity, Computer Methods in Applied Mechanics and Engineering
  48~(1) (1985) 101--118.
\newblock \href {https://doi.org/10.1016/0045-7825(85)90070-2}
  {\path{doi:10.1016/0045-7825(85)90070-2}}.

\bibitem{saye_dg1}
R.~Saye, Implicit mesh discontinuous galerkin methods and interfacial gauge
  methods for high-order accurate interface dynamics, with applications to
  surface tension dynamics, rigid body fluid–structure interaction, and free
  surface flow: Part i, Journal of Computational Physics 344 (2017) 647--682.
\newblock \href {https://doi.org/10.1016/j.jcp.2017.04.076}
  {\path{doi:10.1016/j.jcp.2017.04.076}}.

\bibitem{saye_dg2}
R.~Saye, Implicit mesh discontinuous galerkin methods and interfacial gauge
  methods for high-order accurate interface dynamics, with applications to
  surface tension dynamics, rigid body fluid–structure interaction, and free
  surface flow: Part ii, Journal of Computational Physics 344 (2017) 683--723.
\newblock \href {https://doi.org/10.1016/j.jcp.2017.05.003}
  {\path{doi:10.1016/j.jcp.2017.05.003}}.

\bibitem{saye_gauge}
R.~Saye, Interfacial gauge methods for incompressible fluid dynamics, Science
  Advances 2~(6) (2016).
\newblock \href {https://doi.org/10.1126/sciadv.1501869}
  {\path{doi:10.1126/sciadv.1501869}}.

\bibitem{brown01}
D.~L. Brown, R.~Cortez, M.~L. Minion, Accurate projection methods for the
  incompressible {Navier--Stokes} equations, J. Comput. Phys. 168~(2) (2001)
  464--499.
\newblock \href {https://doi.org/10.1006/jcph.2001.6715}
  {\path{doi:10.1006/jcph.2001.6715}}.

\bibitem{min06}
C.~Min, F.~Gibou, A second order accurate projection method for the
  incompressible {N}avier--{S}tokes equations on non-graded adaptive grids,
  Journal of Computational Physics 219~(2) (2006) 912--929.
\newblock \href {https://doi.org/10.1016/j.jcp.2006.07.019}
  {\path{doi:10.1016/j.jcp.2006.07.019}}.

\bibitem{zhang-2006}
Y.~Zhang, A.~L. Greer, Thickness of shear bands in metallic glasses, Applied
  Physics Letters 89~(7) (2006) 071907.
\newblock \href {https://doi.org/10.1063/1.2336598}
  {\path{doi:10.1063/1.2336598}}.

\bibitem{greer-2013}
A.~Greer, Y.~Cheng, E.~Ma, Shear bands in metallic glasses, Materials Science
  and Engineering: R: Reports 74~(4) (2013) 71--132.
\newblock \href {https://doi.org/10.1016/j.mser.2013.04.001}
  {\path{doi:10.1016/j.mser.2013.04.001}}.

\bibitem{schuh-2007}
C.~A. Schuh, T.~C. Hufnagel, U.~Ramamurty, Mechanical behavior of amorphous
  alloys, Acta Materialia 55~(12) (2007) 4067--4109.
\newblock \href {https://doi.org/10.1016/j.actamat.2007.01.052}
  {\path{doi:10.1016/j.actamat.2007.01.052}}.

\bibitem{maass15}
R.~Maa\ss, J.~F. L\"offler, Shear-band dynamics in metallic glasses, Advanced
  Functional Materials 25~(16) (2015) 2353--2368.
\newblock \href {https://doi.org/10.1002/adfm.201404223}
  {\path{doi:10.1002/adfm.201404223}}.

\bibitem{maas-2014}
R.~Maaß, K.~Samwer, W.~Arnold, C.~A. Volkert, A single shear band in a
  metallic glass: Local core and wide soft zone, Applied Physics Letters
  105~(17) (2014) 171902.
\newblock \href {https://doi.org/10.1063/1.4900791}
  {\path{doi:10.1063/1.4900791}}.

\bibitem{lubliner08}
J.~Lubliner, Plasticity Theory, Dover, New York, 2008.

\bibitem{brown_acc_proj}
D.~L.~Brown, Accuracy of projection methods for the incompressible
  navier-stokes equations (01 2003).
\newblock \href {https://doi.org/10.1142/9789812796837_0006}
  {\path{doi:10.1142/9789812796837_0006}}.

\bibitem{Summers1881}
D.~M. Summers, A.~J. Chorin, Numerical vorticity creation based on impulse
  conservation, Proceedings of the National Academy of Sciences 93~(5) (1996)
  1881--1885.
\newblock \href {https://doi.org/10.1073/pnas.93.5.1881}
  {\path{doi:10.1073/pnas.93.5.1881}}.

\bibitem{Cortez1996}
R.~Cortez, An impulse-based approximation of fluid motion due to boundary
  forces, Journal of Computational Physics 123~(2) (1996) 341 -- 353.
\newblock \href {https://doi.org/https://doi.org/10.1006/jcph.1996.0028}
  {\path{doi:https://doi.org/10.1006/jcph.1996.0028}}.

\bibitem{Buttke1993}
T.~F. Buttke, Velicity Methods: Lagrangian Numerical Methods which Preserve the
  Hamiltonian Structure of Incompressible Fluid Flow, 1993, pp. 39--57.
\newblock \href {https://doi.org/10.1007/978-94-015-8137-0_3}
  {\path{doi:10.1007/978-94-015-8137-0_3}}.

\bibitem{kamrin14a}
K.~Kamrin, E.~Bouchbinder, Two-temperature continuum thermomechanics of
  deforming amorphous solids, Journal of the Mechanics and Physics of Solids 73
  (2014) 269--288.
\newblock \href {https://doi.org/10.1016/j.jmps.2014.09.009}
  {\path{doi:10.1016/j.jmps.2014.09.009}}.

\bibitem{bouch_linear}
E.~Bouchbinder, J.~S. Langer, Shear-transformation-zone theory of linear glassy
  dynamics, Phys. Rev. E 83 (2011) 061503.
\newblock \href {https://doi.org/10.1103/PhysRevE.83.061503}
  {\path{doi:10.1103/PhysRevE.83.061503}}.

\bibitem{bouch_eff_dyn}
E.~Bouchbinder, Effective temperature dynamics in an athermal amorphous
  plasticity theory, Phys. Rev. E 77 (2008) 051505.

\bibitem{cugli1}
D.~Loi, S.~Mossa, L.~F. Cugliandolo, Effective temperature of active matter,
  Phys. Rev. E 77 (2008) 051111.
\newblock \href {https://doi.org/10.1103/PhysRevE.77.051111}
  {\path{doi:10.1103/PhysRevE.77.051111}}.

\bibitem{cugli2}
L.~F. Cugliandolo, The effective temperature, Journal of Physics A:
  Mathematical and Theoretical 44~(48) (2011) 483001.

\bibitem{bouchbinder07b}
E.~Bouchbinder, J.~S. Langer, I.~Procaccia, Athermal shear-transformation-zone
  theory of amorphous plastic deformation. {II}.{A}nalysis of simulated
  amorphous silicon, Phys. Rev. E 75~(3) (2007) 036108.
\newblock \href {https://doi.org/10.1103/PhysRevE.75.036108}
  {\path{doi:10.1103/PhysRevE.75.036108}}.

\bibitem{leveque_fv}
R.~J. LeVeque, Finite Volume Methods for Hyperbolic Problems, Cambridge
  University Press, 2002.

\bibitem{leveque_fd}
R.~J. LeVeque, Finite Difference Methods for Ordinary and Partial Differential
  Equations, Cambridge University Press, 2002.

\bibitem{shu88}
C.-W. Shu, S.~Osher, Efficient implementation of essentially non-oscillatory
  shock-capturing schemes, J. Comp. Phys. 77~(2) (1988) 439--471.
\newblock \href {https://doi.org/10.1016/0021-9991(88)90177-5}
  {\path{doi:10.1016/0021-9991(88)90177-5}}.

\bibitem{openmpi}
E.~Gabriel, G.~E. Fagg, G.~Bosilca, T.~Angskun, J.~J. Dongarra, J.~M. Squyres,
  V.~Sahay, P.~Kambadur, B.~Barrett, A.~Lumsdaine, R.~H. Castain, D.~J. Daniel,
  R.~L. Graham, T.~S. Woodall, Open {MPI}: Goals, concept, and design of a next
  generation {MPI} implementation, in: Proceedings, 11th European PVM/MPI
  Users' Group Meeting, Budapest, Hungary, 2004, pp. 97--104.

\bibitem{xu92}
J.~Xu, Iterative methods by space decomposition and subspace correction, SIAM
  Review 34~(4) (1992) 581--613.
\newblock \href {https://doi.org/10.1137/1034116} {\path{doi:10.1137/1034116}}.

\bibitem{rycroft13}
M.~Theillard, C.~H. Rycroft, F.~Gibou, A multigrid method on non-graded
  adaptive octree and quadtree cartesian grids, Journal of Scientific Computing
  55~(1) (2013) 1--15.
\newblock \href {https://doi.org/10.1007/s10915-012-9619-2}
  {\path{doi:10.1007/s10915-012-9619-2}}.

\bibitem{li-2002}
J.~Li, Z.~L. Wang, T.~C. Hufnagel, Characterization of nanometer-scale defects
  in metallic glasses by quantitative high-resolution transmission electron
  microscopy, Phys. Rev. B 65 (2002) 144201.
\newblock \href {https://doi.org/10.1103/PhysRevB.65.144201}
  {\path{doi:10.1103/PhysRevB.65.144201}}.

\bibitem{hinkle15}
A.~R. Hinkle, C.~H. Rycroft, M.~D. Shields, M.~L. Falk, Coarse graining
  atomistic simulations of plastically deforming amorphous solids, Phys. Rev. E
  95 (2017) 053001.
\newblock \href {https://doi.org/10.1103/PhysRevE.95.053001}
  {\path{doi:10.1103/PhysRevE.95.053001}}.

\bibitem{multi-bands}
B.~Sun, S.~Pauly, J.~Tan, M.~Stoica, W.~Wang, U.~Kühn, J.~Eckert, Serrated
  flow and stick–slip deformation dynamics in the presence of shear-band
  interactions for a zr-based metallic glass, Acta Materialia 60~(10) (2012)
  4160--4171.
\newblock \href {https://doi.org/10.1016/j.actamat.2012.04.013}
  {\path{doi:10.1016/j.actamat.2012.04.013}}.

\bibitem{weld1}
H.-S. Shin, Tool geometry effect on the characteristics of dissimilar friction
  stir spot welded bulk metallic glass to lightweight alloys, Journal of Alloys
  and Compounds 586 (2014) S50 -- S55.
\newblock \href {https://doi.org/https://doi.org/10.1016/j.jallcom.2012.12.031}
  {\path{doi:https://doi.org/10.1016/j.jallcom.2012.12.031}}.

\bibitem{weld2}
C.~Tan, Z.~Jiang, L.~Li, Y.~Chen, X.~Chen, Microstructural evolution and
  mechanical properties of dissimilar al–cu joints produced by friction stir
  welding, Materials \& Design 51 (2013) 466 -- 473.
\newblock \href {https://doi.org/https://doi.org/10.1016/j.matdes.2013.04.056}
  {\path{doi:https://doi.org/10.1016/j.matdes.2013.04.056}}.

\bibitem{weld3}
W.~Li, A.~Vairis, M.~Preuss, T.~Ma, Linear and rotary friction welding review,
  International Materials Reviews 61~(2) (2016) 71--100.
\newblock \href {https://doi.org/https://doi.org/10.1080/09506608.2015.1109214}
  {\path{doi:https://doi.org/10.1080/09506608.2015.1109214}}.

\end{thebibliography}

\end{document}